\documentclass[longauth]{aa}
\usepackage{graphicx}
\usepackage[varg]{txfonts}
\usepackage{academicons}
\usepackage{hyperref}
\usepackage{xcolor}
\usepackage{ulem}
\usepackage{amsmath}

\begin{document} 


\newcommand{\FeII}{[\ion{Fe}{ii}]}
\newcommand{\FeIIp}{\ion{Fe}{ii}}
\newcommand{\TiII}{[\ion{Ti}{ii}]}
\newcommand{\SII}{[\ion{S}{ii}]}
\newcommand{\OI}{[\ion{O}{i}]}
\newcommand{\OIp}{\ion{O}{i}}
\newcommand{\PII}{[\ion{P}{ii}]}
\newcommand{\NI}{[\ion{N}{i}]}
\newcommand{\NII}{[\ion{N}{ii}]}
\newcommand{\NIp}{\ion{N}{i}}
\newcommand{\NiII}{[\ion{Ni}{ii}]}
\newcommand{\CaIIp}{\ion{Ca}{ii}}
\newcommand{\PI}{[\ion{P}{i}]}
\newcommand{\CIp}{\ion{C}{i}}
\newcommand{\HeI}{\ion{He}{i}}
\newcommand{\MgIp}{\ion{Mg}{i}}
\newcommand{\MgIIp}{\ion{Mg}{ii}}
\newcommand{\NaI}{\ion{Na}{i}}
\newcommand{\HI}{\ion{H}{i}}
\newcommand{\brg}{Br$\gamma$}
\newcommand{\Brg}{Br$\gamma$}
\newcommand{\Pab}{Pa$\beta$}

\newcommand{\Ftot}{F$_{tot}$}
\newcommand{\Fres}{F$_{res}$}
\newcommand{\Fcomp}{F$_{comp}$}
\newcommand{\Fhalo}{F$_{halo}$}
\newcommand{\Rres}{R$_{res}$}
\newcommand{\Rcomp}{R$_{comp}$}
\newcommand{\Rhalo}{R$_{halo}$}

\newcommand{\macc}{$\dot{M}_{acc}$}
\newcommand{\lacc}{L$_{acc}$}
\newcommand{\lbol}{L$_{bol}$}
\newcommand{\mjet}{$\dot{M}_{jet}$}
\newcommand{\mh}{$\dot{M}_{H_2}$}
\newcommand{\Ne}{n$_e$}
\newcommand{\h}{H$_2$}
\newcommand{\kms}{km\,s$^{-1}$}
\newcommand{\um}{$\mu$m}
\newcommand{\lam}{$\lambda$}
\newcommand{\msyr}{M$_{\odot}$\,yr$^{-1}$}
\newcommand{\Av}{A$_V$}
\newcommand{\msun}{M$_{\odot}$}
\newcommand{\lsun}{L$_{\odot}$}
\newcommand{\rsun}{R$_{\odot}$}
\newcommand{\Mstar}{M$_{\ast}$}
\newcommand{\Lstar}{L$_{\ast}$}
\newcommand{\Rstar}{R$_{\ast}$}
\newcommand{\RBrg}{R($Br_\gamma$)}
\newcommand{\Rcont}{R$_{cont}$}
\newcommand{\Rtrunc}{R$_{tr}$}
\newcommand{\Rcorot}{R$_{corot}$}
\newcommand{\cm}{cm$^{-3}$}
\newcommand{\ergscm}{erg\,s$^{-1}$\,cm$^{-2}$}

\newcommand{\vsini}{v$sin$i}

\newcommand{\tw}{TW\,Hya}

\newcommand{\bet}{$\beta$}
\newcommand{\alfa}{$\alpha$}


\hyphenation{mo-le-cu-lar pre-vious e-vi-den-ce di-ffe-rent pa-ra-me-ters ex-ten-ding a-vai-la-ble excited}

   \title{The GRAVITY young stellar object survey}

   \subtitle{XII. The hot gas disk component in Herbig Ae/Be stars}

\author{GRAVITY Collaboration(\thanks{GRAVITY is developed in a
collaboration by the Max Planck Institute for Extraterrestrial Physics,
LESIA of Paris Observatory and IPAG of Université Grenoble Alpes / CNRS,
the Max Planck Institute for Astronomy, the University of Cologne, the
Centro Multidisciplinar de Astrofisica Lisbon and Porto, and the European
Southern Observatory.}): {R. Garcia Lopez} \inst{1,2} 
\and A. Natta \inst{3} 
\and R. Fedriani \inst{4,5} 
\and  A. Caratti o Garatti \inst{6,2}
\and J. Sanchez-Bermudez \inst{16,2}
\and K. Perraut \inst{14}
\and C. Dougados \inst{14}
\and Y.-I. Bouarour\inst{1}
\and J. Bouvier \inst{14}
\and W. Brandner \inst{2}
\and P. Garcia \inst{9, 8}
\and M. Koutoulaki\inst{17}
\and L. Labadie \inst{7}
\and H. Linz \inst{2}
%
\and E. Al\'ecian \inst{14}
\and M. Benisty \inst{14}
\and J.-P. Berger \inst{14}
\and G. Bourdarot \inst{10}
\and P. Caselli \inst{10}
\and Y. Cl\'{e}net \inst{11} 
\and P. T. de Zeeuw \inst{13}
\and R. Davies \inst{10}
\and A. Eckart \inst{7, 12}
\and F. Eisenhauer \inst{10}
\and N. M. F\"{o}rster-Schreiber \inst{10}
\and E. Gendron \inst{11}
\and S. Gillessen \inst{10}
\and S. Grant \inst{10}
\and Th. Henning \inst{2}
\and P. Kervella \inst{11}
\and S. Lacour \inst{11}
\and V. Lapeyrère \inst{11} 
\and J.-B. Le Bouquin \inst{14}
\and D. Lutz \inst{10}
\and F. Mang \inst{10}
\and H. Nowacki \inst{14}
\and T. Ott \inst{10}
\and T. Paumard \inst{11}
\and G. Perrin \inst{11}
\and J. Shangguan \inst{10}
\and T. Shimizu \inst{10}
\and A. Soulain \inst{14}
\and C. Straubmeier \inst{7}
\and E. Sturm \inst{10}
\and L. Tacconi \inst{10}
\and E. F. van Dishoeck \inst{13}
\and F. Vincent \inst{11}
\and F. Widmann \inst{10}
}

   \institute{School of Physics, University College Dublin, Belfield, Dublin 4, Ireland\\
   \email{rebeca.garcialopez@ucd.ie}
   \and Max Planck Institute for Astronomy, K\"{o}nigstuhl 17, Heidelberg, Germany, D-69117 
    \and  Dublin Institute for Advanced Studies, 31 Fitzwilliam Place, D02\,XF86 Dublin, Ireland
    \and Instituto de Astrof\'isica de Andaluc\'ia, CSIC, Glorieta de la Astronom\'ia s/n, E-18008 Granada, Spain
    \and Dept. of Space, Earth \& Environment, Chalmers University of Technology, SE-412 93 Gothenburg, Sweden
  \and INAF-Osservatorio Astronomico di Capodimonte, via Moiariello 16, 80131 Napoli, Italy
  \and  I. Physikalisches Institut, Universität zu K\"{o}ln, Z\"{u}lpicher Str. 77, 50937, K\"{o}ln, Germany
  \and  Faculdade de Engenharia, Universidade do Porto, Rua Dr. Roberto Frias, P-4200-465 Porto, Portugal
  \and  CENTRA, Centro de Astrof\'{i}sica e Gravitação, Instituto Superior T\'{e}cnico, Avenida Rovisco Pais 1, P-1049 Lisboa, Portugal
  \and  Max Planck Institute for Extraterrestrial Physics, Giessenbachstrasse, 85741 Garching bei M\"{u}nchen, Germany
  \and  LESIA, Observatoire de Paris, Universit\'{e} PSL, CNRS, Sorbonne Université, Universit\'{e} de Paris, 5 place Jules Janssen, 92195 Meudon, France
  \and  Max-Planck-Institute for Radio Astronomy, Auf dem H\"{u}gel 69, 53121 Bonn, Germany
  \and  Leiden Observatory, Leiden University, PO Box 9513, 2300 RA Leiden, The Netherlands
  \and  Univ. Grenoble Alpes, CNRS, IPAG, F-38000 Grenoble, France
  \and  Universidade de Lisboa - Faculdade de Ciências, Campo Grande, P-1749-016 Lisboa, Portugal
  \and  Instituto de Astronom\'{i}a, Universidad Nacional Aut\'{o}noma de M\'{e}xico, Apdo. Postal 70264, Ciudad de M\'{e}xico, 04510, M\'{e}xico 
  \and School of Physics \& Astronomy, University of Leeds, Woodhouse Lane, LS2 9JT, Leeds, UK
}

   \date{Received ; accepted }

\titlerunning{Herbigs}
\authorrunning{Garcia Lopez, R. et al.}

  \abstract
   {The region of protoplanetary disks closest to a star (within 1-2\,au) is shaped by a number of different processes, from accretion of the disk material onto the central star to ejection in the form of winds and jets. 
   Optical and near-IR emission lines are potentially good tracers of inner disk processes if very high spatial and/or spectral resolution are achieved.
   }
   {In this paper, we exploit the capabilities of the VLTI-GRAVITY near-IR interferometer to determine the location and kinematics of the hydrogen emission line \brg.}
   {We present VLTI-GRAVITY observations of the \brg\ line for a sample of 26 stars of intermediate mass (HAEBE), the largest sample so far analysed with near-IR interferometry. 
   }
   {The \brg\ line was detected in 17 objects. The emission is very compact (in most cases only marginally resolved), with a size of 10-30 \Rstar (1-5 mas). About half of the total flux comes from even smaller regions, which are unresolved in our data. For eight objects, it was possible to determine the position angle (PA) of the line-emitting region, which is generally in agreement with that of the inner-dusty disk emitting the K-band continuum. 
   
   The position-velocity pattern of the \brg\ line-emitting region of the sampled 
   objects is roughly consistent with Keplerian rotation. The exception is HD~45677, which shows more extended emission and more complex kinematics. 
   The most likely scenario for the \brg\ origin is that the emission comes from an MHD wind launched very close to the central star, in a region well within the dust sublimation radius. An origin in the bound gas layer at the disk surface cannot be ruled out, while accreting matter provides only a minor fraction of the total flux. }
   {These results show the potential of near-IR spectro-interferometry to study line emission in young stellar objects. }


\maketitle
%
\section{Introduction}

Protoplanetary disks are ubiquitous around young stellar 
objects of very different mass. 
In recent years, high spatial resolution observations at infrared and millimetre wavelengths have shown that protoplanetary disks are not smooth and homogeneous but show a variety of structures, from gaps to rings to  spirals and vortices \citep{garufi18, andrews20}.
However, very little is known of the inner disk, within 1--2\,au from the central star, due to the high angular resolution required to spatially resolve this region.

The inner disk of young stars is shaped by a number of different processes. As matter in the disk accretes  onto the central star, it is expected that a significant fraction will be ejected in magneto-hydrodynamic (MHD)-driven and/or photoevaporative winds, with specific mass-loss rates peaking in the inner disk \citep[e.g.][]{pascucci22,lesur22}. Observations of evolved stars have demonstrated that planets can be located in the very inner regions of disks. 
Disentangling and characterising these processes is necessary if we want to understand the inner disk properties and their link to planet formation and evolution.

Studying the inner disk requires milli-arcsecond (mas) spatial resolution, which is only achievable with optical interferometry. 
The analysis of available interferometric data has so far shown that the near-IR continuum emission is dominated by the emission of dust near the sublimation radius, with some contribution closer to the star, whose origin remains unidentified \citep{millan-gabet07, kraus08, benisty10, menu15, lazareff17, karine19,davies20,karine21, marcos-arenal21, Ganci21}.


Optical interferometry provides information not only on the continuum emission but also on some emission lines. In particular, the Very Large Telescope Interferometer (VLTI) K-band beam combiner GRAVITY and its precursor, VLTI-AMBER, cover the hydrogen recombination line \brg. The origin of the hydrogen emission lines commonly detected in young stars is usually associated with the presence of circumstellar material such as the surface of the accretion disk itself and/or accreting and outflowing gas. 
In TW\,Hya, the closest T\,Tauri star (d$\sim$60 pc), GRAVITY-VLTI \Brg\ observations have shown that the line traces the emission of the accretion columns and provided a direct confirmation of the magnetic accretion models for T\,Tauri stars, where the size of the accretion region is directly related to the strength of the stellar magnetic field and the mass-accretion rate \citep{rebeca20}. In more distant T\,Tauri star objects, we expect that the current optical interferometers will not reach the spatial resolution required to resolve the magnetosphere \citep{rebeca17, bouvier20B, GRAVITY_TTauri}. Furthermore, existing studies on more luminous Herbig AeBe stars have shown that most of the emission of the \Brg\ line originates in a wind, with a possible minor contribution from unresolved emission from the accreting gas \citep{weigelt11, rebeca15, ale15, rebeca16, hone17, kreplin18, hone19}. So far, only very few objects have been studied in detail (namely, MWC297, HD163296, HD98922, VVSer, HD58647). However, these studies have proved that it is possible to use interferometric data to constrain the wind's physical properties (e.g. mass-loss rate) and kinematics and directly test wind models.

This paper presents the results of a survey of 26  intermediate-mass stars (HAEBE) observed in the \Brg\ line with VLTI-GRAVITY. The objects cover a mass range between $\sim$1.5 and $\sim$7\,\msun\ and represent the largest sample studied so far. The observations and data reduction procedure are described in Sect.~2. Section~3 presents the properties of the stars in the sample. Section~4 describes the results obtained from the interferometric observations, namely, the size of the \brg\ line-emitting region, its position angle on the plane of sky, and the position-velocity diagrams. Section~5 is dedicated to a discussion of the results and of the constraints they set on current models of the line emission. Section~6 follows with a summary and conclusions.

\begin{table*}[t]
\caption{Properties of the sample.}
\centering
\vspace{0.1cm}
\small
\begin{tabular}{c l l c c c c c c c c c c }
\hline
\hline
& Name & $d$\tablefootmark{a}  & $T_{\rm eff}$\tablefootmark{a}& $\log g$\tablefootmark{a}                     & $\log L_\ast$\tablefootmark{a}      & $M_\ast$\tablefootmark{a}           & $\log L_{acc}$\tablefootmark{a}       & V\tablefootmark{b}     & A$_{\rm v}$\tablefootmark{c} & K\tablefootmark{d}     & v$sin$i\tablefootmark{e} & v$_{rad}$\tablefootmark{e} \\
   &   & [pc] & [K]          & [cm/s$^2$]                   &[L$_{\odot}$] & [M$_{\odot}$] &[L$_{\odot}$] &  [mag] & [mag]       & [mag] & [km/s]  & [km/s]       \\
\hline
[1] &HD37806 & 428 & 10500 & 4.0  & 2.17$^{+0.19}_{-0.14}$  & 3.11$^{+0.55}_{-0.33}$ & 1.46$^{+0.14}_{-0.14}$ & 7.9 & 0.13 & 5.6 & 120$^{+10}_{-3}$ & 47 \\ [1ex]
[2]&HD38120 & 405 & 10700 & 4.0 & 1.72$^{+0.31}_{-0.20}$ & 2.37$^{+0.43}_{-0.24}$ & 1.24 $^{+0.17}_{-0.17}$&  9.1 & 0.21 & 7.2 & 97$^{+10}_{-3}$ & 28 \\ [1ex]
[3]&HD45677 & 620 & 16500 & 4.0 & 2.88$^{+0.32}_{-0.17}$   & 4.72$^{+1.19}_{-0.34}$ & 2.07 $^{+0.16}_{-0.15}$& 7.4 & 0.57 & 4.5 & 200 & 22 \\[1ex]   
[4]&HD58647 & 319 & 10500 & 3.3 & 2.44$^{+0.11}_{-0.09}$ & 3.87$^{+0.33}_{-0.19}$ & 1.62 $^{+0.11}_{-0.11}$&  6.8 & 0.37 & 5.4 & 118${\pm 4}$ & \\[1ex]
[5]& HD85567 & 1023 & 13000 & 3.5 & 3.28$^{+0.19}_{-0.20}$ & 6.79$^{+1.16}_{-1.04}$ & 2.68 $^{+0.14}_{-0.13}$& 8.5 &  0.89 & 5.8 & 31$\pm 3$ &  \\[1ex]
[6]&HD95881 & 1168 & 10000 & 3.2 & 3.13$^{+0.12}_{-0.13}$ & 7.02$^{+0.62}_{-0.67}$ & 2.37$^{+0.16}_{-0.14}$ & 8.2 &  0.72 & 5.7 & 50 & 36 \\  [1ex]  
[7]&HD97048 & 185 & 10500 & 4.3 & 1.76$^{+0.12}_{-0.12}$ & 2.40$^{+0.15}_{-0.03}$ & 1.19$^{+0.13}_{-0.13}$ &  8.4 & 0.90 & 5.9 & 140$\pm 20$ & 18 \\ [1ex]  
[8]&HD98922 & 689 & 10500 & 3.6 & 3.20$^{+0.11}_{-0.11}$ & 7.17$^{+0.67}_{-0.69}$ & 2.44$^{+0.13}_{-0.13}$ &  6.7 & 0.09 & 4.4 & 50$\pm3$ & 0 \\ [1ex]
[9]& HD100546 & 110 & 9750 & 4.3 & 1.37$^{+0.07}_{-0.05}$ & 2.06$^{+0.10}_{-0.12}$ & 0.97$^{+0.14}_{-0.14}$ &  6.7 & 0.00 & 5.7 & 60$\pm5$ & 9 \\   [1ex]
[10]&HD114981 & 705 & 16000 & 3.6 & 3.31$^{+0.13}_{-0.12}$ & 6.46$^{+0.68}_{-0.62}$ & 1.86$^{+0.16}_{-0.16}$ &  7.2 & 0.15 & 7.4 & 239$\pm13$ & -50$\pm11$  \\ [1ex]
[11]&HD135344B$^\star$ & 136 & 6375 & 4.0 & 0.94$^{+0.13}_{-0.14}$ & 1.67$^{+0.18}_{-0.16}$ & 0.09$^{+0.21}_{-0.22}$ &  8.7 & 0.53 & 7.6 & 82$\pm$2 & 0 \\ [1ex]
[12]&HD139614$^\star$ & 135 & 7750 & 4.3 & 0.86$^{+0.08}_{-0.09}$ & 1.61$^{+0.02}_{-0.11}$ & 0.04$^{+0.22}_{-0.23}$ & 8.4 & 0.21 & 6.7 & 24$\pm3$ & 0$\pm2$  \\ [1ex]
[13]&HD141569 & 111 & 9500 & 4.2 & 2.34$^{+0.06}_{-0.07}$ & 2.16$^{+0.04}_{-0.15}$ & 0.35$^{+0.21}_{-0.23}$ &  7.1 &  0.38 & 6.6 & 228$\pm10$ & -12$\pm7$ \\ [1ex]
[14]&HD142527$^\star$ & 157 & 6500& 3.9 & 1.39$^{+0.09}_{-0.10}$ & 2.26$^{+0.02}_{-0.13}$ & 0.65$^{+0.18}_{-0.19}$ &  8.3 & 0.00 & 5.0 &  & -3.5 \\[1ex]
[15]&HD142666$^\star$ & 148 & 7500 & 4.0 & 1.14$^{+0.10}_{-0.11}$ & 1.69$^{+0.13}_{-0.11}$ & 0.05$^{+0.25}_{-0.28}$ &  8.8 & 0.95 & 6.1 & 65$\pm$3 & -7$\pm$3 \\[1ex]
[16]&HD144432$^\star$ & 155 & 7500 & 4.0 & 1.18$^{+0.09}_{-0.09}$ & 1.74$^{+0.13}_{-0.10}$ & 0.38$^{+0.20}_{-0.21}$ &  8.2 & 0.48 & 5.9 &  79$\pm$4 & -3$\pm$3 \\[1ex]
[17]&HD144668$^\star$ & 161 & 8500& 3.7 & 1.95$^{+0.08}_{-0.08}$ & 2.97$^{+0.17}_{-0.21}$& 1.33$^{+0.14}_{-0.15}$ &  7.2 & 0.87 & 4.4 & 199$\pm$11 & -10$\pm$8 \\[1ex]
[18]&HD145718$^\star$ & 153 & 8000 & 4.4 & 1.10$^{+0.10}_{-0.10}$ & 1.67$^{+0.06}_{-0.02}$ & -0.04$^{+0.28}_{-0.32}$ & 8.8 & 1.23 & 6.7 & 113$\pm3$ & -4$\pm$2 \\[1ex]
[19]&HD150193 & 151 & 9000 & 4.0 & 1.50$^{+0.13}_{-0.14}$ & 2.05$^{+0.19}_{-0.12}$ & 0.83$^{+0.17}_{-0.18}$ &  8.8 & 1.80 & 5.3 & 108$\pm5$ & -5$\pm 4$\\[1ex]
[20]& HD158643 & 123& 9800 & 3.6 & 2.22$^{+0.26}_{-0.07}$ & 3.35$^{+0.79}_{-0.22}$ & 1.30$^{+0.16}_{-0.15}$ & 4.8 &0.00 & 4.3 & 256 & -11 \\[1ex]
[21]&HD163296 & 102 & 9250 & 4.3 & 1.36$^{+0.08}_{-0.09}$ & 2.04$^{+0.07}_{-0.10}$ & 0.62$^{+01.8}_{-0.19}$ & 6.8 & 0.37 & 4.6 & 129$\pm8$ & -9$\pm6$\\[1ex]
[22]&HD169142$^\star$ & 114 & 10700 & 4.0 & 1.31$^{+0.12}_{-0.34}$ & 2.00$^{+0.13}_{-0.13}$ & 0.59$^{+0.15}_{-0.15}$ & 8.2 & 1.02 & 6.4 & 48$\pm$2 & 0$\pm$2 \\[1ex]
[23]&HD179218 & 266 & 9500 & 3.9 & 1.97$^{+0.07}_{-0.07}$ & 2.86$^{+0.16}_{-0.20}$ & 1.08$^{+0.13}_{-0.13}$ &  7.8 & 0.33 & 5.8 & 69$\pm3$ & 15$\pm2$ \\[1ex]
[24]&HD190073 & 870& 9750 & 3.4 & 2.84$^{+0.16}_{-0.14}$ & 5.62$^{+0.78}_{-0.65}$ &2.31$^{+0.18}_{-0.15}$ & 7.7 & 0.20 & 5.7 & 0-8 & 0 \\[1ex]
[25]&HD259431 & 721 & 14000 & 4.3 &  2.97$^{+0.27}_{-0.40}$ & 5.2$^{+1.8}_{-1.3}$ & 2.43 $^{+0.194}_{-0.13}$&  8.7 & 1.11 & 5.7 & 83$\pm11$ & 26$\pm 8$ \\[1ex]
[26]&V1818Ori$^\star$ & 695 & 13000 &  & 2.96$^{+0.24}_{-0.29}$ & 5.3$^{+1.3}_{-1.1}$ & & 11.0 & 3.72 & 6.0 & 46$\pm10$ & 26  \\ [1ex]
\hline
\end{tabular}
\label{tab:starproperties}
\tablefoot{\tablefoottext{a}{Distance, effective temperature, surface gravity, stellar luminosity and mass, and accretion luminosity from \cite{vioque18, Wichittanakom20}}. \tablefoottext{b}{Visual magnitude from CDS-Simbad}. \tablefoottext{c}{Visual extinction from \cite{vioque18}}. \tablefootmark{d}{2MASS K-band magnitude}. \tablefootmark{e}{v$sin i$ and radial velocity from \cite{Miroshnichenko01, guimaraes06,Kharchenko07, montesinos09, alecian13,ilee14}}.
}
\end{table*}


%
\begin{table*}[t]
\caption{Results from the geometrical fitting and displacement analysis.}
\centering
\vspace{0.1cm}
\small
\begin{tabular}{c l c c c c c c c c c c c}
\hline
\hline
& Name & FWHM\tablefootmark{a}  & FWHM\tablefootmark{a} & $i$\tablefootmark{a}    & PA\tablefootmark{a}&  F$_{res}$\tablefootmark{a} & F$_{comp}$\tablefootmark{a} & F$_{halo}$\tablefootmark{a} & $\chi^2_{r}$\tablefootmark{b} & PA$_{disp}$\tablefootmark{c} & i$_{cont}$\tablefootmark{d} & PA$_{cont}$\tablefootmark{d}\\
&      & [mas] & [au] & [\degr]        & [\degr]   &     &       &            &            & [\degr]     & [\degr]    & [\degr]    \\
%
%
\hline
[1] &HD37806 & 1.4$^{+1.1}_{-0.4}$ & 0.6$^{+0.5}_{-0.2}$ & 40$^{+24}_{-26}$ & 29$^{+27}_{-79}$  & 0.36$^{+0.26}_{-0.21}$  & 0.51$^{+0.16}_{-0.25}$ & 0.12$^{+0.05}_{-0.05}$ & 0.20 & 44$^{+3}_{-2}$ & 60$^{+3}_{-4}$ & 51$^{+4}_{-1}$  \\[1ex] 
[2]&HD38120 & 1.4$^{+1.2}_{-0.8}$ & 0.6$^{+0.5}_{-0.3}$ & 46$^{+29}_{-31}$ & 167$^{+56}_{-61}$ & 0.40$^{+0.27}_{-0.27}$ & 0.54$^{+0.17}_{-0.26}$ & 0.06$^{+0.06}_{-0.04}$ & 0.75 & -- & 49$^{+2}_{-1}$ & 164$^{+2}_{-2}$ \\[1ex]
[3]&HD45677 & 3.6$^{+0.2}_{-0.2}$ & 2.2$^{+0.1}_{-0.1}$& 41$^{+3}_{-3}$ & 55$^{+6}_{-6}$ & 0.55$^{+0.10}_{-0.17}$ & 0.19$^{+0.04}_{-0.06}$ & 0.27$^{+0.06}_{-0.08}$ & 48.30\tablefootmark{e} & -- & 60$^{+1}_{-1}$ & 64$^{+1}_{-1}$\\[1ex]    
[4]&HD58647 & 1.7$^{+0.5}_{-0.5}$ & 0.5$^{+0.2}_{-0.2}$ &  40$^{+10}_{-17}$ & $--$ & 0.46$^{+0.19}_{-0.20}$ & 0.47$^{+0.18}_{-0.24}$ & 0.07$^{+0.03}_{-0.02}$ & 1.00 & 22$^{+3}_{-5}$ & 64$^{+1}_{-1}$ & 15$^{+1}_{-1}$ \\[1ex]
[5]& HD85567 & 2.8$^{+1}_{-1}$ & 2.9$^{+1.0}_{-1.0}$ &  32$^{+7}_{-8}$ & 120$^{+12}_{-13}$ & 0.22$^{+0.26}_{-0.06}$ & 0.76$^{+0.12}_{-0.20}$ & 0 & 0.91 & 119$^{+1}_{-2}$ & 20$^{+1}_{-2}$ & 96$^{+7}_{-7}$ \\[1ex]
[6]&HD95881 & 1.7$^{+1.0}_{-0.7}$ & 2.0$^{+1.2}_{-0.8}$ & 46$^{+28}_{-30}$ & 158$^{+65}_{-56}$ & 0.44$^{+0.24}_{-0.27}$ & 0.53$^{+0.18}_{-0.27}$ & 0.03$^{+0.04}_{-0.02}$ & 1.57 & -- & 51$^{+4}_{-1}$ & 167$^{+3}_{-3}$ \\  [1ex]  
[7]&HD97048 & 1.5$^{+1.8}_{-0.8}$ & 0.3$^{+0.3}_{-0.2}$ &  50$^{+26}_{-33}$ & 173$^{+51}_{-59}$ & 0.35$^{+0.29}_{-0.23}$ & 0.54$^{+0.16}_{-0.25}$ & 0.11$^{+0.09}_{-0.07}$ & 0.66 & -- & 44$^{+1}_{-2}$ & 9$^{+3}_{-3}$\\ [1ex]  
[8]&HD98922 & 2.8$^{+0.3}_{-0.3}$ & 1.9$^{+0.2}_{-0.2}$ & 16$^{+12}_{-11}$ & 92$^{+43}_{-43}$ & 0.37$^{+0.08}_{-0.11}$ & 0.59$^{+0.11}_{-0.18}$ & 0.04$^{+0.02}_{-0.02}$ & 0.43 & (-23--9) & 49$^{+1}_{-1}$& 130$^{+1}_{-1}$ \\ [1ex]
[9]& HD100546 & 1.7$^{+1.2}_{-0.5}$ & 0.2$^{+0.1}_{-0.1}$ &  30$^{+21}_{-20}$ & 108$^{+48}_{-60}$ & 0.38$^{+0.25}_{-0.22}$ & 0.54$^{+0.16}_{-0.26}$ & 0.08$^{+0.04}_{-0.03}$ & 0.22 & 141$^{+1}_{-2}$ & 50$^{+1}_{-1}$& 147$^{+1}_{-1}$ \\   [1ex]
[10]&HD114981 & $<$1.7 & $<$1.2 & -- & -- &  --&  --& -- & --& -- &--\\ [1ex]
[13]&HD141569 & $<$1.7 & $<$0.2 & -- & -- & -- & -- & -- &-- &-- & --\\ [1ex] 
[19]&HD150193 & 4.6$^{+0.3}_{-1.5}$ & 0.7$^{+0.0}_{-0.2}$ & 42$^{+21}_{-25}$ & 127$^{+20}_{-32}$ & 0.33$^{+0.21}_{-0.14}$ & 0.59$^{+0.14}_{-0.21}$ & 0.08$^{+0.03}_{-0.02}$ & 1.04 & -- & 47$^{+2}_{-2}$&  6$^{+3}_{-3}$\\[1ex]
[21]& HD163296 & 3.0$^{+0.4}_{-0.4}$ &0.3$^{+0.0}_{-0.0}$ & 30$^{+13}_{-19}$ & 106$^{+105}_{-20}$ & 0.32$^{+0.11}_{-0.10}$ & 0.67$^{+0.13}_{-0.20}$ & 0 & 1.11 & 137$^{+1}_{-1}$ & 40$^{+1}_{-1}$ &136$^{+2}_{-2}$ \\[1ex]
[22]&HD169142$^\star$ & -- & -- & -- & -- & -- & -- & - & & 35$^{+5}_{-5}$\\ [1ex]
[23]&HD179218 & 1.4$^{+1.6}_{-0.8}$ & 0.4$^{+0.4}_{-0.2}$ & 54$^{+25}_{-35}$ & 106$^{+59}_{-84}$ & 0.38$^{+0.23}_{-0.27}$ & 0.49$^{+0.16}_{-0.25}$ &0.12$^{+0.07}_{-0.06}$ & 1.79 & - & 54$^{+8}_{-10}$ & 49$^{+9}_{-9}$   \\[1ex]
[24]&HD190073 & 1.2$^{+0.9}_{-0.4}$& 1.0$^{+0.8}_{-0.4}$ & 44$^{+30}_{-29}$ & 86$^{+52}_{-45}$ & 0.41$^{+0.28}_{-0.28}$ & 0.55$^{+0.18}_{-0.28}$ & 0.03$^{+0.04}_{-0.02}$ & 0.81 & - & 21$^{+2}_{-1}$ & 64$^{+9}_{-9}$ \\[1ex]
[25]&HD259431 & 2.0$^{+0.4}_{-0.3}$ & 1.4$^{+0.3}_{-0.2}$ &  17$^{+14}_{-11}$ & 70$^{+67}_{-37}$  &  0.45$^{+0.18}_{-0.17}$ & 0.53$^{+0.15}_{-0.21}$ & 0 & 0.61 & 33$^{+3}_{-4}$ & 28$^{+1}_{-1}$ & 50$^{+3}_{-3}$ \\[1ex]

\hline
\end{tabular}
\label{tab:results}
\tablefoot{\tablefoottext{a}{FWHM, inclination ($i$), PA, and relative flux contributions of the \brg\ line-emitting region as estimated from the analysis of the continuum-subtracted visibilities (see Sect.\,\ref{sect:Brg_vis})}. \tablefoottext{b}{Reduced $\chi^2$ from our fitting procedure (see Sect.\,\ref{sect:Brg_vis})}. \tablefoottext{c}{PA of the continuum-subtracted \brg\ line emission as derived from the line displacements (see Sect.\,\ref{sect:line_displacements})}. \tablefoottext{d}{Inclination and PA of the continuum emission from \cite{karine19}}. 
 \tablefoottext{e}{see Sect.\,\ref{sect:HD45567}.} 
}
\end{table*}

\section{Observations and data reduction}
%




The data in this paper  were collected using the European Southern Observatory (ESO) VLTI-GRAVITY \citep{GRAVITY} as part of the Guaranteed Time Observations (GTO) program. Most of our targets were observed using the four 1.8-m Auxiliary Telescopes (ATs), with the only exception being HD\,36917, for which unit telescope (UT) data are presented. The different datasets shown in this paper were collected during a period of four years (2017-2021). A full log of the observations can be found in Table\,\ref{tab:obs}. 

The GRAVITY spectro-interferometric signals were recorded on six baselines simultaneously using the fringe tracker (FT) detector at low spectral resolution (six spectral channels) and the scientific (SC) detector at a resolution of $R\sim$4000. Typical integration times of 10\,s--30\,s were used for the SC data, with each observation file corresponding to 300\,s--360\,s. The FT uses much shorter integration times, with frame rates ranging from $\sim$300\,Hz to $\sim$900\,Hz. To calibrate the atmospheric transfer function, a minimum of one interferometric calibrator has been observed per target. The calibrator was observed immediately after or before the science target at a sky location very close to that of the science target (see Table\,\ref{tab:obs} for a full list of the calibrators used in this study).

The data have been reduced using the GRAVITY data reduction pipeline \citep{GRAVITY_pipeline}. The GRAVITY observations provide the following interferometric observables: the K-band spectrum, six spectrally dispersed visibilities and differential phases, and four spectrally dispersed closure phases. The wavelength calibration of the SC high-resolution (HR) observations was refined by using the many telluric lines present in the K-band spectrum. In addition, the calibrator spectra were used to correct the observed spectra for the atmospheric telluric features, atmospheric transmission, and instrumental response. 

In this paper, we focus on the HR SC observations to constrain the properties of the \brg\ line-emitting region in our sample. The FT observations used to probe the inner dust rim in the K-band continuum have been presented in \cite{karine19}. The four interferometric variables in the \Brg\ region are shown for all the stars in the sample in Fig.\,B1--B8.

\section{Sample}


Our sample consists of 26 Herbig AeBe stars with effective temperatures ranging from $\sim$6000\,K to $\sim$16\,000\,K, and stellar luminosities from $\sim$12\,\lsun\ to $\sim$1900\,\lsun. The main properties of our sample are reported in Table\,\ref{tab:starproperties}. Additional  details  can  be found in \cite{karine19}.

In order to avoid spurious effects due to line and/or continuum variability, datasets as close in time as possible have been selected, ideally including data obtained with different configurations acquired during the same observational campaign (a few weeks apart). In the case of data acquired in different epochs, we excluded datasets showing evident line and/or continuum variability. Variability studies will be performed in separate papers (see e.g. \citealt{Sanchez-Bermudez21}). 

Of the original sample of 26 stars, nine showed no or very faint \brg\ emission, even after correcting for the underlying photospheric absorption line (see Sect.\,\ref{sect:photospheric_correction}). These targets are marked with a star symbol in Table\,\ref{tab:starproperties} and  were omitted from further analysis.


\section{Results}


\subsection{HI Br$\gamma$ line spectral signal}
\label{sect:photospheric_correction}

The spectrum of pre-main sequence intermediate-mass stars is usually characterised by bright \HI\ \brg\ line emission from circumstellar material, namely, accreting and outflowing gas and/or the heated surface of the accretion disk itself.
Independent of the origin of this emission, the interferometric observables across the line also have contributions from the stellar photosphere and the continuum cicumstellar emission. 
The total observed flux across the line can be expressed as
\begin{equation}
F(\lambda)_{tot}=F_{disk}+F(\lambda)_*+F(\lambda)_{circ},
\end{equation}
where F$_{disk}$ is the continuum disk emission, F($\lambda$)$_*$ is the photospheric component, and F($\lambda$)$_{circ}$ is the circumstellar \brg\ line emission. 

Over our narrow wavelength range, the continuum disk circumstellar emission is constant with wavelength. However, the contribution from the stellar phostosphere is wavelength dependent, as Herbig AeBe stars have strong and wide \brg\ line photospheric absorption features \citep{1998AJ....116.2530G}.
%
%
To derive the values of the three components, we followed the procedure described in  \cite{rebeca06}. 
Firstly, synthetic spectra were created for each star using the \cite{stellar_atm} grid of models\footnote{https://wwwuser.oats.inaf.it/castelli/} along with the appropriate values of the effective temperature ($T_{eff}$), surface gravity ($g$), and \vsini\,  as reported in Table\,\ref{tab:starproperties}. For the case of HD\,142527, where no \vsini\ value is reported in the literature, a typical  \vsini\ value of 100\,\kms was assumed. The synthetic spectra were smoothed to the GRAVITY spectral resolution of 4000. The source radial velocity (see Table\,\ref{tab:starproperties}) was also taken into account to avoid shifts between the observed and synthetic spectra for nearly all our sources, with the exceptions being  HD\,58647 and HD\,85567, as no radial velocities were available in the literature. 
Under the further assumption that the stellar photosphere dominates the V-band magnitude of the source, given the spectral type and extinction, the photospheric component was computed for each star, corrected for the K-band continuum,  and  removed from the observed  spectrum. 
The amount of correction to the observed \brg\ line depended on the infrared continuum excess of the source. Sources with little continuum excess are more affected by the correction, whereas sources with a high amount of excess did not show significant changes in the \brg\ line flux.  

The observed \brg\ lines are shown as black solid lines in the top panels of Figs.\,\ref{fig:data1}--\ref{fig:data22}, whereas the photospheric-corrected \brg\ line profiles are shown as blue dashed lines. 
After photospheric correction, most sources show a triangular \brg\ line shape profile, with the exception of four sources (HD\,58647, HD\,114981, HD\,141569, and HD\,158643), which have double-peaked line profiles, indicative of rotating emission regions.


\subsection{Continuum-subtracted \brg\ line visibilities and differential phases}


As in the case of the \brg\ line flux, we computed the line visibility and differential phase by removing the contribution of the other components. The observed visibilities across the line (V($\lambda$)$_{tot}$) have contributions from the circumstellar disk continuum (V$_{disk}$), the photospheric component (V($\lambda$)$_*$), and the circumstellar \brg\ line emission (V($\lambda$)$_{circ}$). The relation between these quantities can be expressed in a simple way in two extreme cases: when the differential phase is negligible (diff. phase $\lesssim$5\degr--10\degr) or when the photospheric flux is negligible with respect to the circumstellar continuum emission (i.e. F$_{tot}$/F$_{disk}\gtrsim$20\%).
In the first case, we have

\begin{equation}
    V(\lambda)_{tot} F(\lambda)_{tot} = V_{disk} F_{disk} + V(\lambda)_* F(\lambda)_* + V(\lambda)_{circ} F(\lambda)_{circ}
    \label{eq:line1}
\end{equation}
and
\begin{equation}
    V(\lambda)_{circ}  = V(\lambda)_{tot} \frac{F(\lambda)_{tot}}{F(\lambda)_{circ}} - V_{disk} \frac{F_{disk}}{F(\lambda)_{circ}} - V(\lambda)_* \frac{F(\lambda)_*}{F(\lambda)_{circ}}.  
\end{equation}
We further assumed that the central star is not spatially resolved (i.e. point-like; $V(\lambda)_*$=1) and measured $V_{disk}$ from the observed visibility adjacent to the line $V_{tot}$:
\begin{equation}
    V_{disk}=V_{tot} \frac{F_{tot}}{F_{disk}}- \frac{F_{*}}{F_{disk}}.
\end{equation}

In the second case, if the photospheric flux is negligible at all wavelengths but the differential phase ($\Phi$) is larger than 5\degr--10\,\degr, Eq.(2) is replaced by \citep{weigelt07}:


\begin{equation}
\begin{split}
     V(\lambda)_{circ} F(\lambda)_\mathrm{{circ}} & = ( \left|V(\lambda)_{tot} F(\lambda)_{tot}\right|^2 + \left|V_{disk} F_{disk}\right|^2 - \\
     & 2 V(\lambda)_{tot} F(\lambda)_{tot} V_{disk} F_{disk} \cos{\Phi}) ^{1/2}
\end{split}
\label{eq:pvsi_phi}.
\end{equation}

%

Figures\,\ref{fig:visibility_fit1}--\ref{fig:visibility_fit2} show the value of $V(\lambda)_{circ}$ calculated for all the lines at zero velocity by averaging the signal over the three central points. In the case of the continuum observables, the visibilities and flux values were measured in the nearest line-free wavelength interval. The continuum-subtracted visibilities were computed only for sources where the flux of the peak of the circumstellar \brg\ line is larger than 10\%  of the continuum flux (16 sources). The errors on the continuum-subtracted visibilities were calculated by propagating Eq.\,\ref{eq:line1} and Eq.\,\ref{eq:pvsi_phi}. As a conservative value, the rms at continuum wavelengths close to the line position was considered as the error for the total and continuum visibilities and differential phases. The level of the continuum visibilities is shown as a dashed line in the figures of Appendix \,\ref{ap:data}.

By examining the dependence of $V(\lambda)_{circ}$ as a function of the spatial frequency some information about the spatial extent of the \brg\ emission can be obtained without any assumption on the geometry of the brightness distribution. 
In general, interferometry deals with two extreme behaviours in the dependence of $V(\lambda)_{circ}$ versus the projected baseline length (PBL). On the one hand, for a given PBL, an object is resolved if its angular size (in units of radians) $\Theta \geq$ 1/2 $\lambda$/PBL. In this case, visibilities will drop with an increasing baseline. The opposite extreme case is that of an unresolved object (i.e. point source). In such a case, the visibilities are equal to one and constant over all baselines. In most cases, however, interferometry deals with marginally resolved objects. This means that there is a drop in the visibility values with baseline but without fully resolving the object. In this case, the size of the object is a fraction of the PBL/$\lambda$ \citep[see e.g. ][for a more formal treatment]{monnier07,lachaume03}.   
This scenario could be even more complex when several morphological components are traced by the visibilities (e.g. the overlap of an over-resolved component, a resolved or marginally-resolved component, and a point source). 

In our case, Figs.\,\ref{fig:visibility_fit1} and \ref{fig:visibility_fit2} show that most of our objects are marginally resolved as $V(\lambda)_{circ}$ drops with baseline length. Only two objects (namely, HD114981 and HD141569) show a constant $V(\lambda)_{circ}$=1, within the errors, indicating that they are unresolved. In order to derive further information about the size and geometry of the marginally resolved \brg\ line-emitting region, geometric modelling is needed. A detailed analysis is discussed in Sect.\,\ref{sect:Brg_vis}.

As for the line visibilities, the continuum contribution can also be subtracted from the differential-phase signal across the line. Following  \cite{weigelt07}, the continuum-subtracted differential phase ($\Delta \Phi$) can be written as
\begin{equation}
\sin(\Delta\Phi(\lambda))=\sin(\Phi(\lambda))\cdot \dfrac{|F(\lambda)_\mathrm{{tot}}V(\lambda)_\mathrm{{tot}}|}{|F(\lambda)_\mathrm{{circ}}V(\lambda)_\mathrm{{circ}}|},
\end{equation}
where $\phi$  is the observed differential phase.  
The continuum-subtracted differential phases were computed for all the sources where a differential-phase signal was detected above the rms noise in the adjacent continuum (eight sources).


\subsection{The size of the \brg\ line-emitting region }

\label{sect:Brg_vis}

The size of the \brg\ line-emitting region was derived using a
Bayesian approach by means of Markov chain Monte Carlo (MCMC)
methods. The MCMC method is preferred over ordinary least square methods when performing multiple parameter fits, as the latter can be biased towards the initial conditions of the fit and end up in local minima. We used
the continuum-subtracted \brg\ line visibilities at the peak of the line (i.e. at zero velocity) for the 16 stars with detected \Brg\ emission.
 In order to interpret the \brg\ line visibilities, a brightness distribution of the emission needs to be assumed.
In our case, we describe the total \Brg\ intensity distribution as the sum of three components: a resolved (or marginally resolved) component with a Gaussian brightness distribution of flux $F_\mathrm{res}$ and visibility V$_\mathrm{res}$; an unresolved component with flux $F_\mathrm{comp}$ and visibility $V_\mathrm{comp}=1$; and an over-resolved component with flux $F_\mathrm{halo}$ and visibility $V_\mathrm{halo}$=0. 
This does not necessarily imply that the emission originates in distinct physical regions (e.g. accretion rather than a wind), but it is simply a result of the limited range of available baselines.
The observed line flux at the peak of the normalised line ($F_{line}$) is therefore $F_\mathrm{line}= F_\mathrm{res} + F_\mathrm{comp} + F_\mathrm{halo}$=1, and the corresponding observed line visibility is given by

\begin{equation}
    V_\mathrm{line}= {{V_\mathrm{res}F_\mathrm{res}+F_\mathrm{comp}}\over{F_\mathrm{res}+F_\mathrm{comp}+F_\mathrm{halo}}}
    \label{eq:vline}.
\end{equation}

We note that $F_\mathrm{line}, V_\mathrm{line}$ are $F(\lambda)_\mathrm{circ}$, $V(\lambda)_\mathrm{circ}$, respectively, as defined in Sec.(4.2), measured at the peak of the line.
    

This simple model has six free parameters (five of them independent), namely, 
the relative fluxes $F_\mathrm{res}$, $F_\mathrm{comp}$, $F_\mathrm{halo}$;
the size (i.e. full width at half maximum, FWHM), and the geometrical parameters (i.e. inclination, $i$, and position angle, PA) of the resolved (or marginally resolved) Gaussian component.
In practice, we used the Python module \textit{emcee} to derive the free parameters in our model \citep[see][for a detailed description of the module]{foreman-mackey13}. We set the prior distribution to be
uniform (i.e. non-informative prior), the log-likelihood to be simply a Gaussian including Eq.\,\ref{eq:vline} in its definition, and the posterior distribution was therefore given by the product between the prior distribution function and the log-likelihood function. We initialised
the MCMC with 1000 walkers and ran them for 1000 steps. We set a
burn-in phase of 10\% of the steps to account for the warm-up period of
the chain \citep[see][for details]{ale20}. Figures~\ref{fig:visibility_fit1} and \ref{fig:visibility_fit2} show the best-fitting visibility as a function of the projected baseline for the 16 objects.

The best-fitting value of the parameters and their uncertainties are given in Table\,\ref{tab:results}. 
Two objects (HD~114981 and HD~141569) with  good quality interferometric data are unresolved for our baselines. Considering our longest baseline of 130\,m, this translates into an angular size of less than 1.7\,mas (i.e. < 1.2\,au and < 0.2\,au for HD\,114981 and HD\,141569, respectively). We note that though not fully discussed in this work, the fact that some objects show evidence of asymmetries (i.e. non-zero differential phases and/or closure phases) indicates that these sources are not point-like.  
Sources with detected differential phases and/or closure phases are HD259431, HD190073, HD163296, HD158643, HD150193, HD100546, HD98922, HD85567, HD58647, HD45677, and HD37806. 

\begin{figure}
   \centering
       \includegraphics[width=12cm]{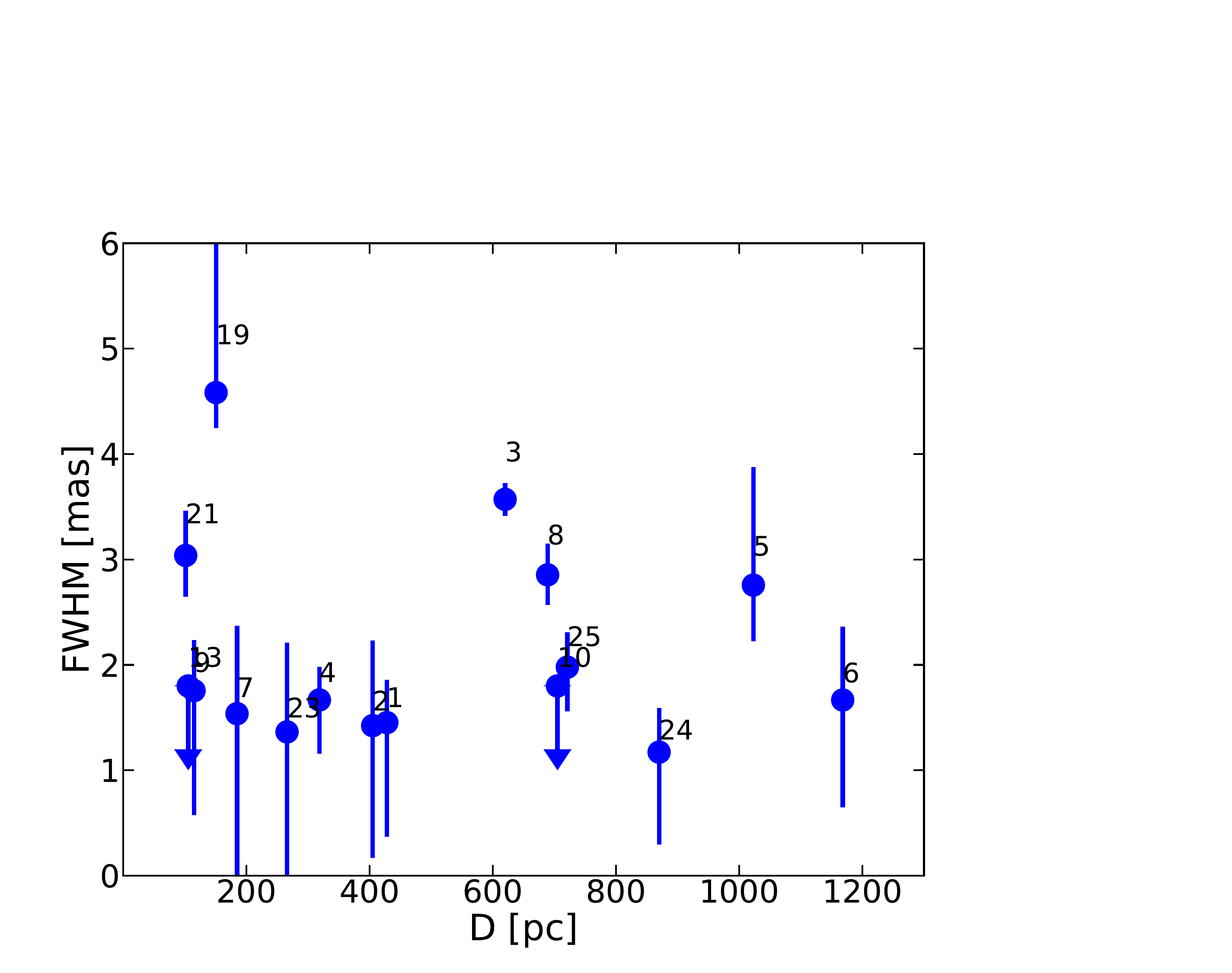}
      \caption{FWHM of the marginally resolved component as a function of the distance. The numbers identify different  stars (see first column in Table 2). 
      }
         \label{Fig:fwhm_dist}
   \end{figure}

\begin{figure}
   \centering
         \includegraphics [width=9cm]{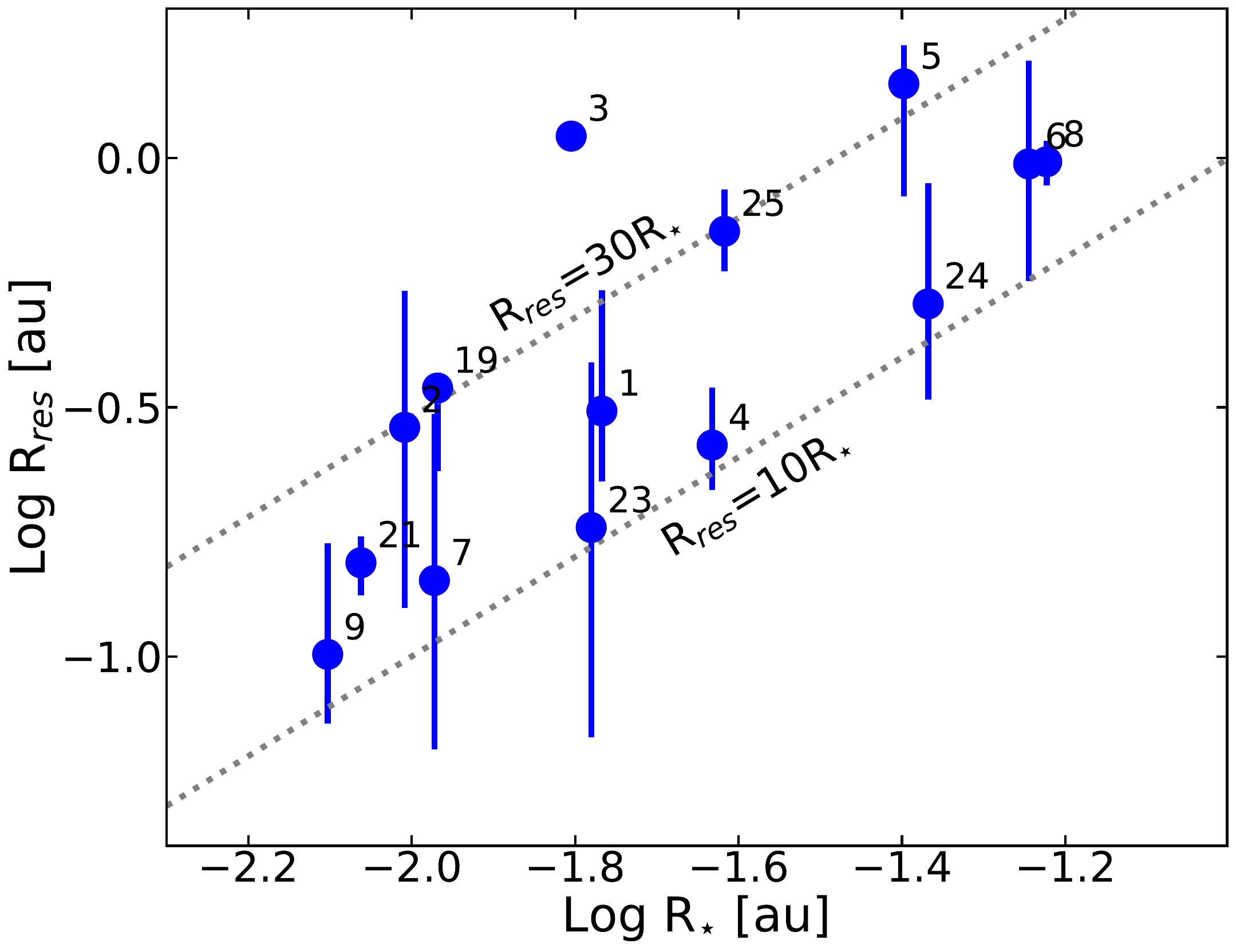} 
      \caption{Radius of the marginally resolved \brg\ component as a function of R$_{star}$. When not visible, the errors are smaller than the size of the dots.
          The number labels are as in Fig.~\ref {Fig:fwhm_dist}.   }
         \label{Fig:RBrg-Rstar}
   \end{figure}

Due to the limited and sparse $uv$ coverage and the large errors associated with the continuum-subtracted line visibilities, the errors on all  parameters are in general very large. 
We note that, as it is true in general, these errors are just the errors provided by the fitting procedure, and they do not take into account further sources of uncertainty (e.g. degeneracy between the inclination, PA, and size in marginally resolved objects).
Nevertheless, there is clear evidence that the \Brg\ line emission is very compact.
The  FWHM of the Gaussian brightness distribution as described above is plotted in Figure~\ref{Fig:fwhm_dist} as a function of the stellar distance reported in Table\,\ref{tab:starproperties}. As shown in the figure, the FWHM of the Gaussian brightness distribution component of the \brg\ line-emitting region has a typical size of a few milli-arcsecond. Moreover, the unresolved component 
accounts for about half of the total flux,  while the halo contributes significantly to the line flux in one source only (HD~45677). In addition, it should be noted that none of our objects are fully resolved and that in most cases, {the \brg\ line emission is} only marginally resolved with our current spatial resolution. 
Figure~\ref{Fig:RBrg-Rstar} shows the size of the resolved and marginally resolved components  as a function of the radius of each star. The values of $R_{res}$ are typically 10--30 \Rstar (see Fig.\,\ref{Fig:RBrg-Rstar}).


These results are usually in agreement with previous \brg\ line spectro-interferometric studies done with VLTI-AMBER and the Keck Interferometer~\citep{kraus08, eisner_herbig07, rebeca15, ale15,kurosawa16}. Indeed, although limited in number of baselines and sensitivity, those studies showed that the \brg\ line emission in early-type HAEBE stars is usually very compact, with an extent of a few milli-arcseconds, which can usually be modelled with a disk wind~\citep[see, e.g.][]{rebeca15, ale15,kurosawa16,hone17} or a gaseous disk in Keplerian rotation~\citep[][]{hone19}.

\subsection {\brg\ line displacements}
\label{sect:line_displacements}


The size of the \brg\ line-emitting region obtained from the analysis of the continuum-subtracted line visibilities is limited by the maximum angular resolution determined by our longest baselines. However, analysis of the \brg\ line differential phases can give us information on smaller angular scales through the study of the line astrometric displacements. This technique is the spectro-interferometric equivalent to the analysis of spectro-astrometric signals in standard spectroscopic observations \citep[e.g.][]{lachaume03, whelan08, lebouquin09}.

\begin{figure}
    \centering
        	\includegraphics[width=0.49\textwidth]{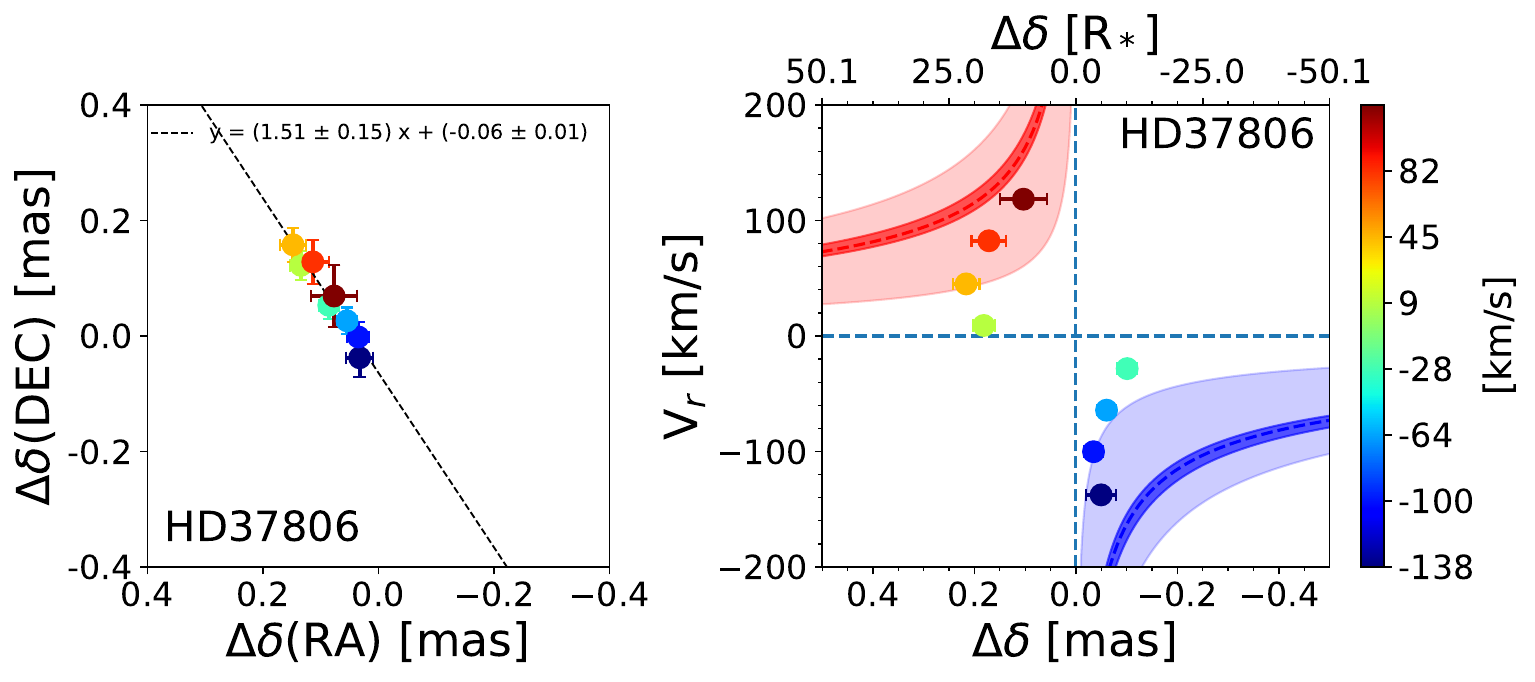}
            \includegraphics[width=0.49\textwidth]{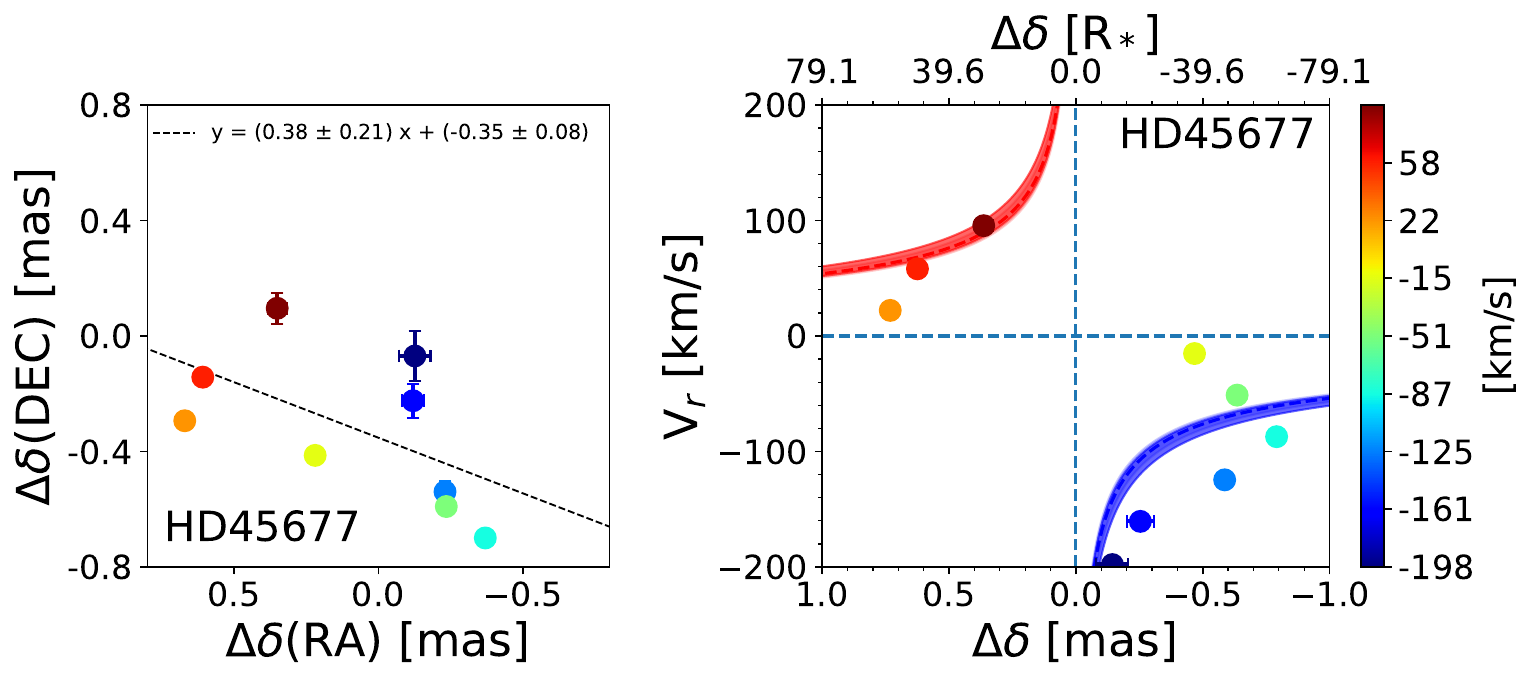}
            \includegraphics[width=0.49\textwidth]{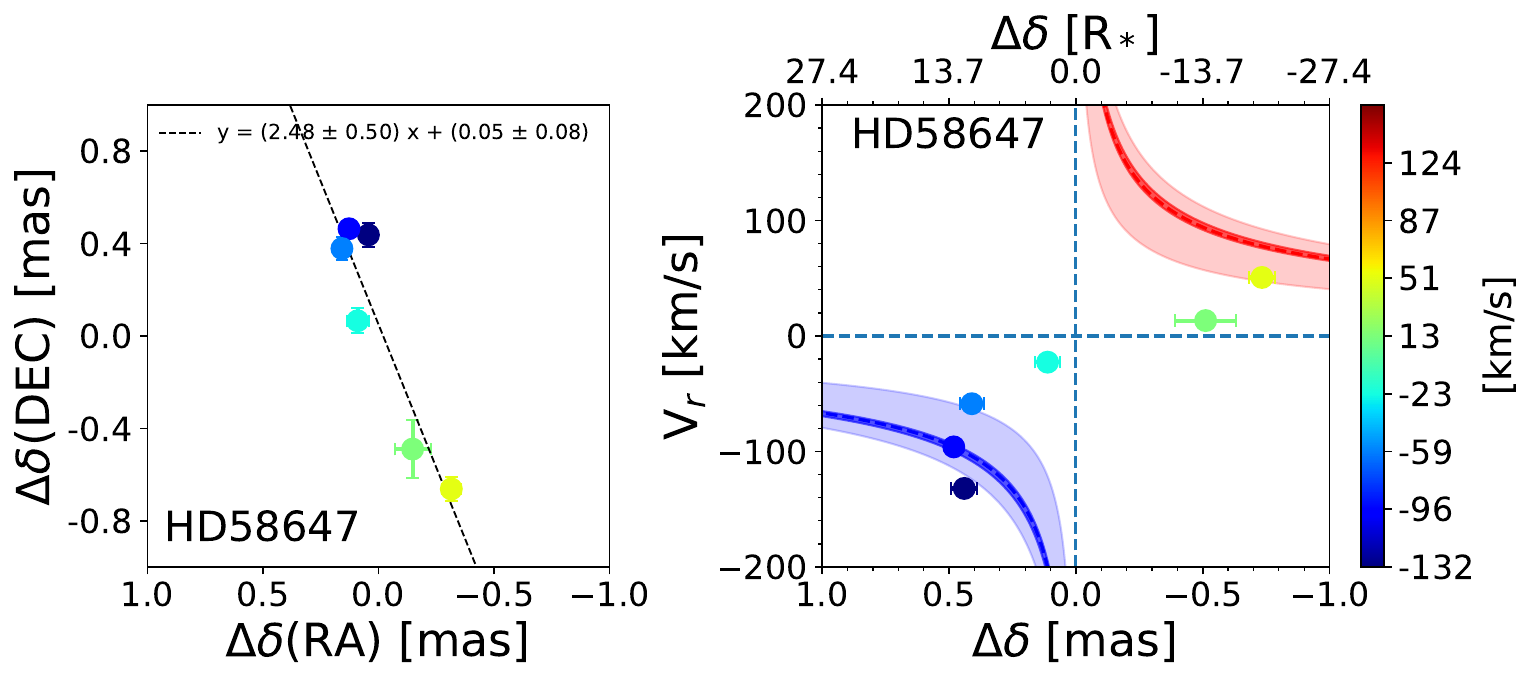}
            \includegraphics[width=0.49\textwidth]{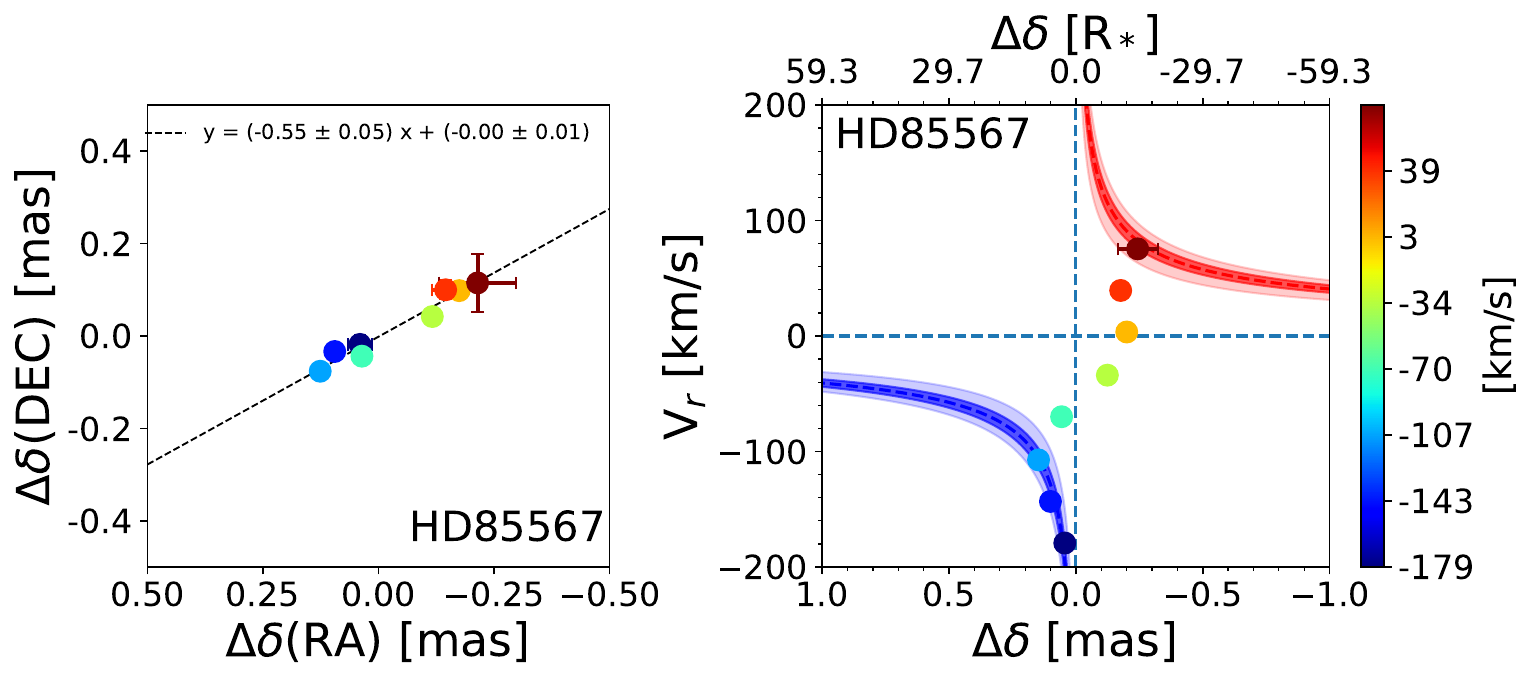}
            \caption {Two-dimensional displacements (left panel) and PVD (right panel) for HD~37806, HD~45677, HD~58647, and HD85567 as computed from the differential phase and described in Sec.\,\ref{sect:line_displacements}. The data are colour-coded according to the radial velocity (scale  on the  right).  In the 2D-displacement figures, the dashed lines show the best linear fit reported at the top of the panel. In the PVD panels, the blue and red shaded regions show the Keplerian rotation pattern (see text for more details). }
            \label{Fig:PVDs1}
   
\end{figure}

\begin{figure}
   \centering
      \includegraphics[width=0.49\textwidth]{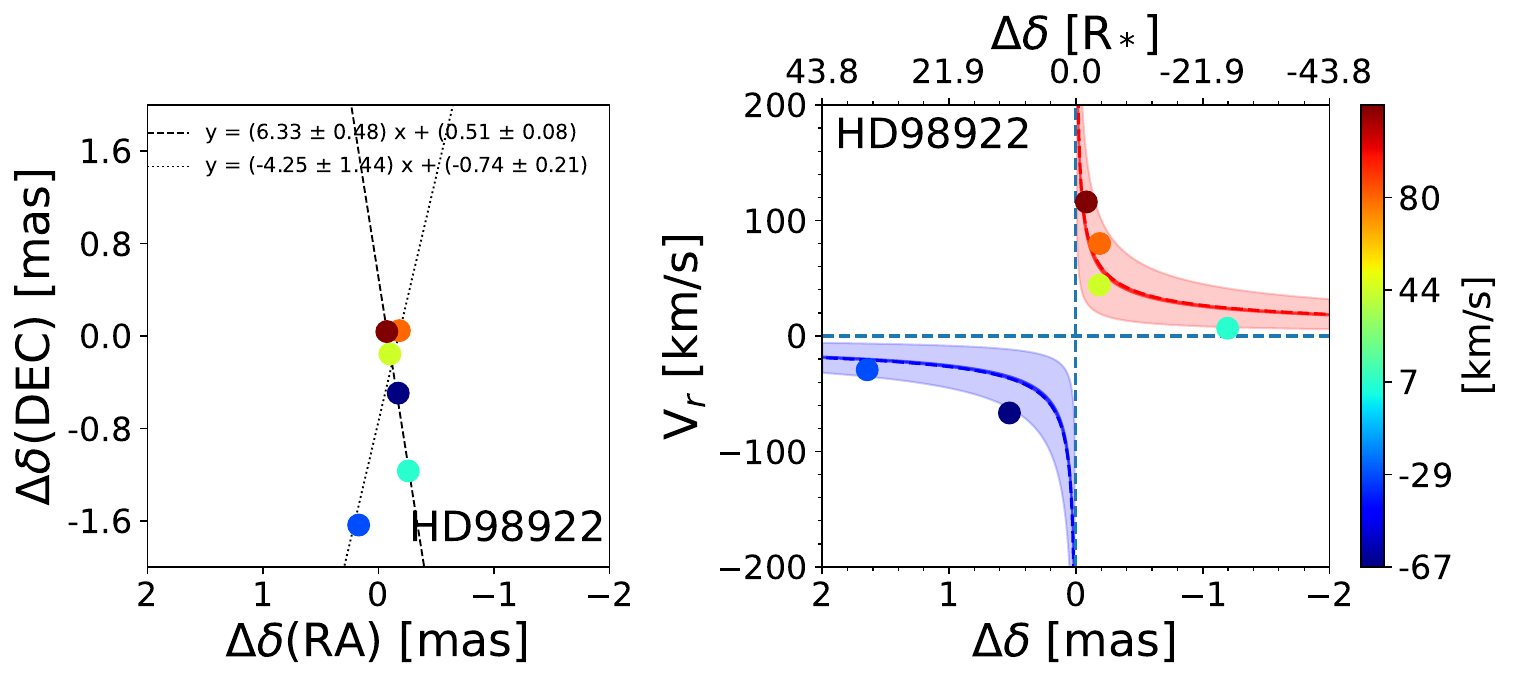}
        \includegraphics[width=0.49\textwidth]{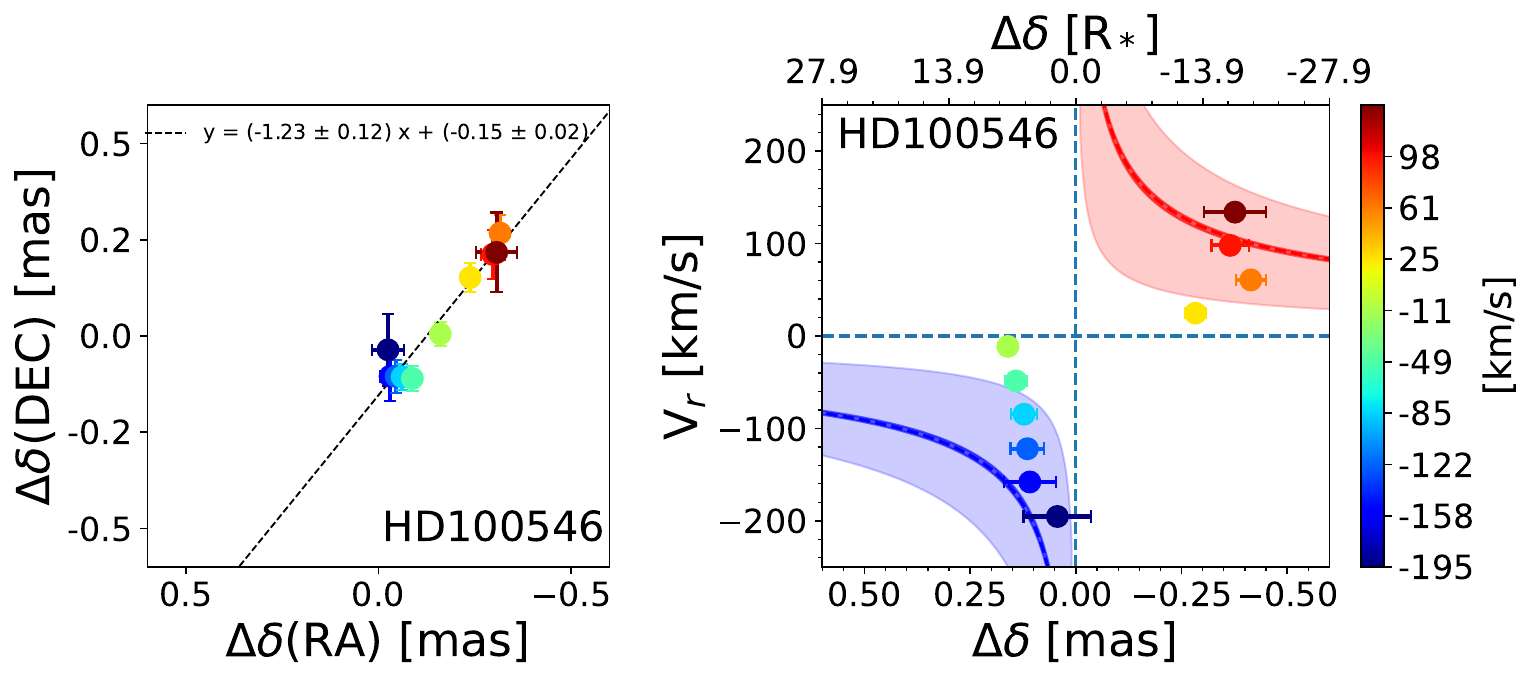}
        \includegraphics[width=0.49\textwidth]{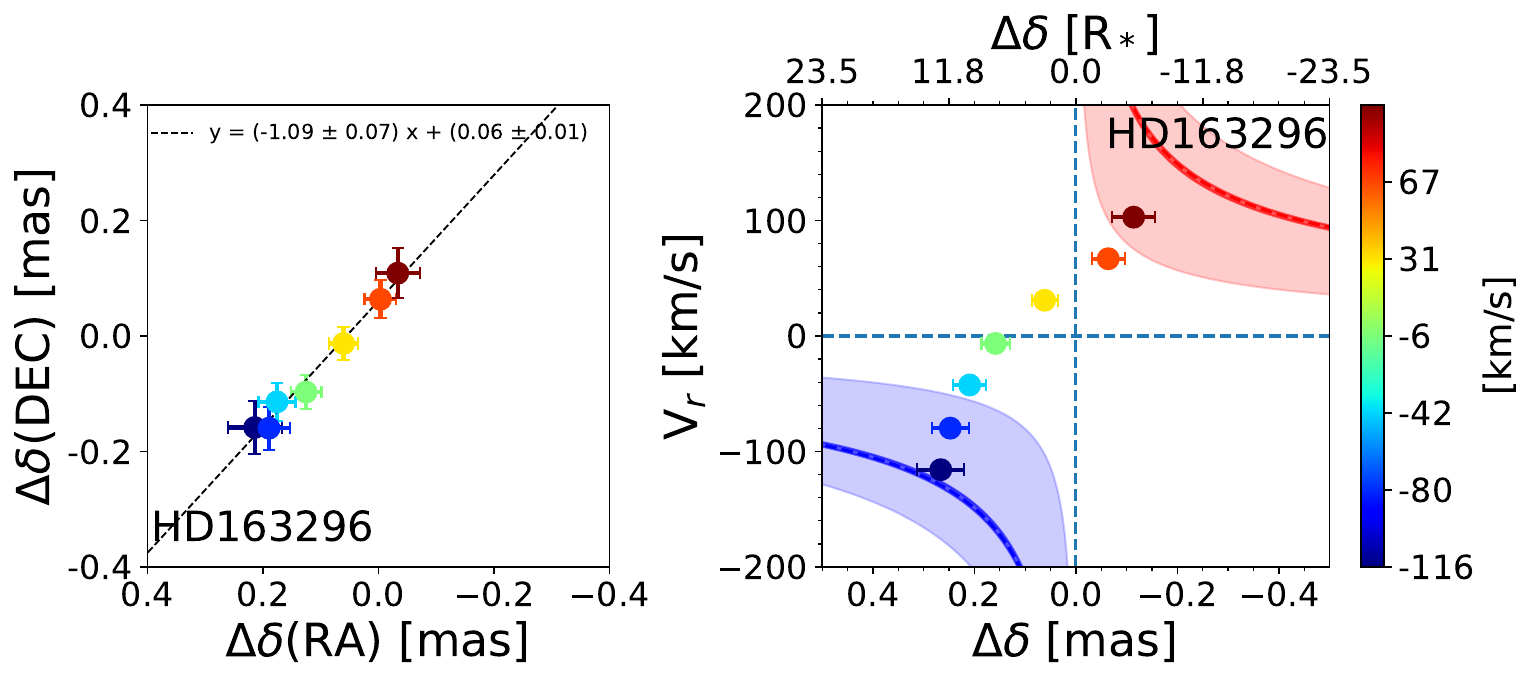}
        \includegraphics[width=0.49\textwidth]{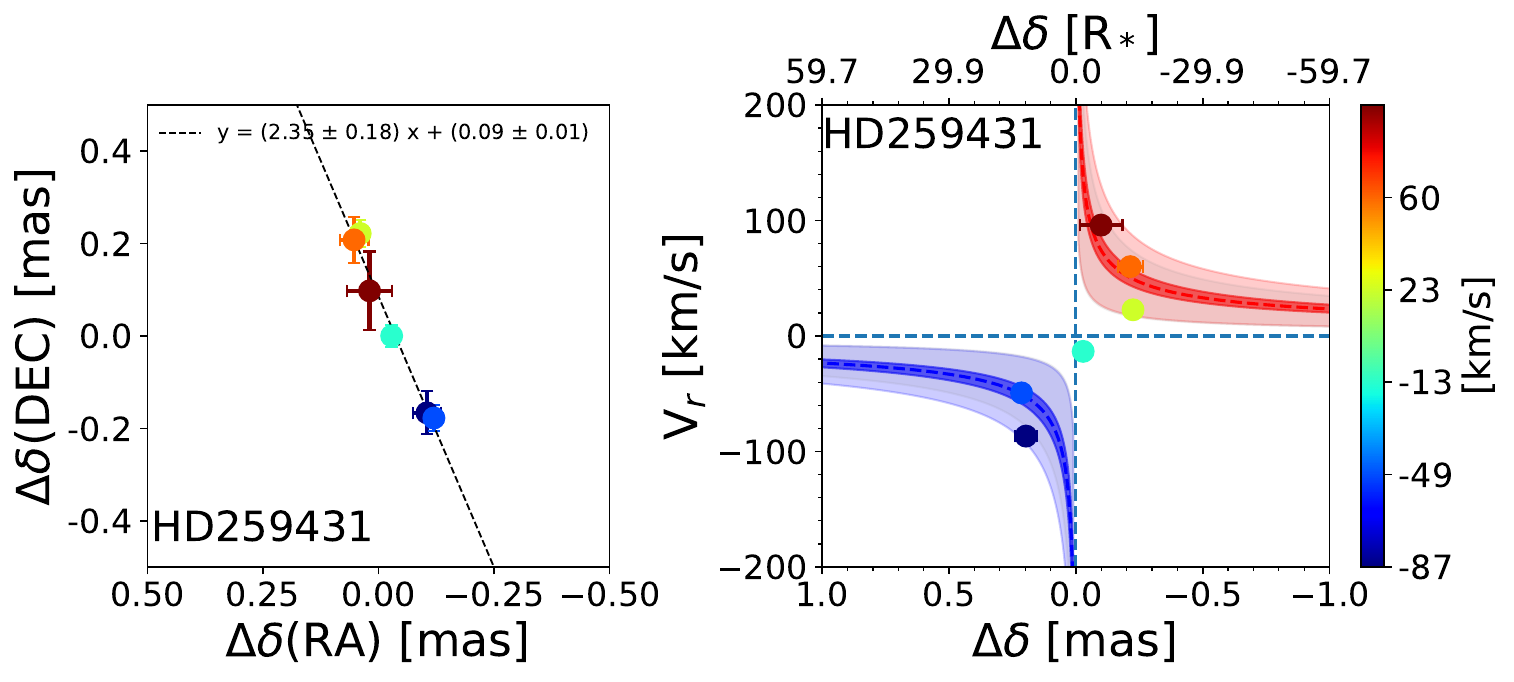}
            \caption {Same as in Fig.\ref{Fig:PVDs1} but for HD~98922, HD~100546, HD~163296, and HD~259431.}
            \label{Fig:PVDs2}
\end{figure}

At any wavelength, the displacement of the photo-centre of the continuum-subtracted line emission with respect to the photo-centre of the continuum ($\delta$) is computed as
\begin{equation}
\delta = -\Delta \Phi \frac{\lambda}{2 \pi B},
\end{equation}
where $B$ is the length of the baseline \citep[e.g.][]{lachaume03}.  In this expression $\delta$ represents the projection, in the baseline direction, of the  photo-centre in the plane of sky. A single spectro-astrometric solution was found by performing a regression fit to all  
%
the baselines at different velocity channels projected in the plane of sky. The results are shown in the left panels of Figs.\,\ref{Fig:PVDs1} and \ref{Fig:PVDs2} for the eight stars with sufficiently good data (half of the sample). Indeed, in order to detect differential-phase signals across the line, a high S/N at the line position is required, along with baselines long enough to be able to at least partially resolve the line emission.

\subsubsection{ The position angle of the \Brg\ emitting region}

From the astrometric line displacements, information about the projected PA of the line emission in the plane of sky and on the centro-symmetry of the brightness distribution can be obtained. The left panels in Figs.\,\ref{Fig:PVDs1}--\ref{Fig:PVDs2} show that for seven out of eight stars, the displacements are roughly  aligned along one direction in the plane of sky (with the exceptions of HD~45677 and HD~98922; \citealt{Sanchez-Bermudez_HD45677, Ganci23} in prep.). Therefore, a linear fit to the \brg\ line displacements provides the PA of the line-emitting region.

The resulting PAs of the \brg\ line emission (PA$_{disp}$) are reported in Table\,\ref{tab:results}, Column~10. The linear fit is also shown as a dashed line overplotted onto the line displacements in Figs.\ref{Fig:PVDs1}--\ref{Fig:PVDs2}.
The PA$_{disp}$ values have much smaller uncertainties than those derived from the line visibility fits (Table\,\ref{tab:results}, Column~6; PA), but the agreement  is quite good when both values are available. The derived values are also in rough agreement with previous results found in the literature for some of our sources \citep[e.g.][]{kurosawa06, kluska20}. In addition, the PA of the \Brg\ emitting region is in good agreement with the PA of the continuum emission in the K-band (PA$_{cont}$ in Table\,\ref{tab:results}, Column~12), as computed in \cite{karine19}. It should be noted that in all cases, the errors on the PA reported in Table\,\ref{tab:results} are just formal errors resulting from the fitting procedures. As for the visibility fitting, these errors do not fully take into account other factors such as the goodness of the $uv$ coverage or the presence of non-centrally peaked emission.

Some information on the centro-symmetry of the brightness distribution can be obtained by inspecting whether the \brg\ line displacements are centred at ($\Delta\delta$(RA)=0, $\Delta\delta$(DEC)=0) (see left panels in Figs.\,\ref{Fig:PVDs1}--\ref{Fig:PVDs2}). Centro-symmetric brightness distributions produce line displacements centred at (0,0), whereas continuum and/or line brightness distributions, not centro-symmetric ones,  produce displacements shifted from (0,0).\footnote{It should be noted that although the displacements were computed from the continuum-subtracted line differential phases, the obtained displacements are still differential, that is, with respect to the photo-centre of the brightness distribution of the continuum.} Examples of centro-symmetric objects are HD~163296 and HD~259431, whereas objects such as HD~98922 and HD~37806 show clear deviations from centro-symmetry. Our simple analysis did not allow us to discern which of the two components (i.e. the line and/or the continuum emission) is not centro-symmetric (i.e. point-symmetric about the (0,0) position). Further information can only be obtained from the analysis of continuum and/or line reconstructed images  
\citep{Sanchez-Bermudez18a, Sanchez-Bermudez18b, kluska20}.
%
However, it should be noted that two of our objects (namely, HD\,45677 and HD\,98922) show continuum closure-phase signals of over 25\degr. Such huge closure-phase signals are difficult to explain with simple inclination effects, and therefore they likely indicate an asymmetric continuum structure \citep[see][for further details]{karine19}.

%
%
%
%


\subsubsection {The position-velocity diagrams}

Additional information about the \brg\ line kinematics can be obtained from position-velocity diagrams (PVDs).
These diagrams are well-known visualisation tools in spectroscopy. They generally display the absolute line displacement with respect to the central source versus radial velocities measured with respect to the source local standard of rest (LSR) and are corrected for the source radial velocity when available from the literature (see Table\,\ref{tab:starproperties}). The PVDs for our sample are shown in the right panel of Figs.\,\ref{Fig:PVDs1}--\ref{Fig:PVDs2}. The 2D line displacements were converted to absolute displacements by computing the module of the vector in RA and DEC. 

The PVDs in Figs.\,\ref{Fig:PVDs1}--\ref{Fig:PVDs2} show evidence of motions dominated by a Keplerian rotation pattern. If the emission is fully spatially and spectroscopically resolved, the PVD of a Keplerian rotating disk is characterised by a typical "butterfly" pattern, with the low-velocity emission located  further from the (0,0) position. In the other extreme case (namely, spatially unresolved emission), the PVD is a straight line with the highest velocity further from the (0,0) position. In the case of partially resolved emission, the PVD has an S shape, with the low-velocity emission projected at decreasing displacements (see, e.g., the PVDs of HD\,163296 in Fig.\,\ref{Fig:PVDs2}). 
Of the eight objects for which PVDs could be computed, only HD~98922 displays the butterfly pattern expected when the emission is fully resolved, while others, such as HD~163296, HD~58647,  HD~85567, and HD~259431, have the S-shape PVD of marginally resolved sources. The sources HD~45677, HD~37806, and HD~100546 show evidence of complex motions that are more difficult to interpret.

Detailed models are necessary to investigate the properties of individual sources but are beyond the scope of this paper. However, a caveat must be stated. Namely, the compactness of the \Brg\ emission and the relatively low spectral resolution of VLTI-GRAVITY limit the potential capability of this tool.

\subsection {HD~45677}
\label{sect:HD45567}

The source HD\,45677 (also known as FS CMa) has properties that differ from the other HAEBEs in the sample. The evolutionary state of this object is debated \citep{Miroshnichenko17}. Very recent MATISSE interferometry at thermal infrared wavelengths has already hinted at a rather complex geometry regarding the circumstellar matter distribution \citep{hofmann22}. In our GRAVITY data, HD 45677 has the largest extension of the \brg\ emitting region ($\sim$ 140 \Rstar), has the lowest contribution from the unresolved component ($\sim$ 20\% of the total flux), and is the only object with clear evidence of an extended, over-resolved component that contributes about 30\% of the total flux. The displacements and the PVD in Fig.\,\ref{Fig:PVDs1} show non-orderly motions that are difficult to interpret in terms of simple kinematics. A paper analysing, in detail, the complex circumstellar environment of this object is currently in preparation \citep{Sanchez-Bermudez_HD45677}, and we do not discuss the topic further in this paper.

\section {Discussion}
\label{Sec:Discussion}

  \begin{figure}
   \centering
       \includegraphics[width=9cm]{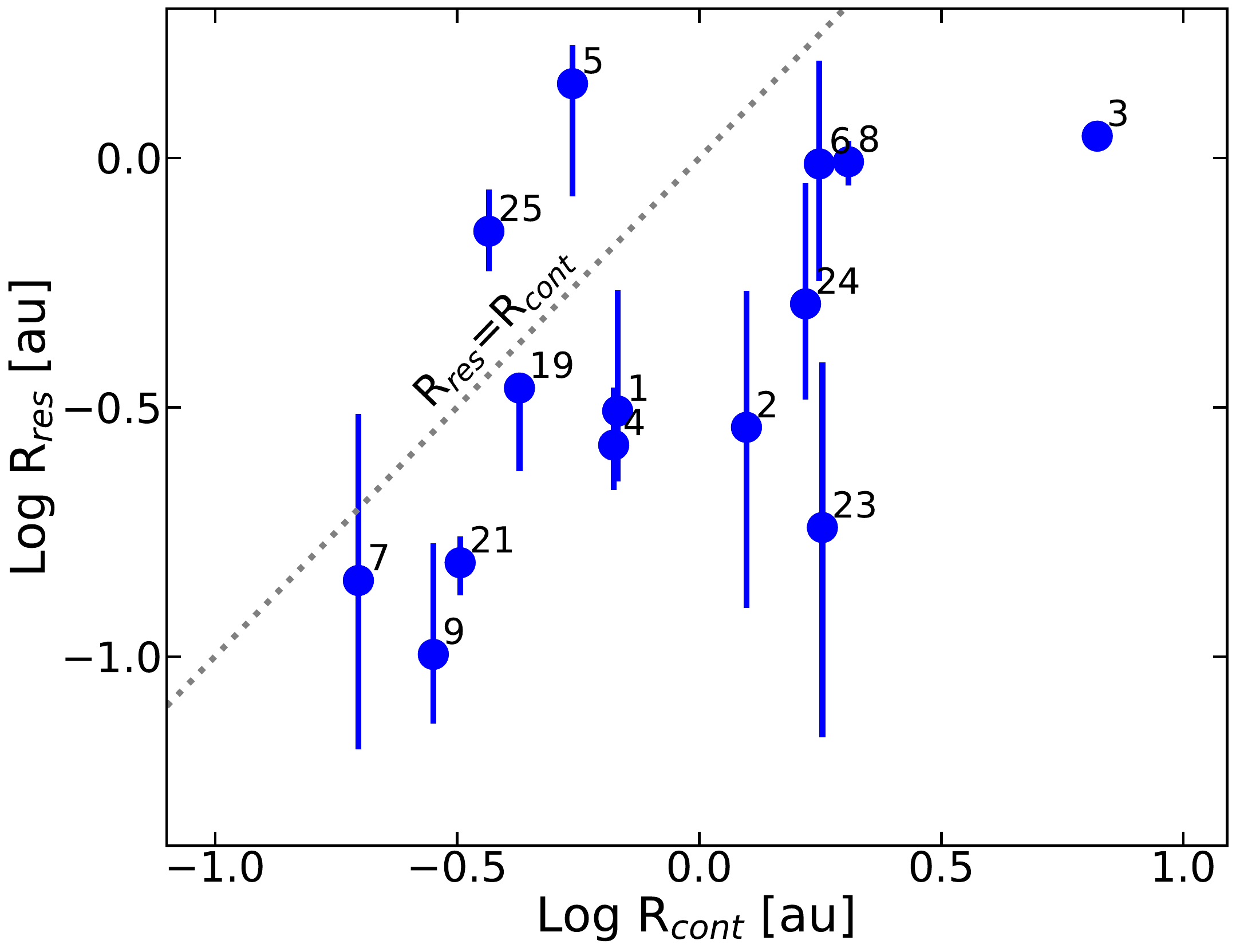}
      \caption{ Comparison of the radius of the \Brg\  line marginally resolved component to that of the  circumstellar continuum emission in the K-band (R$_{cont}$).
          The number labels are as in Fig.~\ref {Fig:fwhm_dist}.       }
         \label{Fig:RBrg-Rcont}
   \end{figure}

   \begin{figure}
   \centering
       \includegraphics[width=9cm]{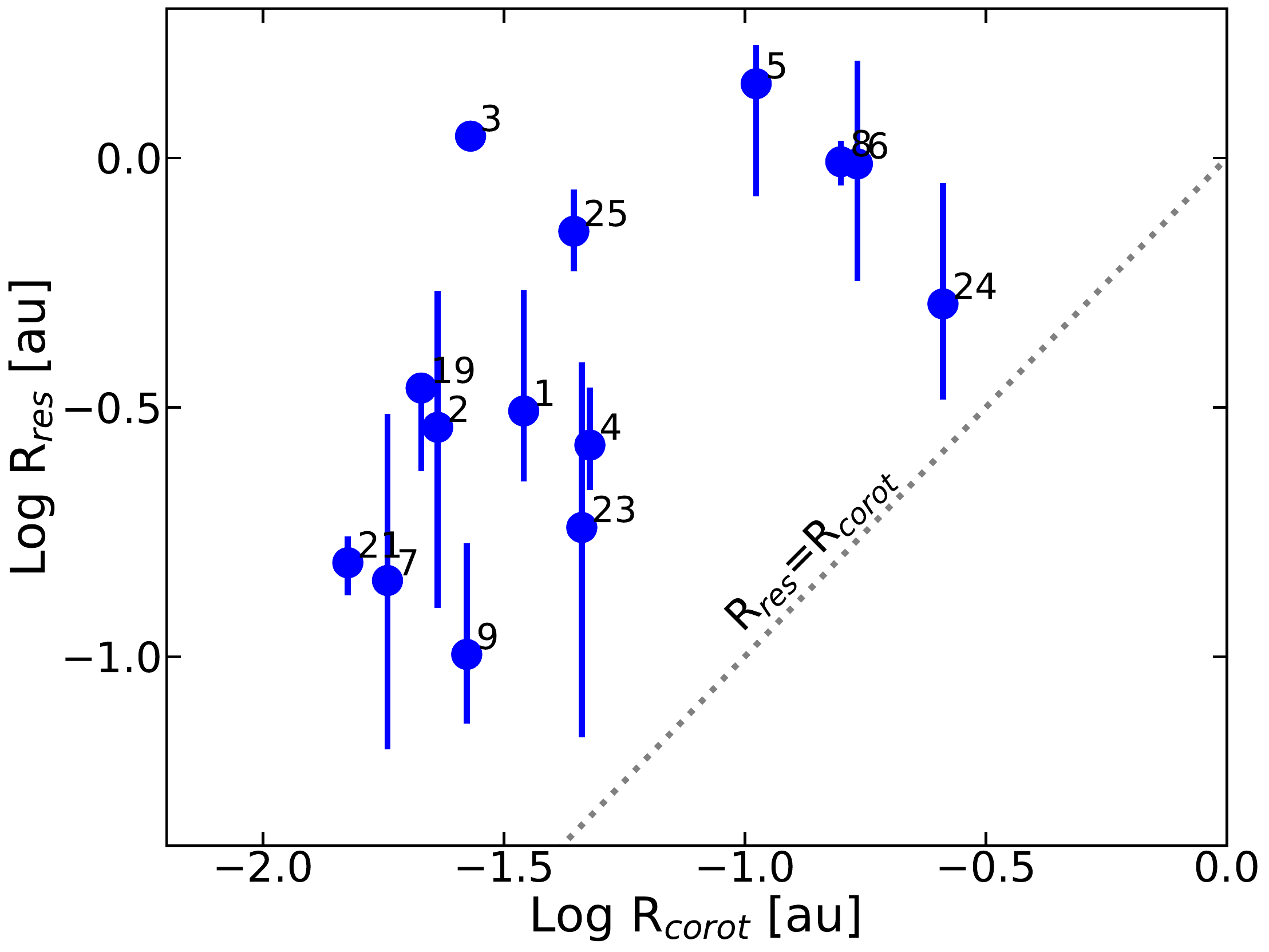}
      \caption{ Radius of the \Brg\ line marginally resolved component versus the co-rotation radius for all sources for which R$_{corot}$ is available (see Table\,\ref{tab:starproperties}). The number labels are as in Fig.~\ref {Fig:fwhm_dist}.  }
         \label{Fig:RBrg-Rcorot}
   \end{figure}  
   
 
    \begin{figure}
   \centering
       \includegraphics[width=9cm]{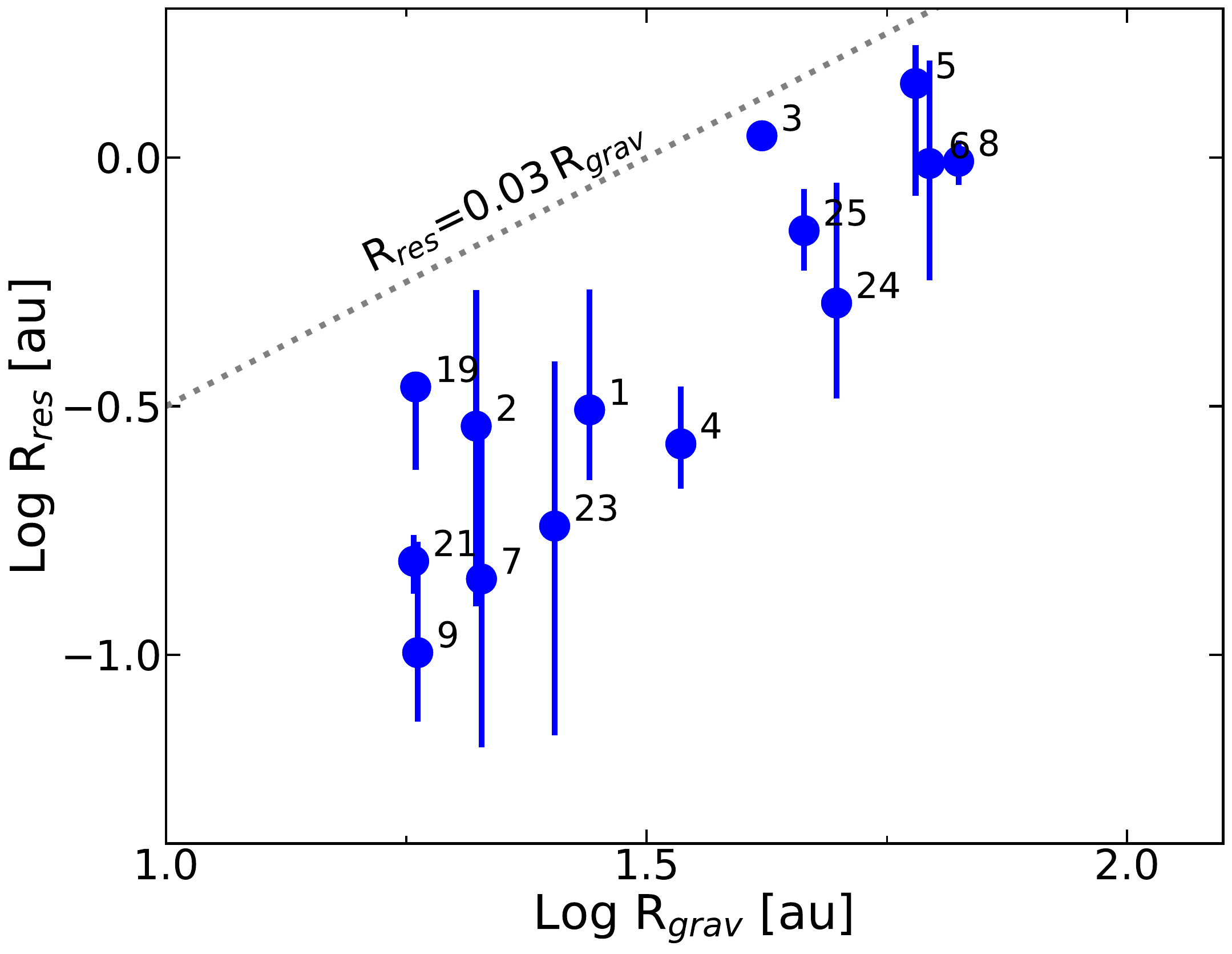}
      \caption{ Radius of the marginally resolved \Brg\ component versus the gravitational radius of the star computed for a sound speed of 10 km/s. The dashed line shows the locus of the points R$_{res}$=0.03$\,$R$_{grav}$ (see text for details). The number labels are as in Fig.~\ref {Fig:fwhm_dist}.  }
         \label{Fig:RBrg-Rgrav}
   \end{figure}

\subsection {Relation to the dusty disk}
The size of the \Brg\ emitting region is often compared to the size of the K-band continuum emission, which traces, in general, the silicate sublimation radius (\citep{karine19}). 
This, in itself,  is not immediately related to the line origin, which we discuss in Sect.\,\ref{sect:brg_origin}; however, it is an important factor to consider in modelling the properties of the line-emitting region as grains control the thermal gas properties. 

In general, the size of the \brg\ line extended component is smaller than the $K$-band continuum emission (see Fig.~\ref{Fig:RBrg-Rcont}). The exceptions are HD~259431 (\#25) and HD~85567 (\#5), for which R$_{res}$/R$_{cont}\sim$2.0--2.4. These two objects are among the hottest (T$_\mathrm{eff}\sim$13\,000\,K--14\,000\,K), and most luminous (log\,L$_*\sim$3--3.3) of our sample. 

Previous studies have shown that the size of the $K$-band continuum emission in Herbig Ae and T Tauri stars tightly correlates with the stellar luminosity as L$_{*}^{1/2}$ and is close to the location of silicate sublimation \citep{monnier02, karine19, karine21}. This relationship breaks for some very luminous Herbig Be stars, for which the  size of the K-band continuum emission is smaller than predicted by the above relationship. The sources HD~85567 and HD~259431 are the objects where this effect is stronger (together with PDS~27; \citealt{karine19}). A possible explanation for this is the presence of an inner gaseous disk inside the dust sublimation radius, 
which produces a smaller continuum emitting region than expected from simple dust sublimation physics. If this is indeed the case, then the \brg\ line would also be emitted in a dust-free region in HD~259431 and HD~85567. On the other hand, if the effect of a gaseous inner disk is simply to screen the stellar radiation and move the dust sublimation radius closer to the star, then grains could be present in the line-emitting region. 

As generally found for the K-band continuum emission, there is evidence in our data that the size of the \Brg\ line-emitting region is also related to the stellar luminosity (see Fig.\,\ref{Fig:RBrg_Lstar_all}). The relation R$_{res} \propto$ \Lstar$^{0.5}$ is expected if the physical conditions required to emit hydrogen recombination lines are controlled by the strength of the geometrically diluted stellar radiation field, as in the case of the dust sublimation temperature. 
We note that the fact that the two processes (i.e. the hydrogen line emission and the dust sublimation) are both controlled by the stellar radiation field does not mean that they occupy the same spatial region, as they indeed do not. It should be noted, however, that the correlation between the size of the marginally resolved \brg\ line emission and the stellar luminosity is most likely due to a strong distance dependence of these two quantities. 

 

\subsection{The kinematics of the innermost region}

As shown in Sect.\,\ref{sect:Brg_vis}, about 50\% of the \Brg\ flux is not spatially resolved by our observations. Therefore, we did not derive any of its properties from the visibilities alone, and we can only give an approximate upper limit to its size. However, further information on the kinematics of the gas on very small scales can be extracted if the continuum-subtracted line differential phases can be computed from the data. As discussed in  Sects.\,\ref{sect:Brg_vis}, and \ref{sect:line_displacements}, this was possible for eight objects for which 
the PA of the major axis of the emission (displacements) and the spectro-astrometric displacements (PVDs) could be computed. 

In most cases, the PVDs show the typical signature of gas in rotation, either spectro-astrometrically resolved (HD~98922)  or partially resolved (HD~163296, HD~58647, HD~85567, and HD~259431). The PVD of HD~98922 is consistent with the Keplerian rotation pattern of a star of HD~98922's mass, as given in Table\,\ref{tab:starproperties}, and the inclination derived from the visibility fits (solid curves in Fig.\,\ref{Fig:PVDs2}).
The predicted Keplerian rotation pattern is also shown for comparison for the partially resolved sources.
 
A comparison of the line PA (PA$_{disp}$) with that of the K-band continuum disk (PA$_{cont}$; Table\,\ref{tab:results}) showed that in several cases, both values coincide within the errors. In partially resolved sources, this indicates that the \brg\ emitting region is aligned with the disk, that is, the line is emitted close to the disk midplane, where Keplerian rotation dominates the gas motion. A wind farther from the launching region, where poloidal motion dominates the kinematics, would have a PA strongly inclined with respect to the disk, as seen, for example, in jets \citep{pascucci22, lesur22}. 
 Two objects (HD~58647 and HD~85567) show evidence of partially resolved Keplerian rotation. Objects that are partially resolved may have a complex behaviour that needs detailed and more complex models in order to be understood \citep[e.g. the presence of variable rotating disk asymmetries][]{kobus20, varga21, Sanchez-Bermudez21, Ganci23}. 
 
We also note that several objects have displacements not centred at (0,0) coordinates. This is likely a consequence of the fact that the observed continuum-subtracted differential phases measure displacements from the adjacent continuum assumed to be symmetric around its photo-centre. In practice, any asymmetric structure in the continuum emitting region, such as a warp in the inner disk or a difference in disk flaring, could be at the origin of the shift from the (0,0) position in the displacement and PVDs, even when the line-emitting region itself is symmetric around the photocenter. For instance, the presence of uneven disk-continuum illumination has been invoked to account for the non S-shaped profile of the observed \brg\ line differential phases in HD~98922 \citep{ale15}. Such a scenario could also explain the \brg\ line displacements measured in this source, where both the blue- and red-shifted velocity displacements are oriented towards the south direction (see Fig.\,\ref{Fig:PVDs2}). 
 
In contrast, \cite{kluska20} computed H-band continuum reconstructed images of five of our sources (HD~144668, HD~98922, HD~100546, HD~163296, and HD~150193). The authors classified these objects as symmetric or asymmetric sources. Of those,  
HD~37806 was classified as a symmetric object, whereas HD~45677, HD~98922, HD~100546, and HD~163296 were classified as asymmetric objects. These results are in agreement with our displacements with the exception of HD~37806, for which a clear displacement offset was observed. This might indicate that the compact \brg\ line emission is indeed asymmetric or that the H-band continuum is asymmetric on spatial scales smaller than those sampled in \cite{kluska20}, i.e. $\lesssim$1.7\,mas.

\subsection{The origin of the \HI\,\brg\ line emission}
\label{sect:brg_origin}

\subsubsection {Magnetospheric accretion}

The \HI\ line emission, and in particular the \brg\ line emission, is often considered a tracer of accretion. This is because of the well-established empirical relationship between the \HI\ line luminosity and the accretion luminosity \citep[see e.g.][]{muzerolle04, alcala14}. This correlation holds for a wide range of stellar masses, from brown dwarfs up to HAEBE stars \citep[see e.g.][]{natta04, gatti08, calvet04, rebeca06}. 

{This empirical relationship is expected in the current accretion paradigm, as it is well established for T Tauri stars,} where hydrogen lines are emitted by matter accreting onto the star along stellar magnetic field lines \citep[see e.g.][and references there in]{hartmann16}. The size of this region depends on the strength of the magnetic field and on the mass-accretion rate \citep[see e.g.][]{bessolaz08}, and it is typically of $\sim$5\,\Rstar\ in T\,Tauri stars. Herbig AeBe stars have much weaker magnetic fields than T\,Tauri sources \citep[e.g.][]{alecian09, hubrig11, alecian13, hubrig15} and comparatively high mass-accretion rates. Therefore, the magnetospheric emitting region is much smaller, with an upper limit  of $\sim$1-2\,\Rstar. Any emission from such a region will be unresolved in our observations.
Furthermore, accretion occurs if the magnetosphere is roughly confined within the disk co-rotation radius, that is, within the distance from the star where the Keplerian angular velocity equals the stellar angular velocity (i.e. $R_{co}=(G M_{*})^{1/3} \Omega_{*}^{-2/3}$).  This is not the case for the HAEBE in our sample.
Figure\,\ref{Fig:RBrg-Rcorot} clearly shows that the size of the marginally resolved component is significantly larger than the value of the co-rotation radius, with \Rres /\Rcorot $\sim$5--20. We then concluded that magnetospheric accreting columns are not the source of the marginally resolved \brg\ line emission. 

On the other hand, the small size of the magnetosphere in HAEBE is, in principle, consistent with the dimensions of the unresolved component but not with the Keplerian rotation pattern that we detected in some objects, as most line emission in magnetospheric columns occurs far from the disk surface, where the gas motion is mostly radial \citep[e.g.][]{muzerolle04, hartmann16}. 
Recent studies have questioned the validity of the magnetospheric accretion scenario in weakly magnetised Herbig AeBe stars and point towards a change of the accretion paradigm within the Herbig AeBe mass regime \citep{vink02, vink15, patel17, ababakr17, marcos-arenal21} from magnetospheric to boundary layer accretion \citep{lynden-bell74, takasao18}. A boundary layer is expected to form very close to the stellar surface and to have a very small radial extension. Any emission from this region will be by far too small when compared to the 
size of the extended \brg\ line emission \citep[e.g.][]{kley96}.  It may be possible that the unresolved \brg\ component traces the boundary layer emission. Although, when the Keplerian pattern is clear, it seems to indicate that the velocity increases at smaller radii, rather than decreasing, as expected in boundary layer accretion. Detailed models and better data are required if this hypothesis is to be explored further.

\subsubsection {Disk winds}

A more convincing scenario that we can infer from our sample is that the \brg\ line mostly forms in outflowing gas. In particular, two different kinds of winds have been extensively studied, namely, photoevaporative winds and the so-called magneto-centrifugally launched winds.

Photoevaporative winds are thermally driven winds 
launched  where  the radiative pressure of the gas at the disk surface is larger than the gravitational pressure. In this context, a fundamental parameter is the so-called gravitational radius (R$_{grav} $= G M$\ast$/$c_s^2$, with $c_s$ the sound speed),  where a gas parcel becomes unbound from the central star \citep{hollenbach00, alexander14, picogna21}. 
The specific mass-loss rate generally peaks at 0.1--0.5\,R$_{grav}$ \citep[see e.g.][]{alexander14}. Depending on the adopted model and assuming a stellar mass of $\sim$2\,\msun\, this translates into a peak of the mass-loss rate around $\sim$1.5--2\,au for the case of EUV-driven photoevaporation. This radius moves even further out for X-ray-driven and FUV-driven winds \citep{alexander14, kunitomo21}. Our measurements do not support these models, as the size of the \Brg\ line-emitting region is by far too small. This is clearly demonstrated in Fig.~\ref{Fig:RBrg-Rgrav}, which shows that R$_{res}$ is  30 to 100 times smaller than the gravitational radius. As a consequence, photoevaporative winds cannot account for the \brg\ line emission. 

A more convincing interpretation is that the \brg\ line traces a magneto-centrifugally launched wind. A fundamental ingredient of these models is a magnetic field anchored to the disk, possibly a residual of  the original collapsing cloud magnetic field. In the simplest version, a wind is launched if the disk magnetic field lines are inclined by more than 30\degr\ from the polar axis \citep{blandford82}. In recent years, more sophisticated MHD winds have received growing attention thanks to the improvement of simulations and the suspected role of the winds in removing angular momentum and thus allowing accretion to occur \citep[see e.g.][]{bai16,zhu18,Jacquemin21}. Depending on the disk region from which these winds originate, different launching mechanisms are involved 
\citep[see e.g.][for recent reviews]{pascucci22,lesur22}. 
 
In our case, to reproduce the extent of the \brg\ line-emitting region, the wind emission needs to occur within a region of  $\sim$10--30\,\Rstar\ from the central source. Due to the high luminosity of our sources and their weak magnetic fields, several wind launching mechanisms might be in place within this region. A weakly magnetised disk rotating at Keplerian velocity and with high enough levels of ionisation is known to be unstable against magneto-rotational instability \citep[MRI; ][]{balbus91}. Recent simulations have shown that the levels of ionisation around typical Herbig AeBe stars up to $\sim$1\,au from the sources are sufficient to sustain  MRI and thus launch an MRI wind \citep{suzuki09,bai13_MRI,flock16,flock17,Jacquemin19}. These winds are launched above the disk mid-plane, with their base rotating at the Keplerian velocity of the disk footpoint at which the magnetic field lines are anchored. As matter blows up following the magnetic field lines, the wind accelerates, and the poloidal velocity becomes dominant over the tangential component.

Previous interferometric studies have successfully modelled the \brg\ emission properties using simplified models of  disk winds launched centrifugally  from the disk surface \citep{weigelt11, rebeca15, ale15, rebeca16, kurosawa16, kreplin18}. The launching regions extend from 1--3\,\Rstar\ to $\sim$10--30\,\Rstar, consistent with the location of the \brg\ emission observations in this paper. The models assume that the line is emitted well above the disk mid-plane (the so-called warm disk-wind models), where an originally cold wind is heated by ambipolar diffusion as it moves above the disk and reaches a roughly constant temperature of around 10\,000\,K  \citep[e.g.][]{safier93,garcia01a}. In these regions, the poloidal velocity component may be large in comparison to the rotational one. Specifically, this approach was used to reproduce the single-peaked \brg\ line profiles and spectro-interferometric observations of several Herbig AeBe stars using VLTI-AMBER observations at R$\sim$12\,000 \citep{weigelt07,ale15,rebeca15,kurosawa16}.
In fact, even at GRAVITY's modest spectral resolution, magneto-centrifugally launched winds should produce double-peaked line profiles for disk inclinations larger than $\gtrsim$40\degr\  if the line emission starts at the wind base \citep{larisa14,larisa16}. Although all the objects with double-peaked \brg\ line profiles in our sample (HD\,58647, HD\,114981, HD\,141569, and HD\,158643) have continuum and/or \brg\ line inclinations $\gtrsim$40\degr, not all the sources with inclinations $\gtrsim$40\degr\ have double-peaked line profiles (see line profiles in Appendix \,\ref{ap:data}). 

Wind models can then reproduce the basic features of the interferometric observations presented here, in particular, the extent of the launching region and the characteristics of the line profiles. Winds are ejected at all scales between the inner and outer radius, with no gaps separating the resolved or marginally resolved component from the unresolved component defined in our visibility analysis. In fact, these separations are indeed arbitrary, justified only by the length of the baselines available with VLTI-GRAVITY. Support of the common origin of the two components is provided  by the fact that, as mentioned in Sect.\,\ref{sect:line_displacements}, the PA derived from our line displacements is generally in agreement with the PA obtained for the marginally resolved \brg\ line-emitting region in Sect.\,\ref{sect:Brg_vis}.

\begin{figure}
   \centering
       \includegraphics[width=9cm]{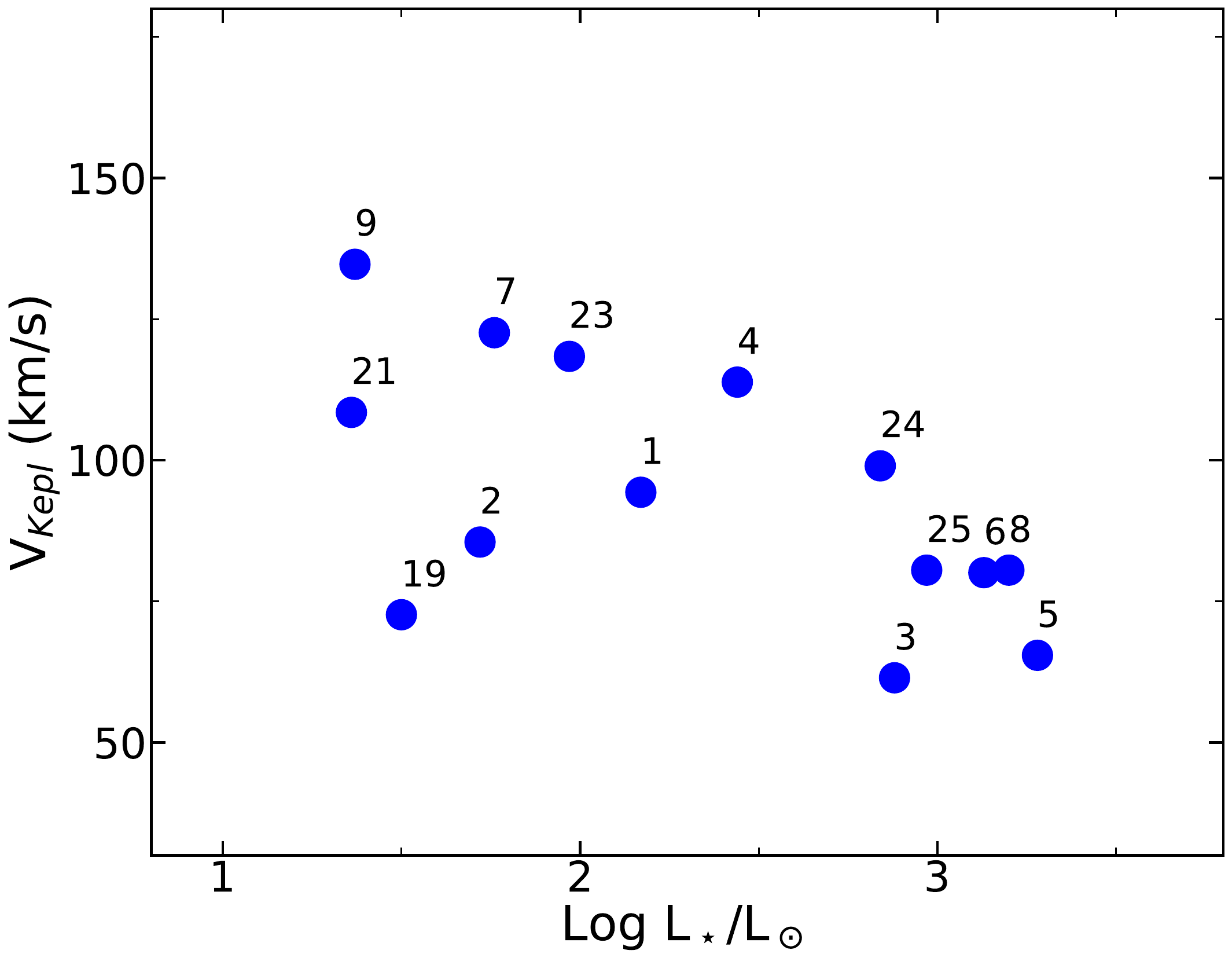}
      \caption{Keplerian velocity at the radius of the marginally resolved \brg\ line emission R$_{res}$ shown as a function of \Lstar. The number labels are as in Fig.~\ref {Fig:fwhm_dist}.  }
         \label{Fig:Vkepl-Lstar}
   \end{figure}

The models can be revisited to include the results on the unresolved component discussed above. 
The Keplerian pattern detected in the innermost regions and the similarity between the line and continuum PA suggest that at small radii the line emission originates close to the launching region, where the Keplerian rotation still dominates the wind kinematics.  
On the other hand, the \brg\-resolved component has a relatively low rotational velocity, even close to the launching point, as the Keplerian velocity at R$_{res}$ is always $<$150\,\kms\ (see Fig.\,\ref{Fig:Vkepl-Lstar}), and its component along the line of sight is  $<$80\,\kms\ in all objects. This is likely to be unresolved in our GRAVITY spectra, given their low spectral resolution  ($\sim$75\,\kms). This kind of data show the potential of future high-resolution optical spectro-interferometers to constrain MHD wind models and explore wind properties such as thermal structure and dependence on the radius of the mass loading.

%

\subsubsection{Disk surface}

An additional possibility, which so far has been very little investigated, is that some of the  \brg\ emission comes from the disk surface, heated by the stellar radiation. This region has similar physical properties to the base of a disk wind, and thus separating gravitationally bound from unbound material may be observationally very challenging. In particular, realistic models of the gas temperature and density at the disk surface in regions completely dust depleted are still lacking, and therefore, we lack accurate line profile modelling accounting for the total flux expected from this region. 
Nevertheless, as in wind models, one expects that the lines will display a double-peaked profile, although the relative contribution to the line emission at each radius will depend on the physical properties of the disk. 
Therefore, it is also possible that the contribution of the bound gas is relevant only at small scales, close to the source, where, indeed, we detect the signature of Keplerian rotation in the line emission.


%

\subsubsection{The \brg\ line luminosity versus accretion luminosity correlation}

After our previous discussion, it may be surprising that in young stars, the luminosity of \brg\ (and, more in general, of other emission lines) shows a very good correlation with the accretion luminosity.  We find it worth stressing that this is an "empirical" correlation well established over a large range of stellar and accretion properties~\citep[see, e.g.][]{rebeca06,alcala14,alcala17,fairlamb17}.

However, we have shown that a significant fraction of the \brg\ emitting region in Herbig AeBe stars is extended and that around 50\% or more of the emission cannot be ascribed to magnetospheric or boundary layer accretion.
Moreover, most recent interferometric studies of the \brg\ line towards classical T\,Tauri stars (CTTs) have shown that in only two cases so far~\citep[TW\,Hya and DoAr44; see][respectively]{rebeca20,bouvier20B} the size of the \brg\ emitting region is consistent with magnetospheric accretion alone. For a larger sample of seven CTTs, which also includes those two sources, \citet{GRAVITY_TTauri} find that, while magnetospheric accretion is the primary driver of \brg\ emission in those two weak accretors, the remaining CTTs have partial or strong contributions from spatially extended \brg\ components, similar to the HAEBE sample.

\begin{figure}
   \centering
       \includegraphics[width=9cm]{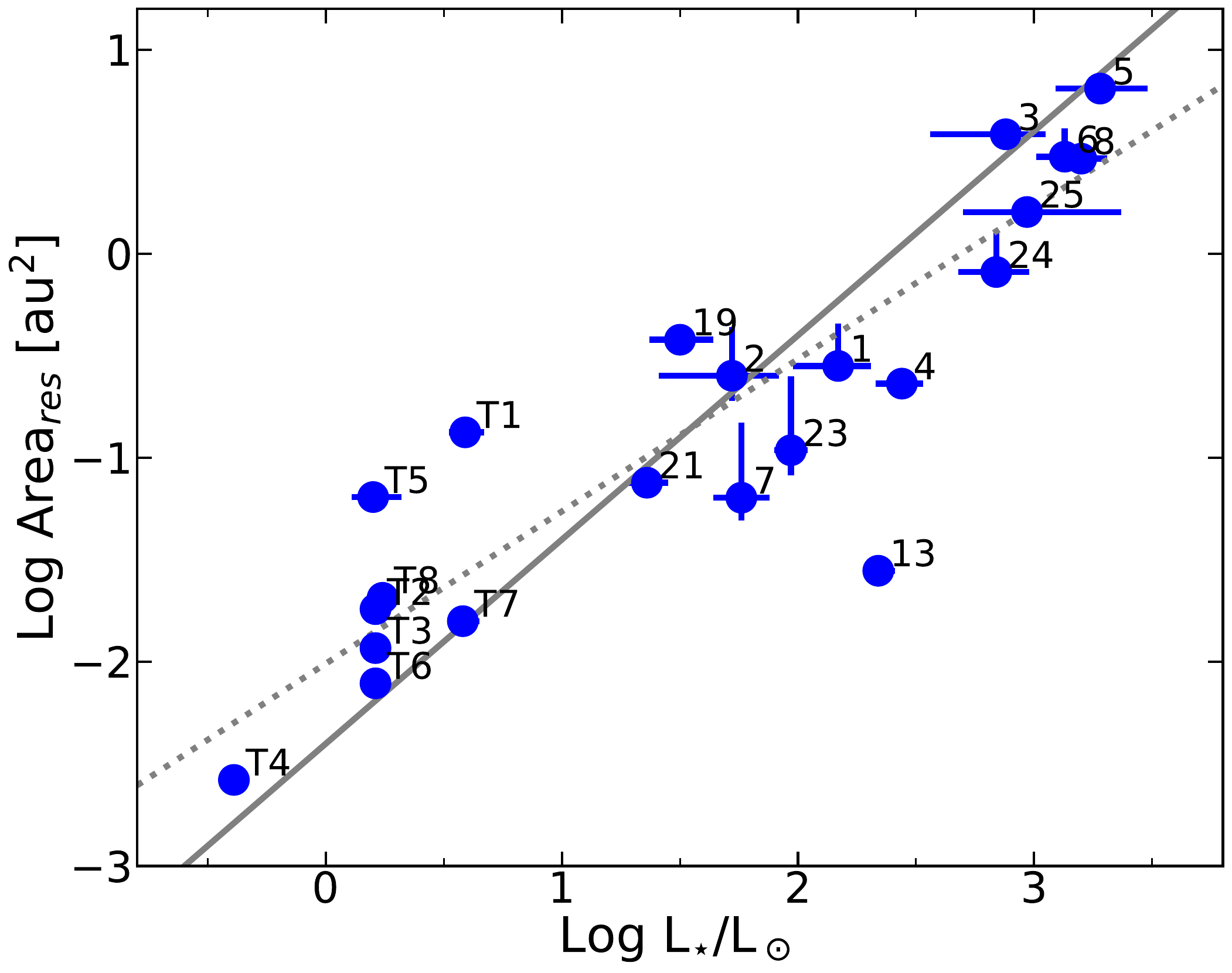}       
      \caption{Area of the region emitting the marginally resolved  \brg\ component ($\pi \,R_{rses}^2$)  plotted as a function of \Lstar for both the Herbig AeBe and CTTs samples. The number labels of the Herbigs are as in Fig.~\ref {Fig:fwhm_dist}. CTTs data are taken from \citet{wojtczak}. The number labels of the CTTs are the following: T1: AS\,353; T2: RU\,Lup (epoch 2018); T3: RU\,Lup (epoch 2021); T4: TW\,Hya; T5: DG\,Tau; T6: DoAr\,44; T7: S\,CrA; T8: VV\,CrA.
      The dotted and solid lines show the best-fitting line of slope $0.75 \pm 0.04$ (reduced $\chi ^2$ of 0.14) and slope 1.0 (reduced $\chi ^2$ of 0.25), respectively.  }
         \label{Fig:RBrg_Lstar_all}
   \end{figure}

\begin{figure}
   \centering
       \includegraphics[width=9cm]{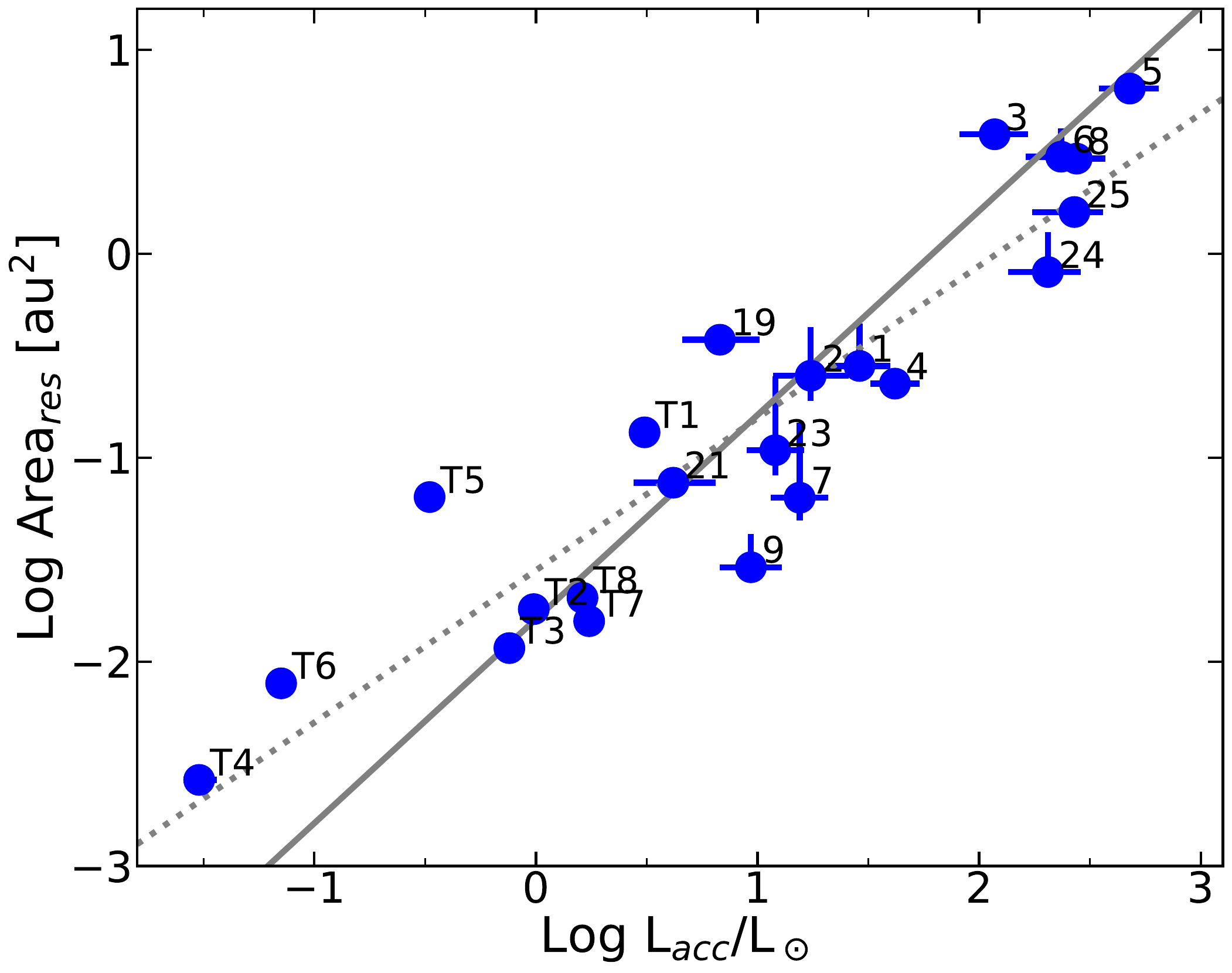}       
      \caption{Area of the region emitting the marginally resolved \brg\ component as a function of $L_{acc}$ for both the Herbig AeBe and CTTs samples. The number labels are as in Fig.~\ref {Fig:RBrg_Lstar_all}. The dotted and solid lines show the best-fitting line of slope $0.75 \pm 0.04$ (reduced $\chi^2$ of 0.15) and slope 1 (reduced $\chi^2$ of 0.24), respectively. 
              }
        \label{Fig:RBrg_Lacc_all}
   \end{figure}

Figures~\ref{Fig:RBrg_Lstar_all} and \ref{Fig:RBrg_Lacc_all} show the area of the \brg\ emitting region, defined as $\pi \, R_{res}^2$, against both stellar and accretion luminosity, respectively. 
The sample shown in these figures includes the HAEBE sample presented in this paper and the CTTs sample presented in \citet{GRAVITY_TTauri} (see their Table\,1 and 3 for stellar and \brg\ parameters). The Herbig AeBe stars are labelled as in Table\,1, and CTTs labels are provided in the caption of Figure~\ref{Fig:RBrg_Lstar_all}. We chose to plot the area of the emitting region rather than its radius, as  both areas and luminosities have the same dependence on the distance of the source ($\propto D^2$). This is particularly important for the HAEBE sample, which is spread over a large range of distances. Given the limited range of projected sizes available with VLTI-GRAVITY (see Fig.\ref{Fig:fwhm_dist}), this introduces a strong correlation between $R_{res}$  and the distance of the star so that, to zero order, $R_{res}\propto$ \Lstar$^{0.5}$.
Figure~\ref{Fig:RBrg_Lacc_all} suggests that there is a correlation between the \brg\ emitting area and L$_{acc}$, as the best-fitting slope is different than one (0.75$\pm$0.04). 
If the area of the emitting region is a good proxy of the line luminosity, then the wind (or extended) component of the \Brg\ line is directly correlated with the accretion luminosity. We note, however, that similar to the results obtained for emission lines in general, the area of the  marginally resolved \brg\ component is equally related to \Lstar\ (see Fig.~\ref{Fig:RBrg_Lstar_all}) and has an identical slope (0.75$\pm$0.04). Indeed, similar empirical relationships between the line luminosity of wind and/or jet tracers (e.g. [\ion{O}{i}]) and the stellar and accretion luminosity have been observed in HAEBE and T Tauri samples~\citep[see, e.g.][]{fairlamb17, nisini18, banzatti19}.
 
For now, we can only hypothesise that hydrogen recombination lines trace hot gas located at the disk surface that is being accreted and/or ejected. All of these processes are related and interdependent, and they correlate somehow with the accretion luminosity. In order to further probe the origin of the empirical correlation between \lacc\ and \Lstar and the \brg\ line luminosity, additional interferometric surveys of complete samples located at similar distances (i.e. same star-forming region) should be performed. Such surveys would likely be possible with the advent of more sensitive intereferometers, such as GRAVITY+, giving us the opportunity to study more homogeneous and unbiased samples.

\section{Summary and conclusions}

This paper presents the results of an interferometric study using VLTI-GRAVITY of the region emitting the hydrogen \brg\ line in a sample of 26 intermediate mass stars. Of these, nine did not show any line emission at the time of the observations, even after correcting for the underlying absorption photospheric lines. For the remaining 17 targets, we derived continuum-subtracted visibilities and differential phases when possible.

We derived the size of the line-emitting region by performing a 2D Gaussian fit to the continuum-subtracted visibilities. The results show that 15 sources are marginally resolved, with typical radii of 10--30\,\Rstar, and two were not resolved for our baselines. A significant fraction of the line flux, ranging from $\sim$20\% to 100\%, comes from regions that are not resolved even with the longest available baselines (B$\sim 130$\,m). In one case only (HD~45677), there is significant contribution ($\sim$30\%) from a more extended component that is over-resolved in our data. Further insight into the properties of the innermost  emitting region is provided by the continuum-subtracted line differential phases that could be inferred for eight sources. These data were used to compute the line displacement with respect to the continuum at different velocities and the PA of the line-emitting region in the plane of sky as well as to derive some kinematic information.

When compared to the dusty disk traced by the K-band continuum emission \citep{karine19}, the \brg\ emission is generally confined to a region inside the dust sublimation radius. The visibility fits provided rough information on the inclination and PA of the marginally resolved emitting region. However, for the eight objects for which continuum-subtracted differential phases could be extracted, we measured the PA with a much higher accuracy. These values are consistent with those found from the visibility fits. When compared to the continuum disk, the continuum and \brg\ emission PAs are roughly in agreement in all cases.  

The PVDs were computed from the continuum-subtracted differential phases for the same eight objects. They provide kinematic information on the innermost disks on scales of a few \Rstar.  Within the uncertainties, the diagrams are typical of disks in Keplerian velocity, with the only clear exception being HD~45677, which shows evidence of very complex kinematics. The source HD~98922 shows the typical "butterfly" pattern expected for disks in Keplerian rotation when the emission is fully resolved, with values consistent with the stellar mass and the inclination of the line-emitting region. The sources HD~163296, HD~58647, HD~85567, and HD~259431 have the typical S shape of a disk in Keplerian rotation shown in the PVD of marginally resolved sources. 

A discussion on the possible origin of the \Brg\ emission favours a disk-wind origin. 
Photoevaporative winds are not a possible explanation, as they are expected to originate at much larger scales than the 10--30\,\Rstar\ typical of the resolved \brg\ component. The MHD disk winds are more likely, as they can be ejected over a wider region, extending very close to the star. The Keplerian rotation pattern, revealed by the PVDs in the innermost regions of the few objects where this information is available, is expected if the emission originates at the base of the wind, still close to the disk surface. However, we cannot exclude that the \brg\ emission in the innermost disk comes from gas on the surface of the disk itself, namely, gaseous layers that are still "bound" by the system's gravity. It is worth stressing that the data lack the required spectral and spatial resolution for more detailed studies,which could reveal more complex kinematical patterns. 

Finally, by combining the HAEBE sample presented here with the T Tauri sample shown in \citet{GRAVITY_TTauri}, we noted a correlation between the area of the \brg\ emitting region and both the accretion and stellar luminosities. Although more homogeneous and unbiased samples are required to prove whether such correlations are real, they might explain the existence of the well-known empirical correlation between the \brg\ line luminosity and the stellar accretion luminosity. 

The results outlined in this paper show the potential of near-IR interferometry to aid in understanding the phenomena that occur in the innermost disk of young stars. The results also reveal the need to obtain higher-quality data, both in terms of spectral and spatial resolution. Moreover, the current effort to reconstruct images of both continuum and line emission needs to be pushed farther. At the same time, models of the inner disk's physical properties, of the winds launched from the disk surface at very small scales, and of the accretion region need to be improved so that predictions of the emission line spectrum can be carried out and compared with the data.

\begin{acknowledgements}
This  material  is  based  upon  works  supported  by  Science  Foundation  Ireland under Grant No. 18/SIRG/5597. R.F. acknowledges support from the grants Juan de la Cierva FJC2021-046802-I, PID2020-114461GB-I00 and CEX2021-001131-S funded by MCIN/AEI/ 10.13039/501100011033 and by ``European Union NextGenerationEU/PRTR''. R.F. also acknowledges funding from the European Union’s Horizon 2020 research and innovation programme under the Marie Sklodowska-Curie grant agreement No 101032092. A.C.G. has been supported by PRIN-INAF MAIN-STREAM 2017 “Protoplanetary disks seen through the eyes of new generation instruments” and from PRIN-INAF 2019 “Spectroscopically tracing the disk dispersal evolution (STRADE)”. 
T.P.R. has received funding from the European Research Council (ERC) under the European Union's Horizon 2020 research and innovation programme (grant agreement No.\ 743029). A.N. acknowledges the kind hospitality of DIAS. A.A., M.F. and P.J.V.G acknowledge funding by Fundação para a Ciência e a Tecnologia, with grants reference UID/FIS/00099/2013 and SFRH/BSAB/142940/2018. T.H. acknowledges support from the European Research Council under the Horizon 2020 Framework Program via the ERC Advanced Grant Origins 832428. 
We thank the technical, administrative, and scientific staff of the participating institutes and the observatory for their extraordinary support during the development, installation, and commissioning of GRAVITY. This research has made use of the Jean-Marie Mariotti Center \texttt{Aspro} and \texttt{SearchCal} services,  \footnote{Available at http://www.jmmc.fr/} and of CDS Astronomical Databases SIMBAD and VIZIER \footnote{Available at http://cdsweb.u-strasbg.fr/}.
\end{acknowledgements}

\bibliographystyle{aa} 
\bibliography{references.bib} 

\begin{appendix}

\section{Log of the observations}

\begin{table*}[h]
\label{tab:obs}

\caption{Observation log of VLTI-GRAVITY observations.}
\centering
\vspace{0.1cm}
\begin{tabular}{c c c c c c}
\hline \hline
HD & Date & Configuration & $\lambda$/PBL$_{max}$ & N & Calibrator \\ 
\hline 
37806 & 2019-03-19 & D0-G2-J3-K0 & 4.7 & 5 & HD~64215\\
      & 2020-01-22 & A0-G1-J2-K0 & 3.6 & 10 & HD~31623\\ 
\hline
38120 & 2019-03-18 & D0-G2-J3-K0 & 5.1 & 8 & HD~37356, HD~38225\\
\hline
45677 & 2020-01-27 & A0-G2-J2-J3 & 3.5 & 7 & HD~43847, HD~49794\\ 
      & 2020-01-29 & K0-G2-D0-J3 & 4.5 & 6 & HD~44423\\
      & 2020-02-04 & A0-B2-D0-C1 & 14.2 & 8 & HD~43847, HD~45420\\
 \hline
58647 & 2020-01-27 & A0-G1-J2-K0 & 3.5 & 9 & HD~57939, HD~60325\\
      & 2020-01-28 & A0-G2-J2-J3 & 4.4 & 7 & HD~60325\\
      & 2020-02-04 & A0-B2-D0-C1 & 13.7 & 12 & HD~65810\\
\hline
85567 & 2021-02-09 & A0-G1-J2-J3 & 3.8 & 18& HD~85313\\
      & 2021-02-10 & A0-G1-J2-J3 & 3.5 & 9 & HD~85313\\
      & 2021-02-20 & A0-B2-D0-C1 & 15.6 & 8 & HD~85313\\
\hline
95881 & 2018-03-02 &  D0-G2-J3-K0 & 5.2 & 5 & HD~90452\\
\hline
97048 & 2017-03-20 & A0-G1-J2-K0 & 3.5 & 6 & HD~82554\\
\hline
98922 & 2019-05-24 & A0-G1-J2-J3 & 3.7 & 6 & HD~103125, HD~141977\\
      & 2020-01-27 & A0-G2-J2-J3 & 3.5 & 5 & HD~103125, HD~99311\\
      & 2020-02-04 & A0-B2-D0-C1 & 14.6 & 11 & HD~103125\\
\hline
100546 & 2020-01-28 & K0-G2-D0-J3 & 5.4 & 7 & HD~101531, HD~99556\\
\hline
114981 & 2019-03-19 & D0-G2-J3-K0 & 4.4 & 6 & HD~113776 \\
\hline
135344B & 2018-03-05 & A0-G1-J2-J3 & 3.5 & 8 & HD~132763\\
\hline
139614 & 2019-03-19 & D0-G2-J3-K0 & 6.6 & 3 & HD~148974\\
\hline
141569 & 2019-07-12 & D0-G2-J3-K0 & 4.5 & 8 & HD~137006\\
\hline
142527 & 2017-03-18 & A0-G1-J2-K0 & 3.5 & 5 & HD~143118\\
\hline
142666 & 2018-06-16 & D0-G2-J3-K0 & 4.4 & 9 & HD~148605\\
\hline
144432 & 2018-03-05 & A0-G1-J2-J3 & 3.4 & 5 & HD~132763\\
\hline
144668 & 2017-03-19 & A0-G1-J2-K0 & 3.5 & 7 & HD~143118 \\ 
\hline 
145718 & 2019-03-18 &D0-G2-J3-K0 & 4.4 & 8 & HD~145809\\
\hline
150193 & 2018-06-15 & D0-G2-J3-K0 & 4.7 & 7 & HD~148605 \\
       & 2018-07-07 & D0-G2-J3-K0 & 5.0 & 3 & HD~181240\\
       & 2019-06-05 & A0-B2-D0-C1 & 15.1 & 14 & HD~148974\\
\hline
158643 & 2017-05-29 & B2-D0-J3-K0 & 3.4 & 6 & HD~163955\\ 
       & 2017-05-30 & B2-D0-J3-K0 & 3.4 & 14 & HD~163955\\
       & 2017-05-31 & B2-D0-J3-K0 & 3.7 & 12 & HD~163955\\
       & 2017-08-15 & A0-G1-J2-K0 & 4.8 & 7 & HD~163955\\
\hline
163296 & 2019-05-24 & A0-G1-J2-J3 & 3.5 & 4  & HD~163495\\  
       & 2019-06-05 & A0-B2-D0-C1 & 18.9 & 13 & HD~148974\\
       & 2019-07-14 & D0-G2-J3-K0 & 5.0 & 7 & HD~172052\\
\hline
169142 & 2019-05-24 & A0-G1-J2-J3 & 3.4 & 6 & HD~169830\\
\hline
179218 & 2018-07-07 & D0-G2-J3-K0 & 5.3 & 9 & HD~181240\\
\hline
190073 & 2018-06-15 & D0-G2-J3-K0 & 4.4 & 8 & HD~183936 \\
\hline
259431       & 2018-03-05 & A0-G1-J2-J3 & 3.5 & 12 & HD~43386, HD~50277\\
             & 2019-01-13 & D0-G2-J3-K0 & 5.0 & 10 & HD~262137\\
\hline
V1818 Ori & 2019-03-19 & D0-G2-J3-K0 & 4.4 & 6 & HD~34045\\
\hline
\end{tabular}
\tablefoot{ ( Column 1: HD name of the target. Column 2: Date of the observations. Column 3: Configuration of the observations. Column 4: $\lambda$/PBL$_{max}$ in units of mas; PBL$_{max}$ is the maximum projected baseline available for that configuration. Column 5: Number of five-minute long files that have been recorded on the target. Column 6: Name of the calibrator.)}
\end{table*}

\section{GRAVITY data}
\label{ap:data}
\begin{figure*}
\centering
\includegraphics[width=0.75\textwidth]{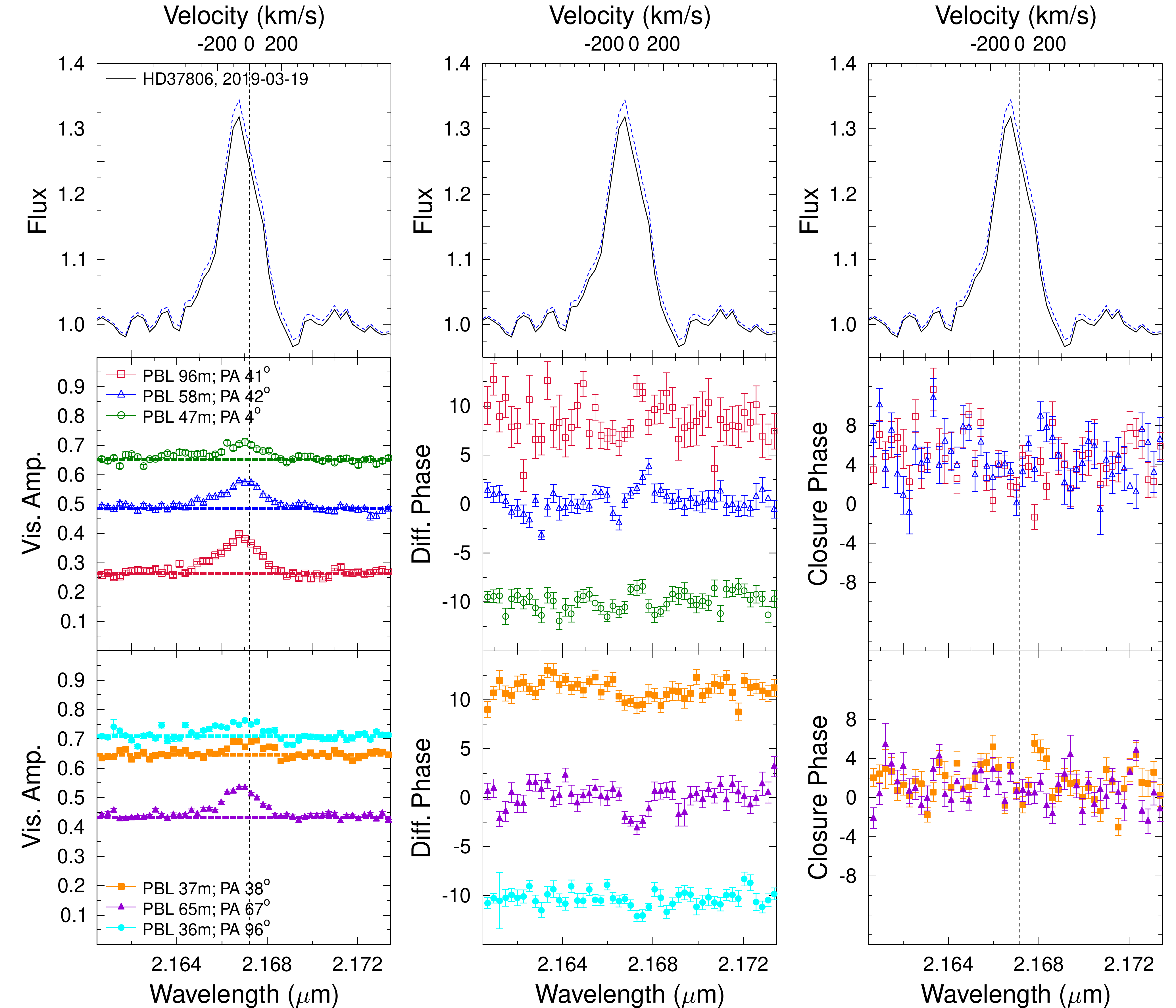}\\
\includegraphics[width=0.75\textwidth]{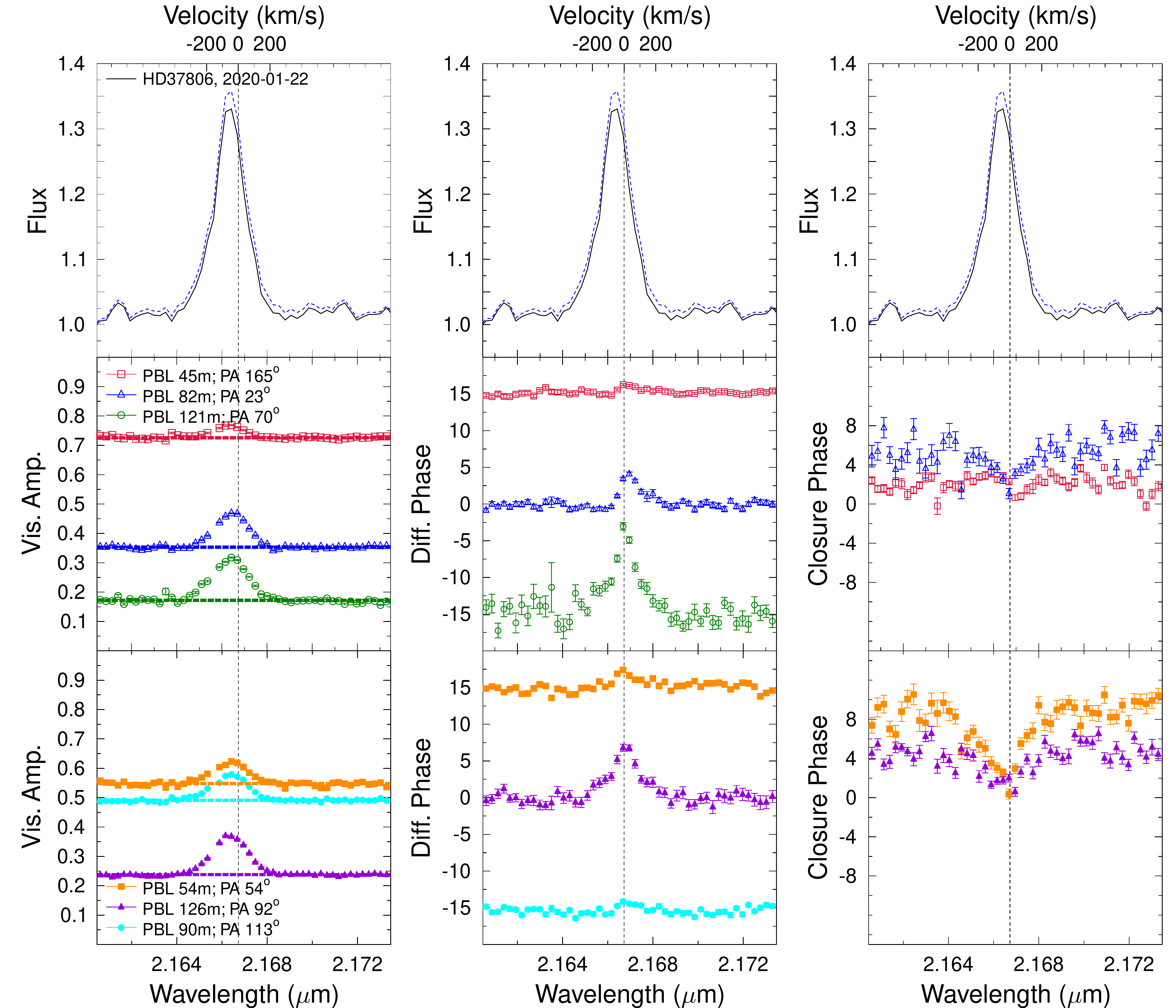}
\caption{VLTI-GRAVITY HR interferometric observables for the star HD~37806.  Each set of figures shows observations for a different date, indicated in the top-left panel. The three columns in each figure show the visibility amplitude (left), differential phase (middle), and closure phase (right). As reference, the observed (black solid line) and photospheric corrected (dahsed blue line) \brg\ line profiles are shown at the top of each panel.  Vertical dashed lines indicate the location of the zero velocity with respect to the local standard of rest.}
    \label{fig:data1}
\end{figure*}

\begin{figure*}
\centering
    \includegraphics[width=0.75\textwidth]{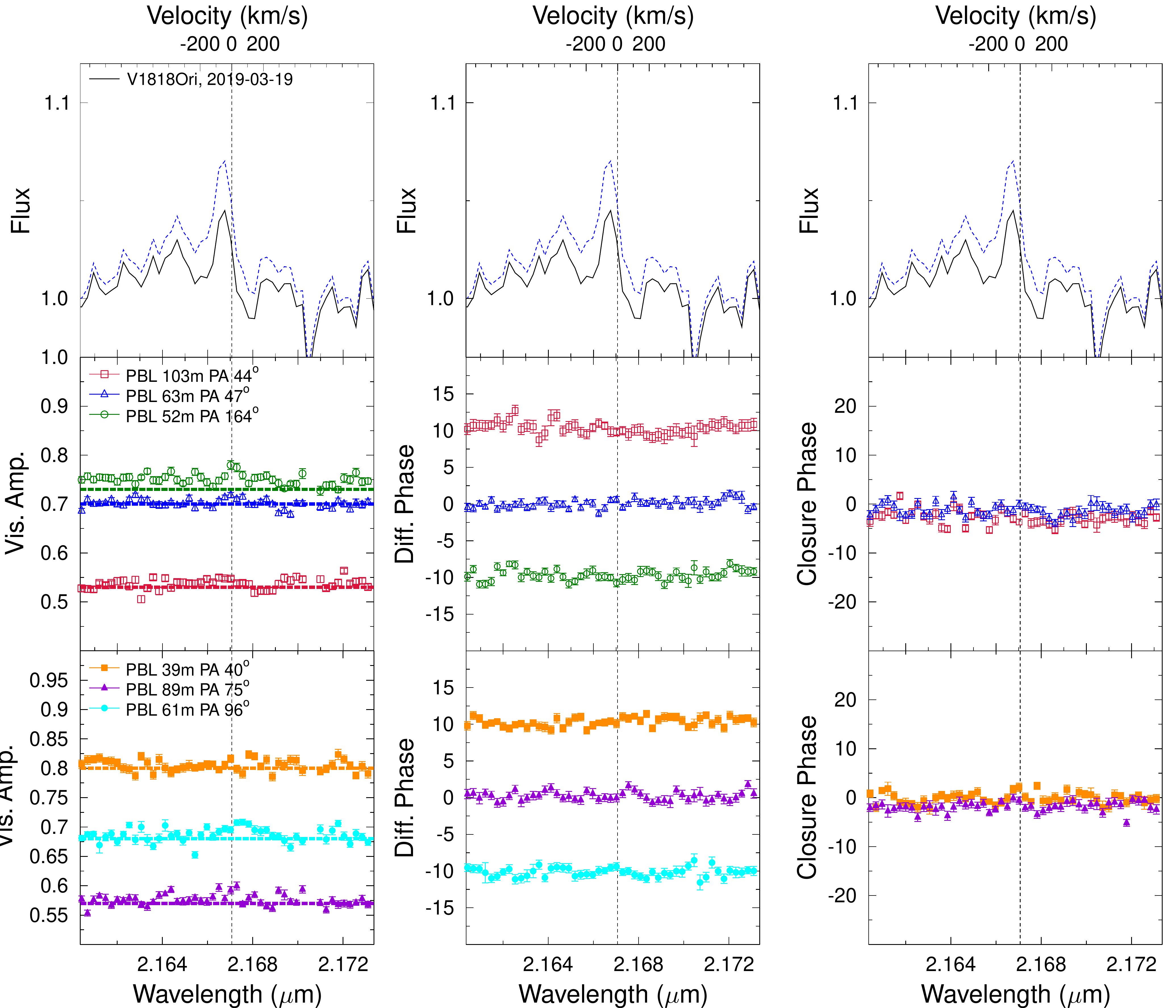}\\
    \caption{Same as Fig.\,\ref{fig:data1} but for V1818\,Ori.}
    \label{fig:data22}
\end{figure*}

The VLTI-GRAVITY observations of the full sample around the wavelength of the \brg\ line are shown from Fig.\,\ref{fig:data1} to Fig.\,\ref{fig:data22}. Each set of figures shows the results for a different star and date, as labelled in the top-left panel. There are three columns in each set of figures showing the visibility amplitude (left), differential phase (middle), and closure phase (right). The top panels show the observed (solid line) and the photospheric-corrected spectrum (dashed blue line), which is the same in all cases. The length and PA of the projected baselines (PBL) are given in the left panels. For reference, the vertical dashed line indicates the location of the zero velocity channel with respect to the local standard of rest.


\begin{figure*}
 \centering
 \includegraphics[width=0.75\textwidth]{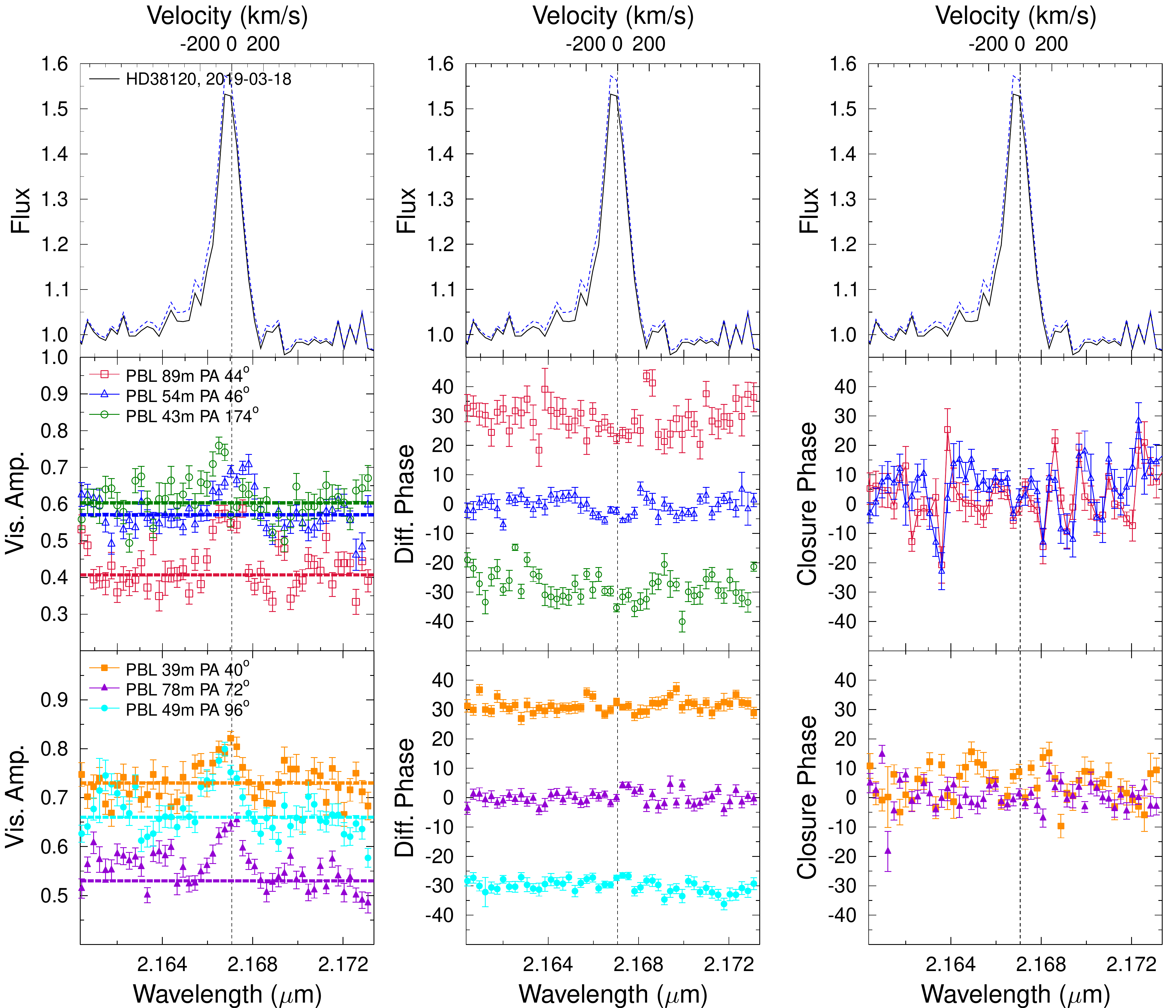}\\
 \includegraphics[width=0.75\textwidth]{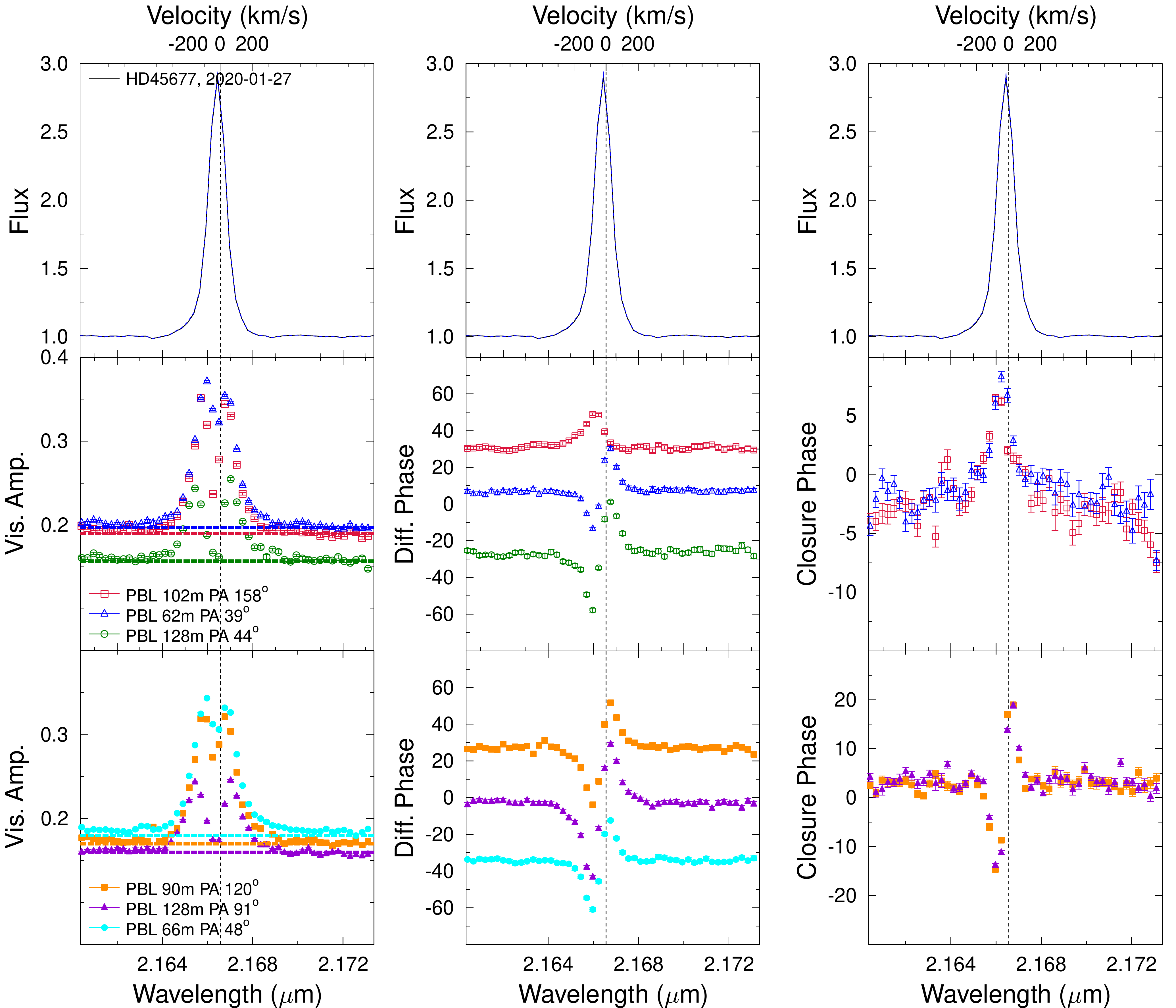}
 \caption{Same as Fig.\,\ref{fig:data1} but for HD\,38120 and HD\,45677.}
 \label{fig:data2}
\end{figure*}


\begin{figure*}
\centering
    \includegraphics[width=0.75\textwidth]{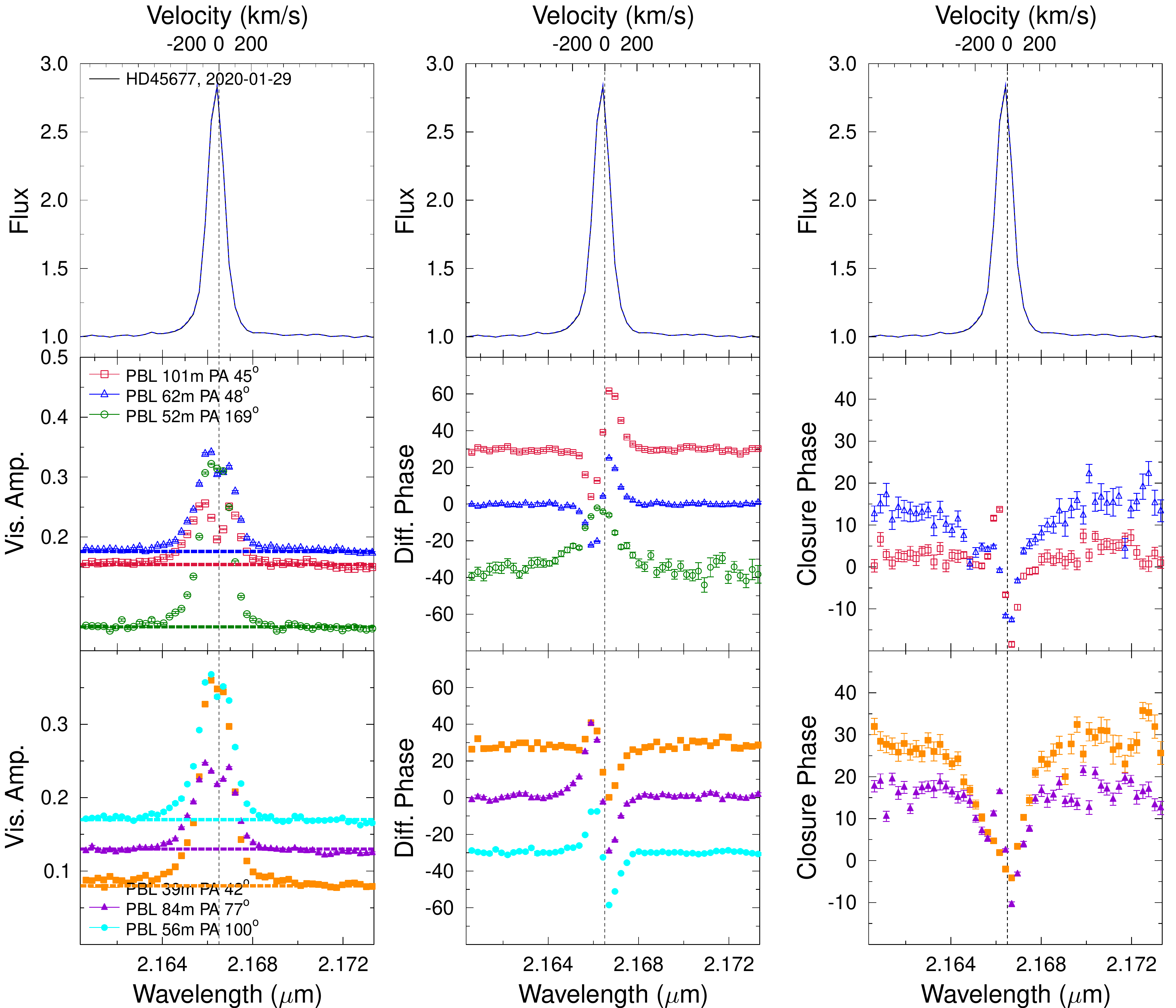}\\
    \includegraphics[width=0.75\textwidth]{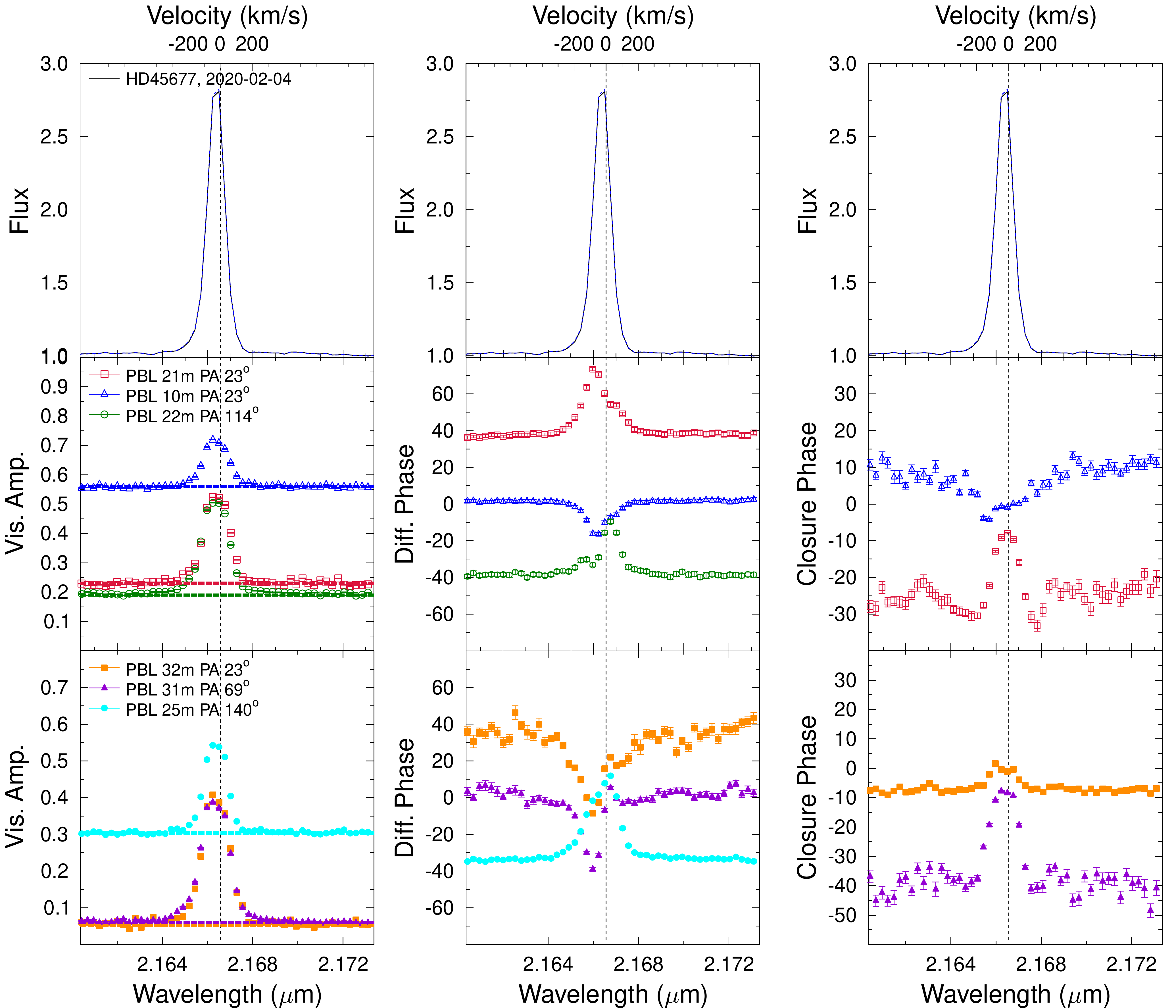}
    \caption{Same as Fig.\,\ref{fig:data1} but for HD45677.}
    \label{fig:data3}
\end{figure*}

\begin{figure*}
\centering
    \includegraphics[width=0.75\textwidth]{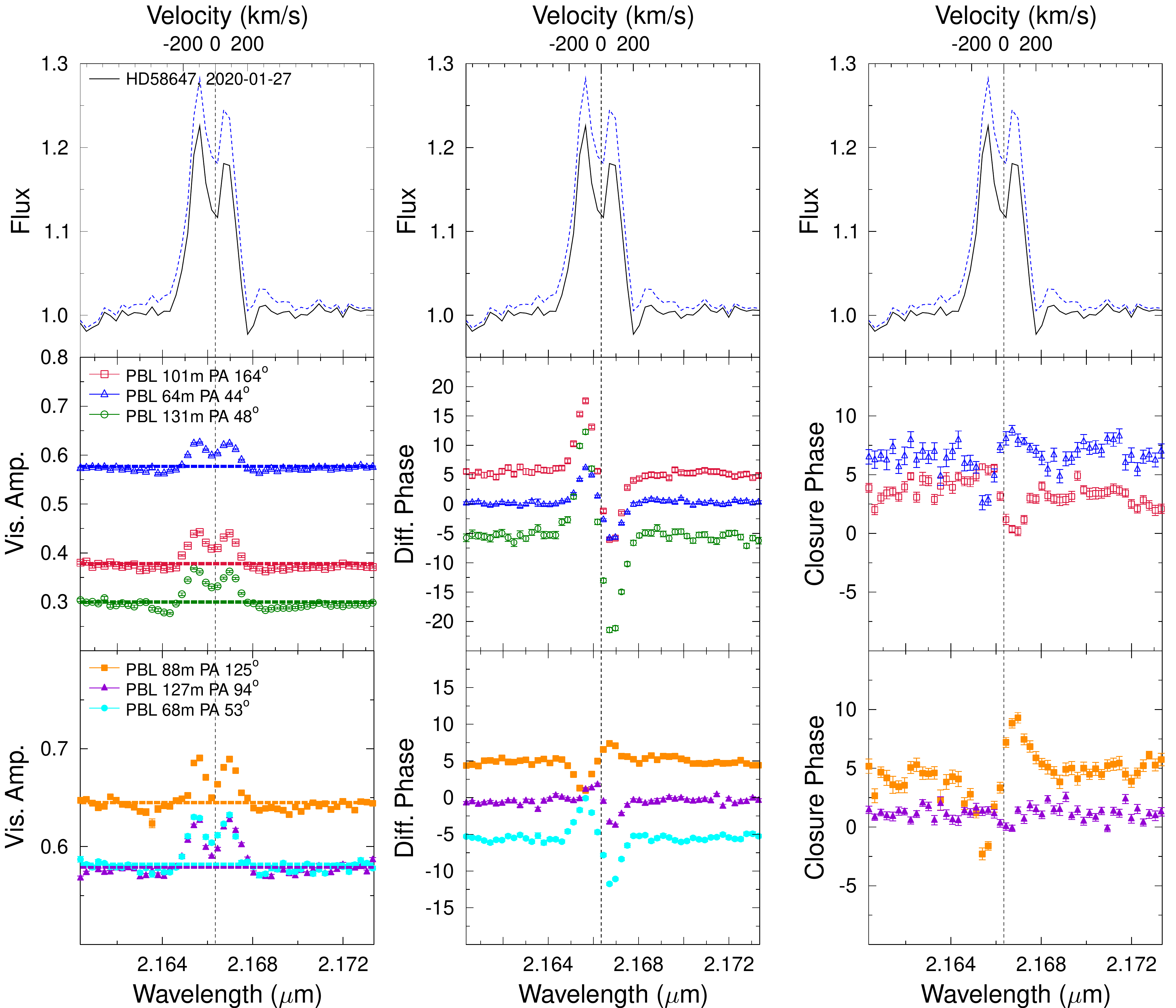}\\
    \includegraphics[width=0.75\textwidth]{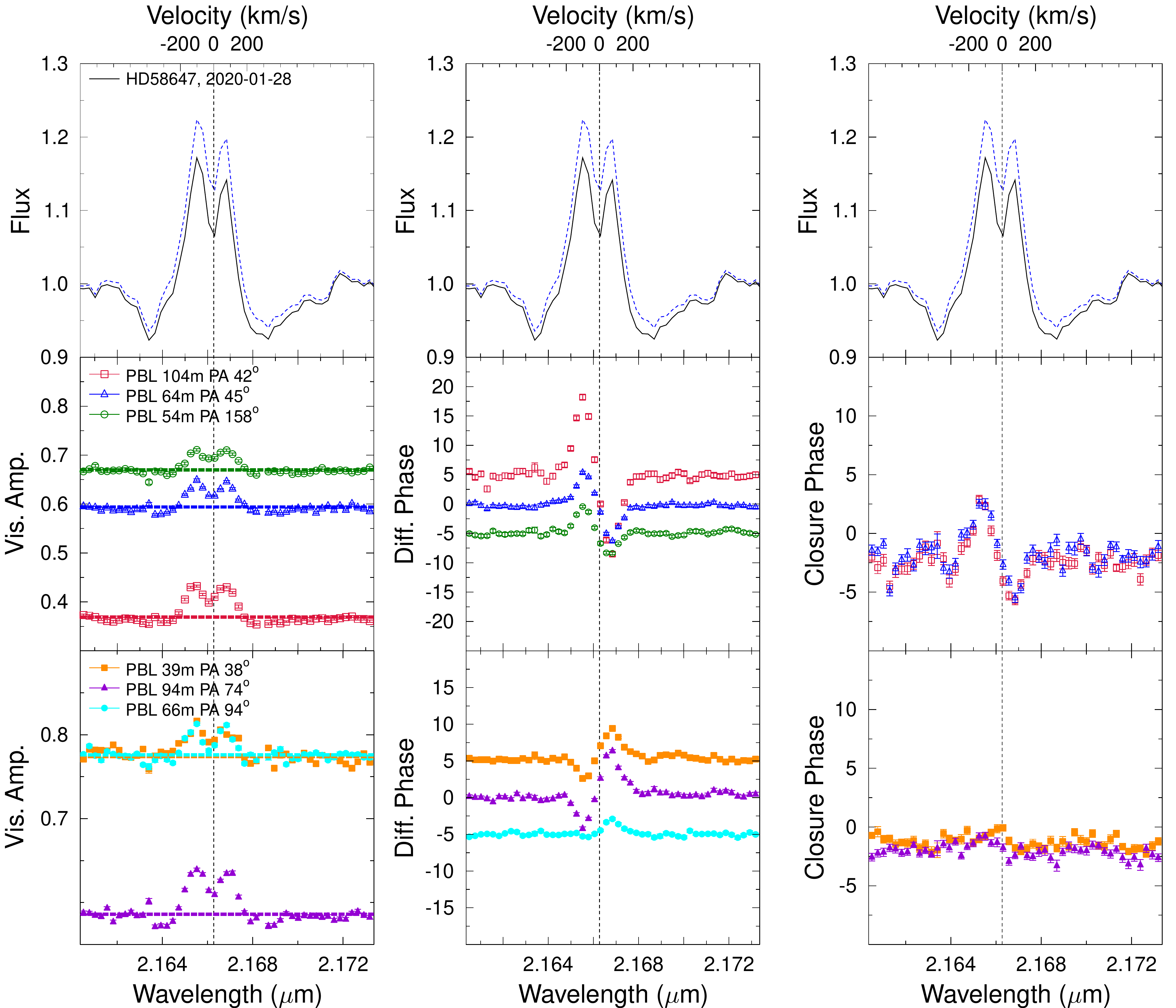}
    \caption{Same as Fig.\,\ref{fig:data1} but for HD58647.}
    \label{fig:data4}
\end{figure*}

\begin{figure*}
    \centering
    \includegraphics[width=0.75\textwidth]{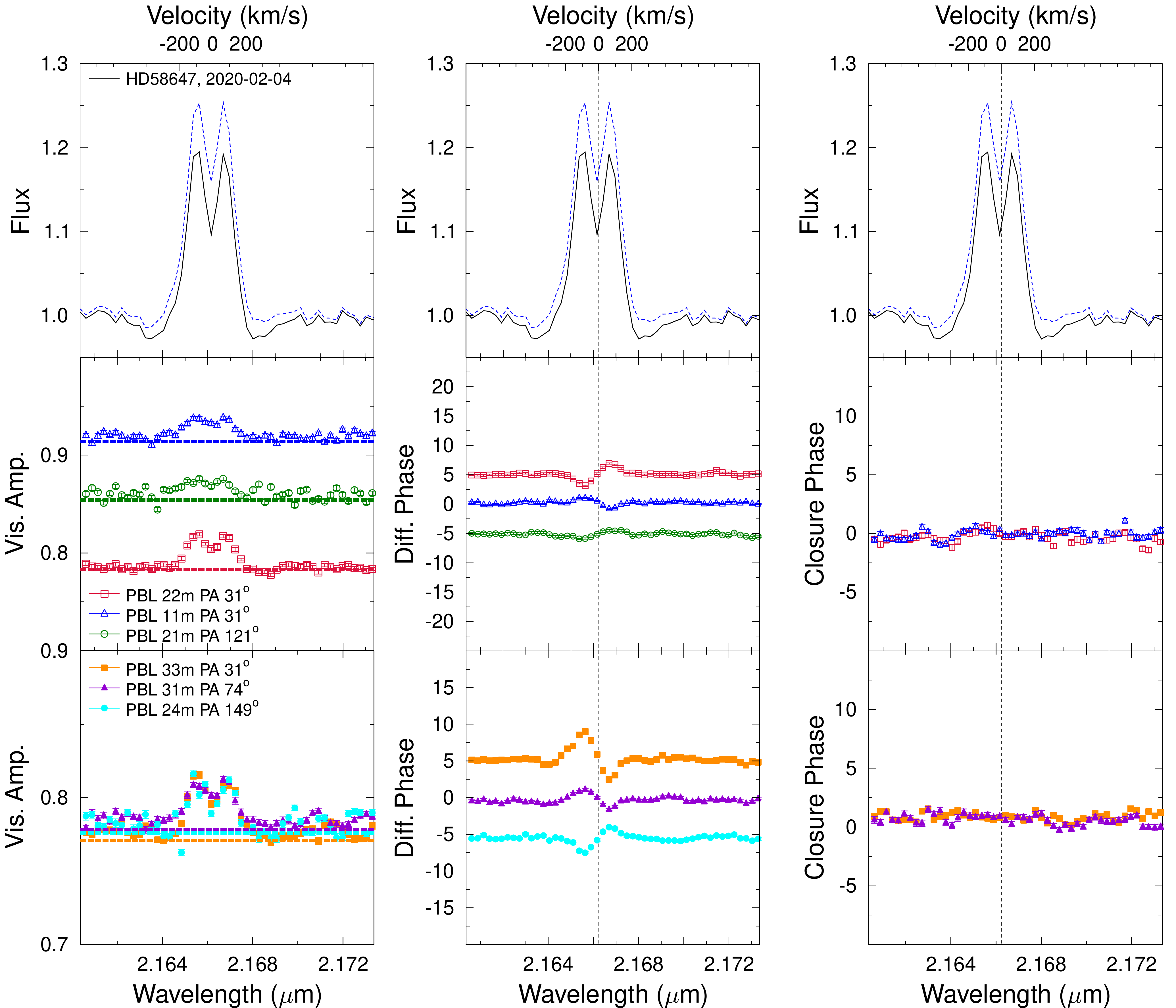}\\
    \includegraphics[width=0.75\textwidth]{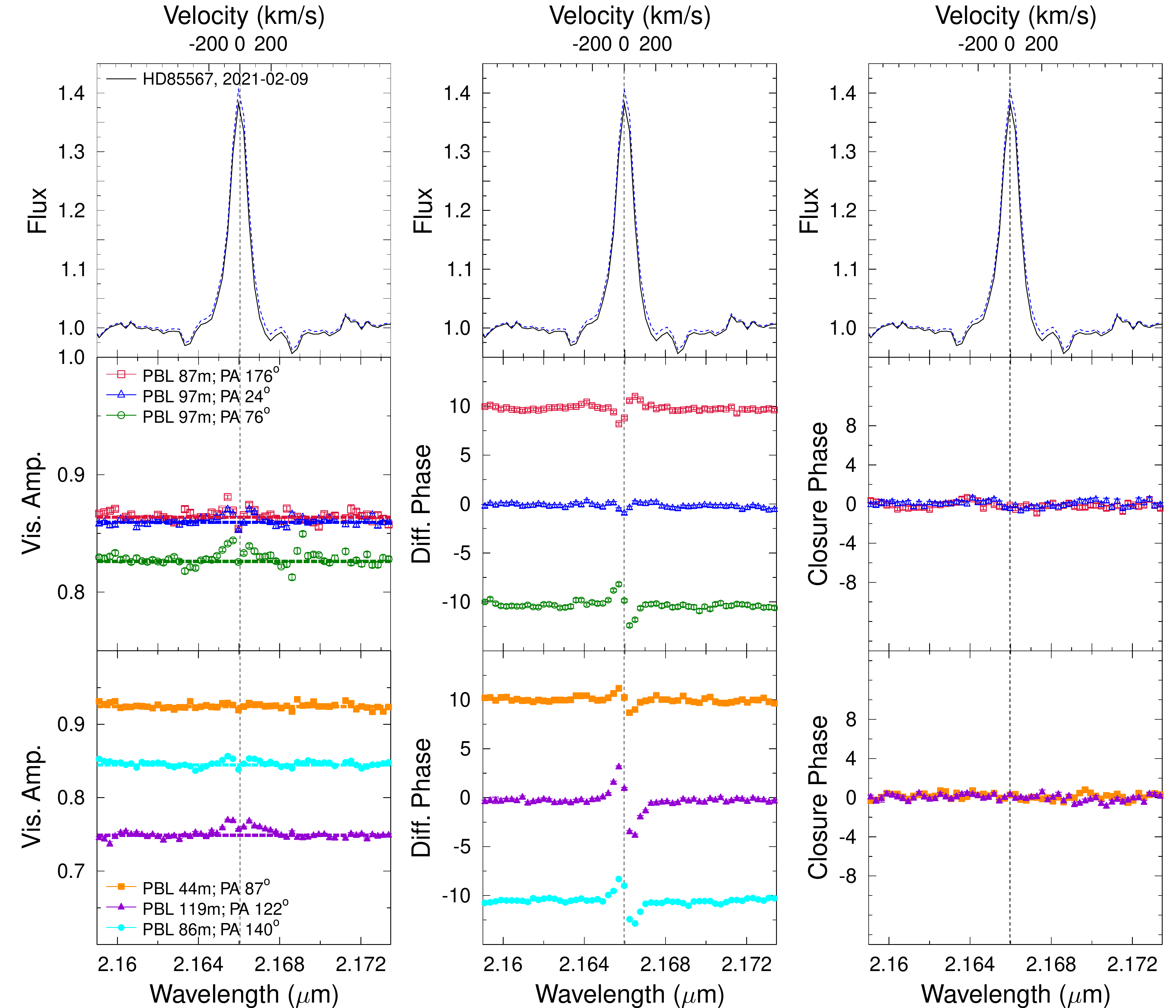}
      \caption{Same as Fig.\,\ref{fig:data1} but for HD58647 and HD85567.}
      \label{fig:data5}
\end{figure*}

\begin{figure*}
 \centering   
    \includegraphics[width=0.75\textwidth]{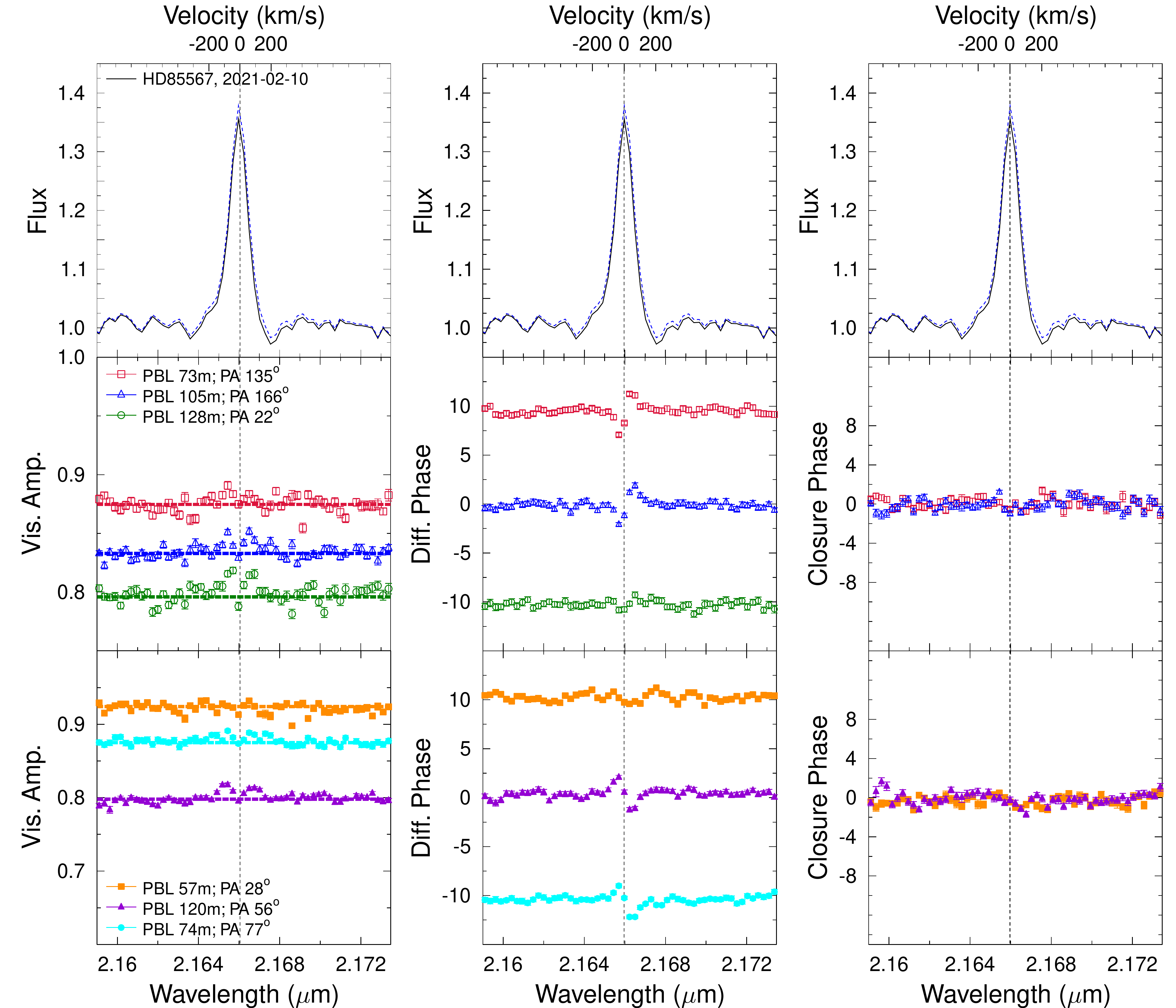}\\
    \includegraphics[width=0.75\textwidth]{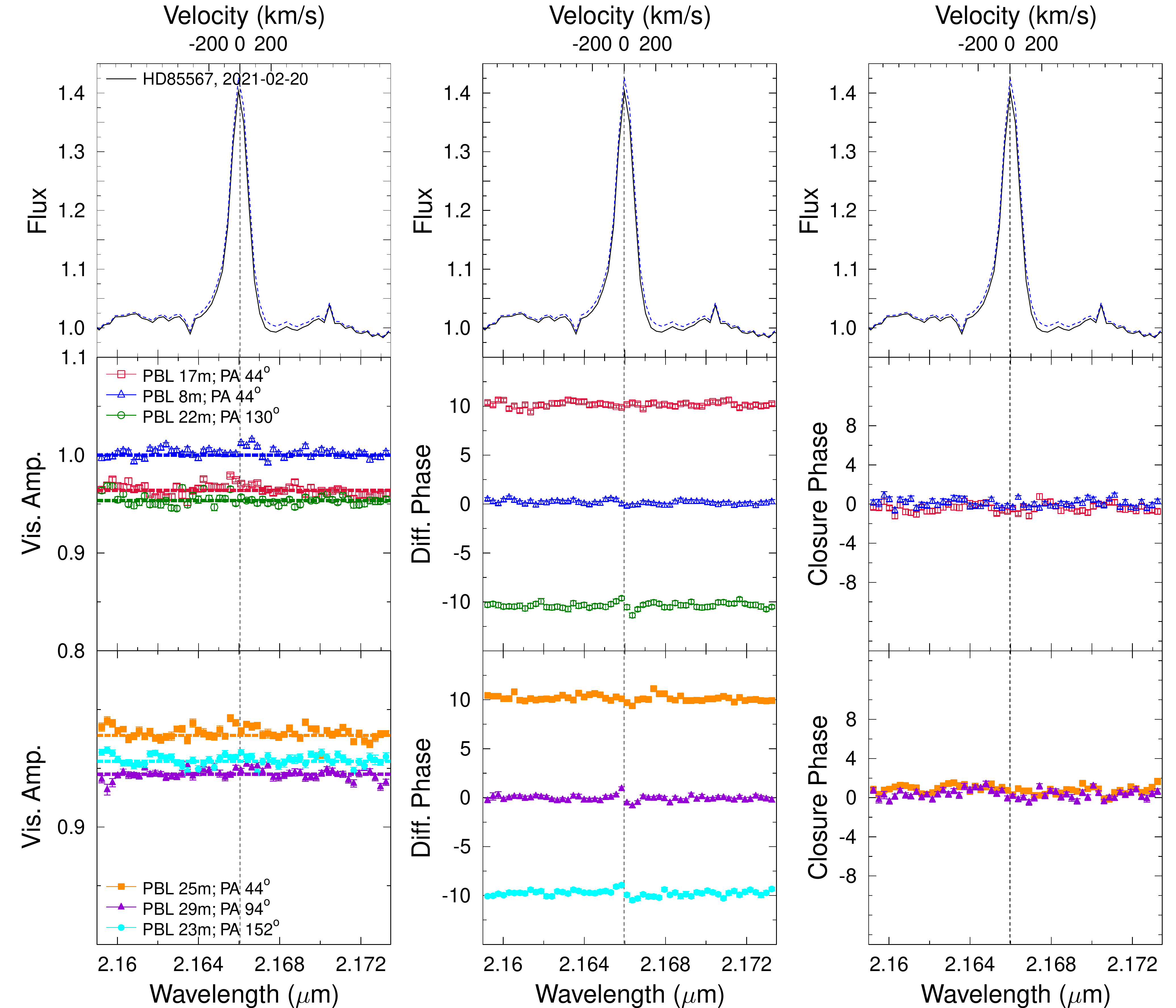}
    \caption{Same as Fig.\,\ref{fig:data1} but for HD85567.}
    \label{fig:data6}
\end{figure*}

\begin{figure*}  
    \centering
    \includegraphics[width=0.75\textwidth]{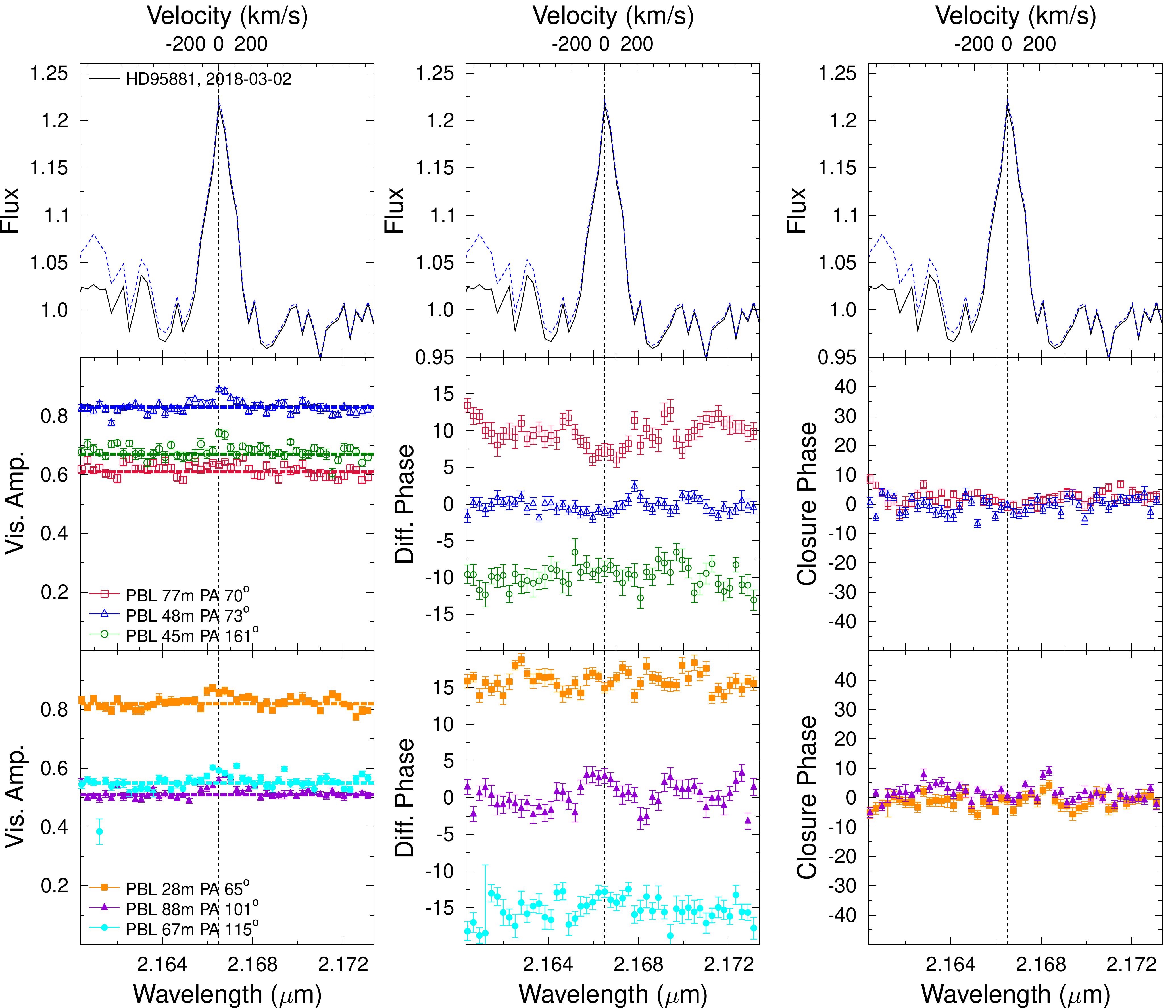}\\
    \includegraphics[width=0.75\textwidth]{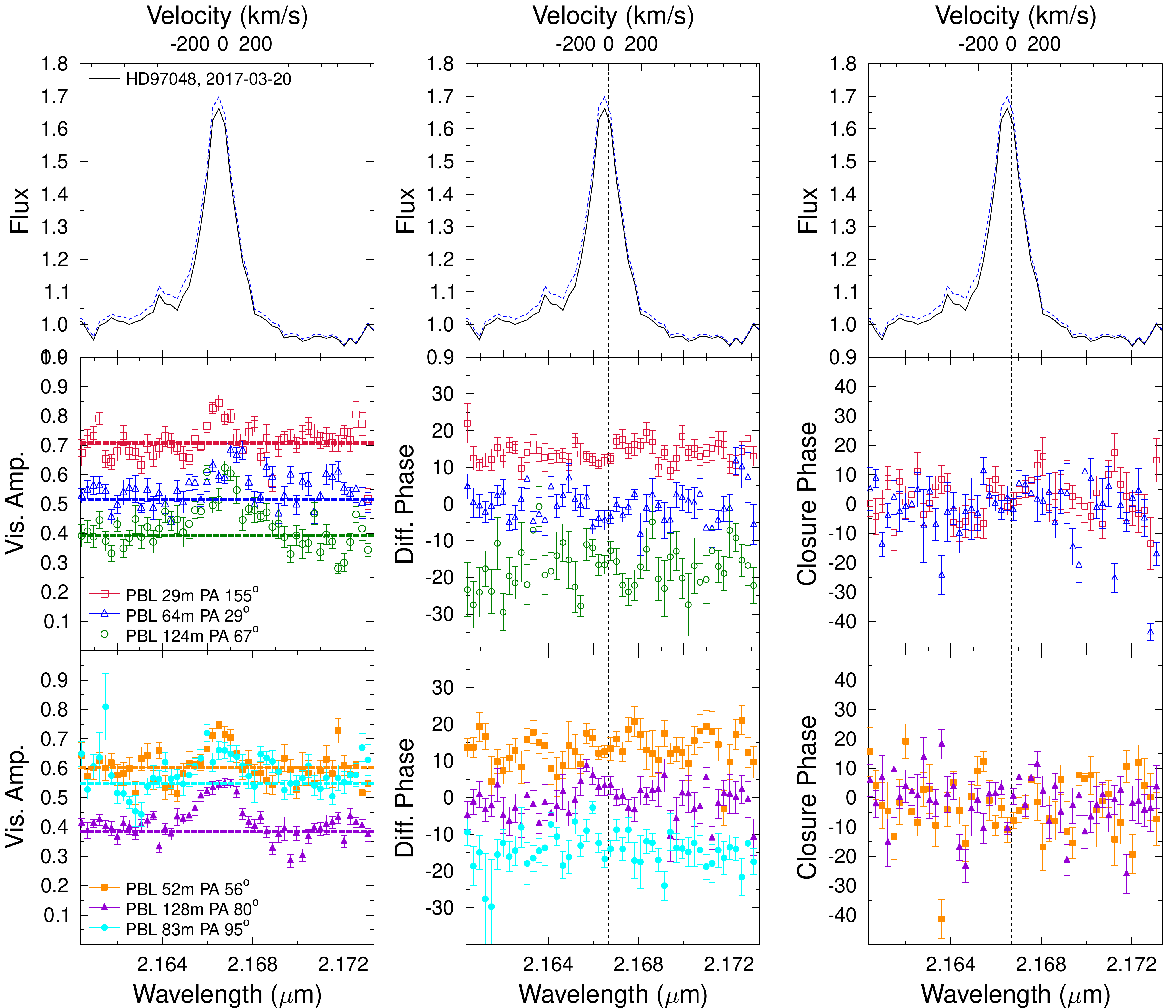}
     \caption{Same as Fig.\,\ref{fig:data1} but for HD\,95881 and HD\,97048.}
     \label{fig:data7}
\end{figure*}

\begin{figure*}
    \centering
    \includegraphics[width=0.75\textwidth]{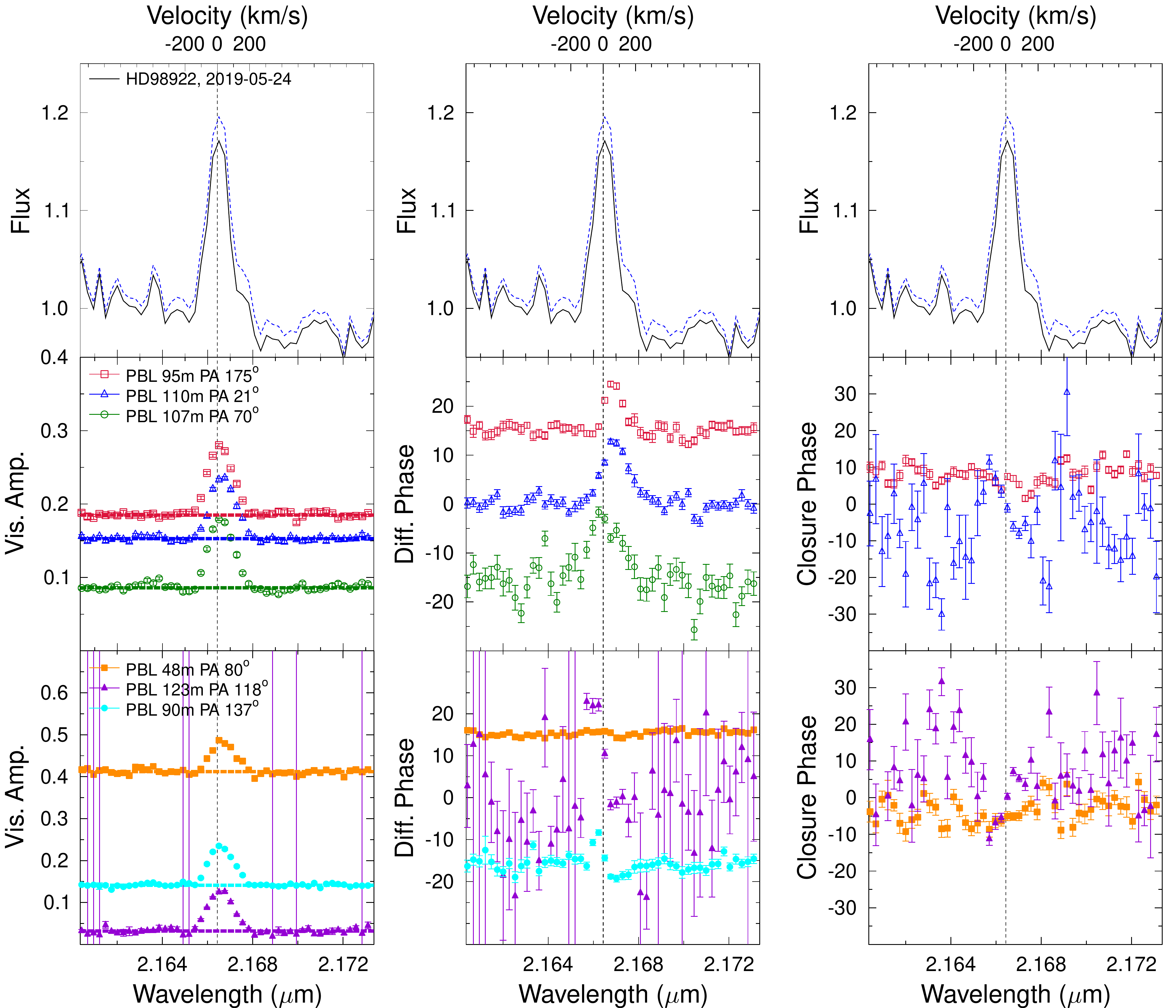}\\
    \includegraphics[width=0.75\textwidth]{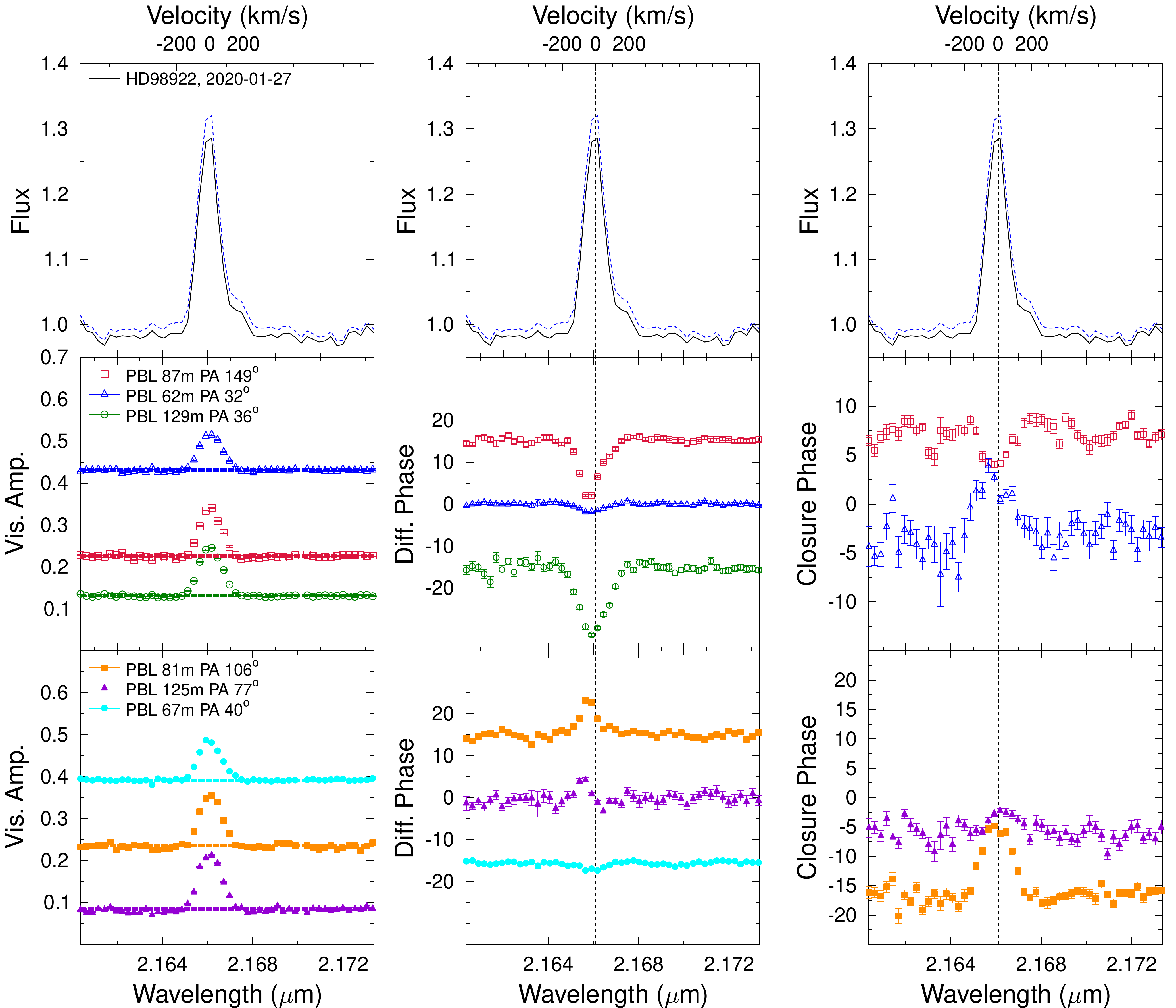}
  \caption{Same as Fig.\,\ref{fig:data1} but for HD\,98922.}   
  \label{fig:data8}
\end{figure*}

\begin{figure*}
    \centering
    \includegraphics[width=0.75\textwidth]{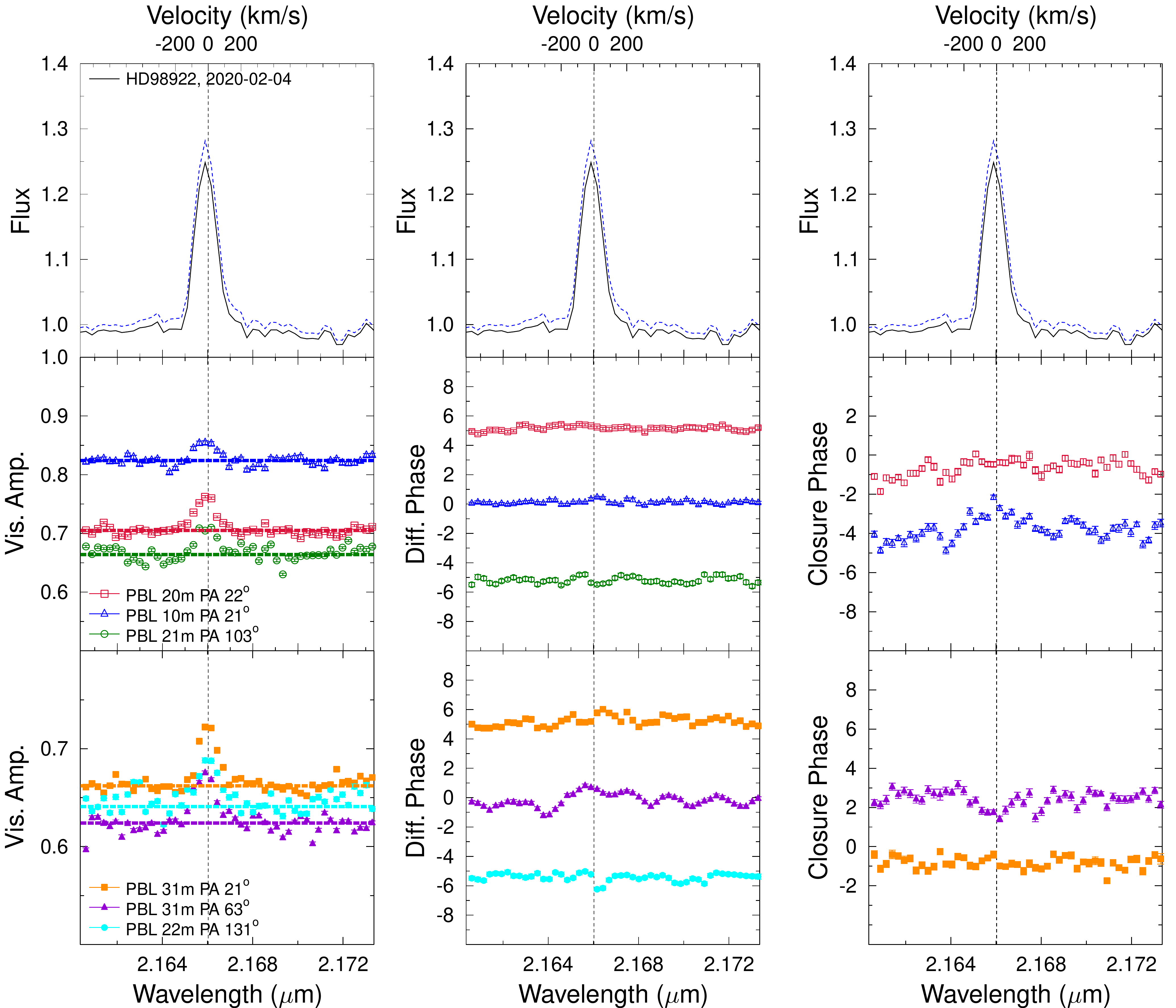}\\
    \includegraphics[width=0.75\textwidth]{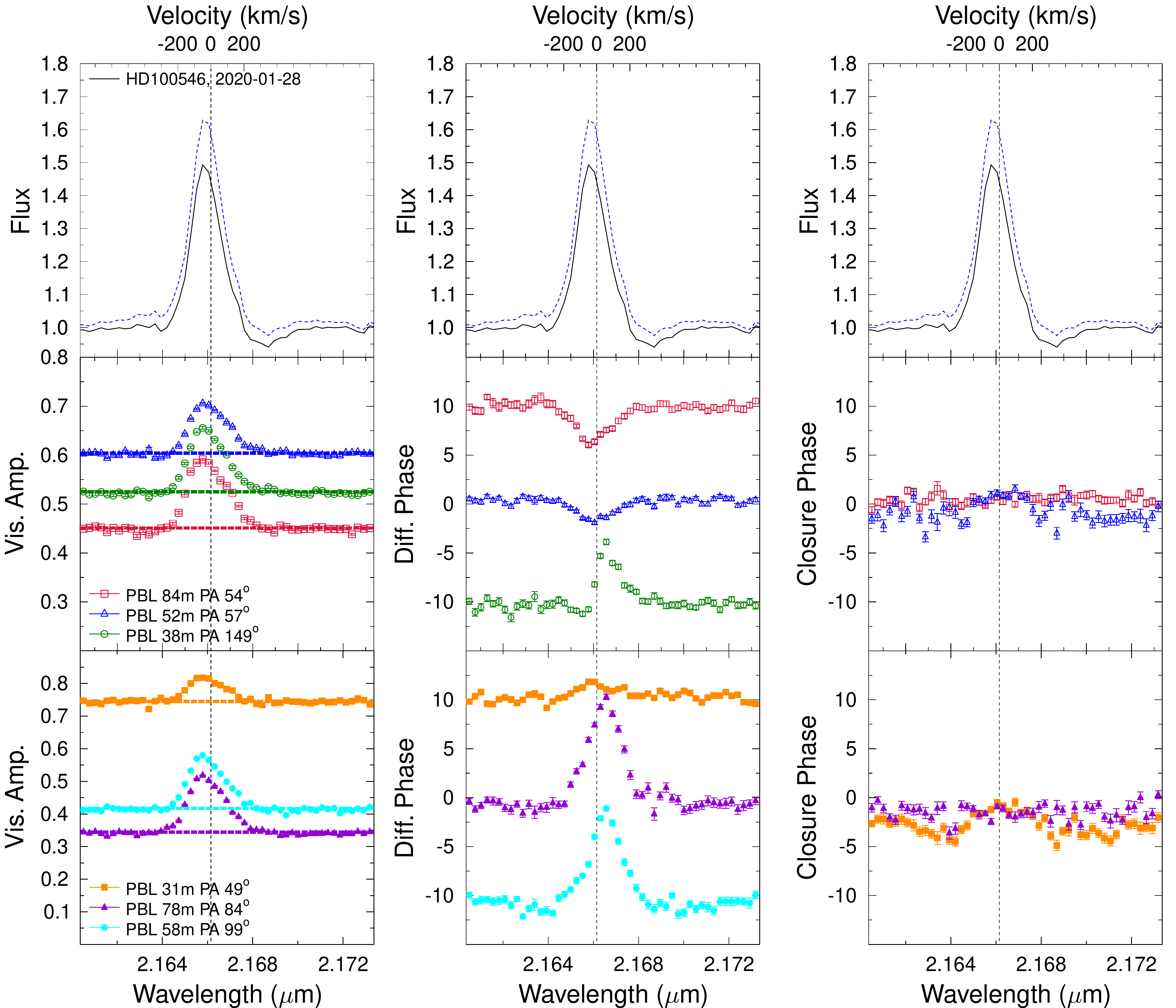}
    \caption{Same as Fig.\,\ref{fig:data1} but for HD\,98922 and HD\,100546.}
    \label{fig:data9}
\end{figure*}


\begin{figure*}
    \centering
    \includegraphics[width=0.75\textwidth]{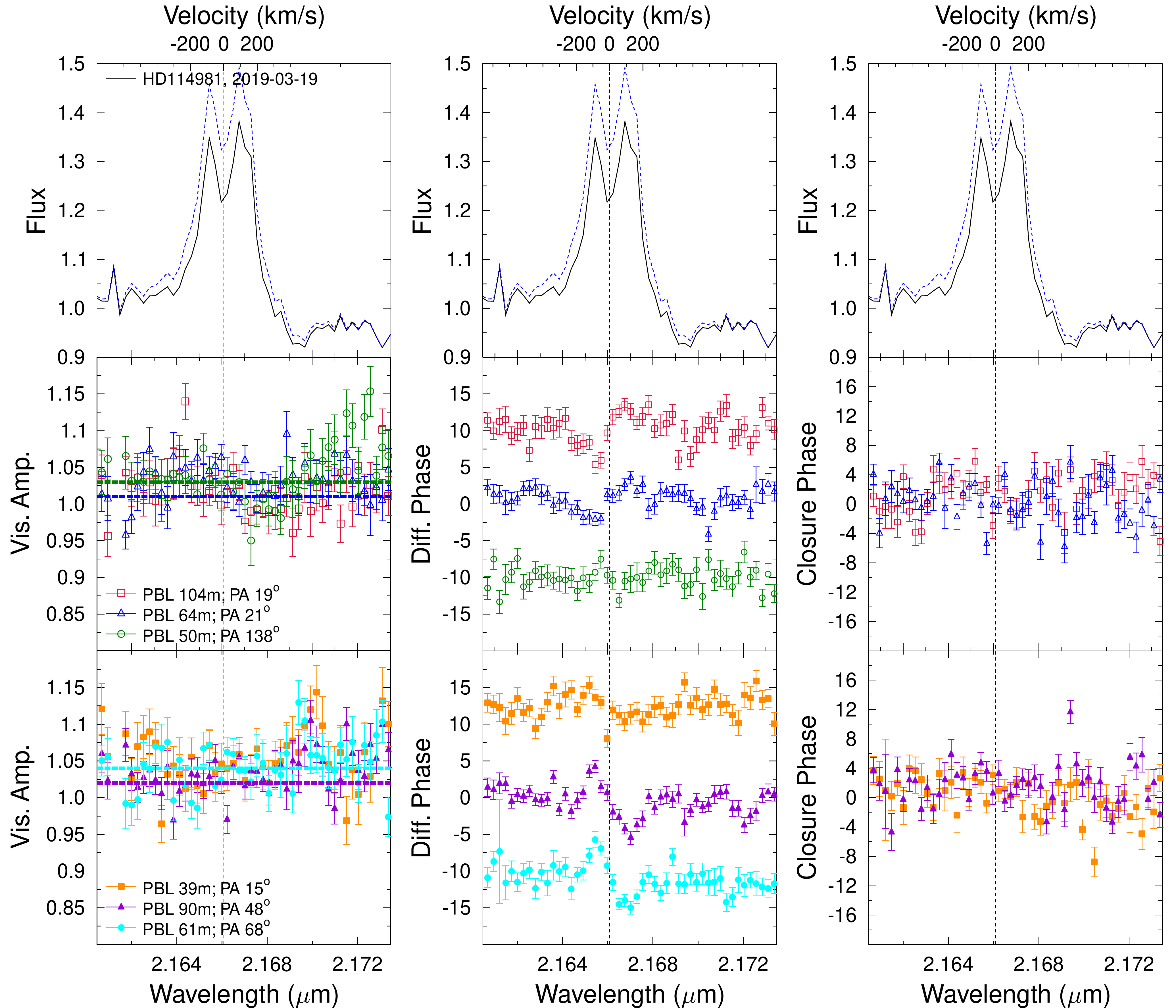}\\
    \includegraphics[width=0.75\textwidth]{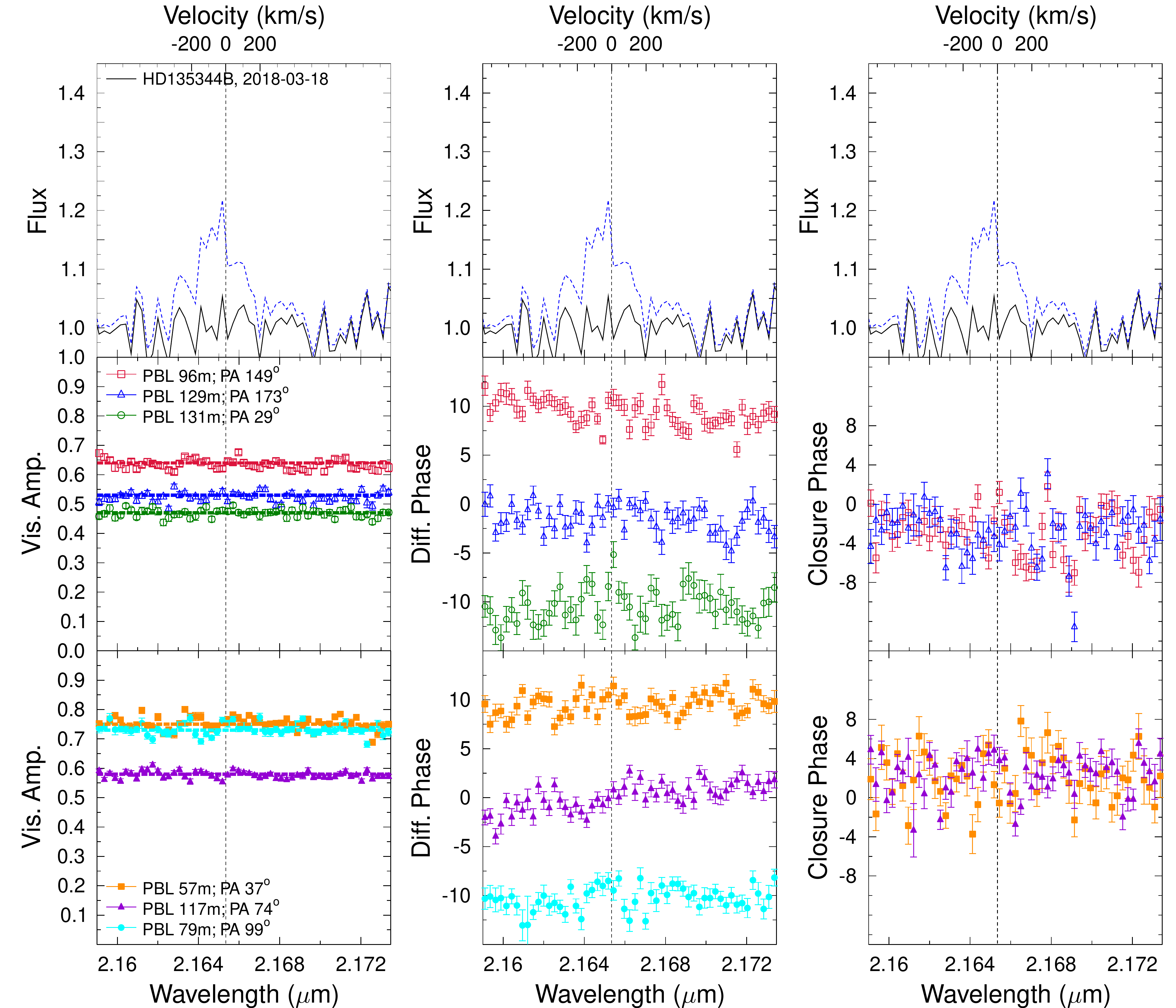} 
     \caption{Same as Fig.\,\ref{fig:data1} but for HD\,114981 and HD\,135344B.}
     \label{fig:data10}
\end{figure*}

\begin{figure*}
    \centering
	\includegraphics[width=0.75\textwidth]{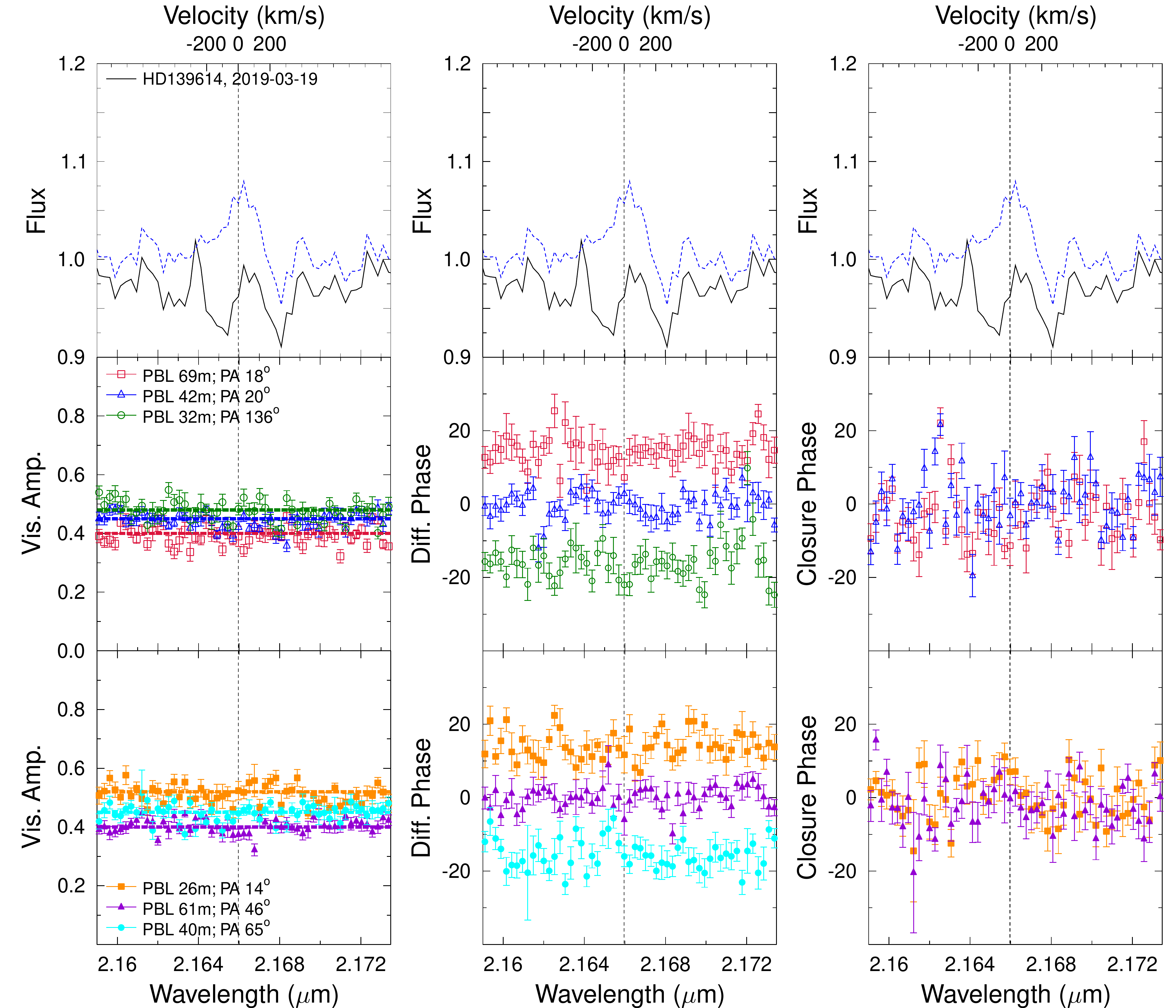}\\ 
 \includegraphics[width=0.75\textwidth]{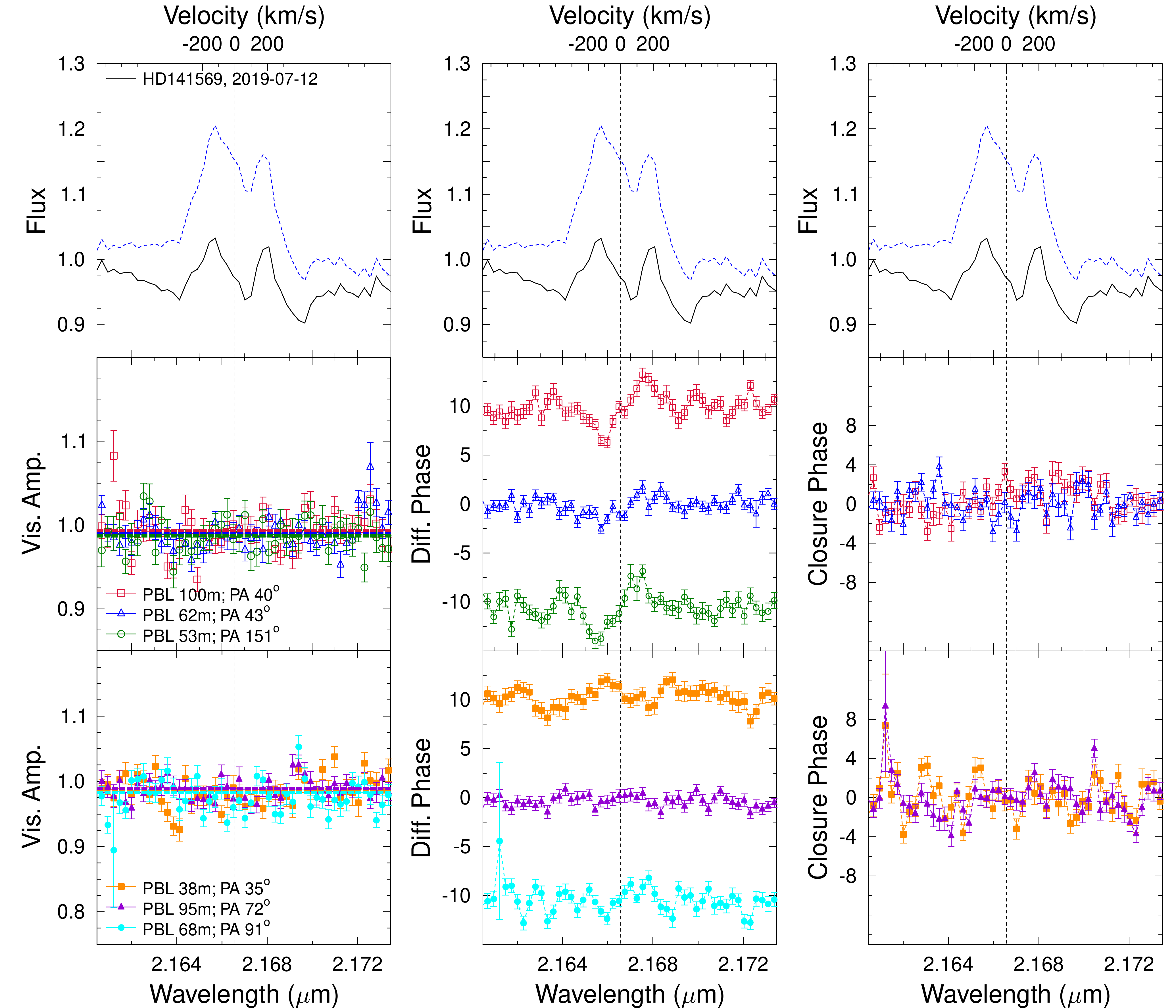}
 \caption{Same as Fig.\,\ref{fig:data1} but for HD\,139614 and HD\,141569.}
   \label{fig:data11}
\end{figure*}

\begin{figure*}
    \centering
    \includegraphics[width=0.75\textwidth]{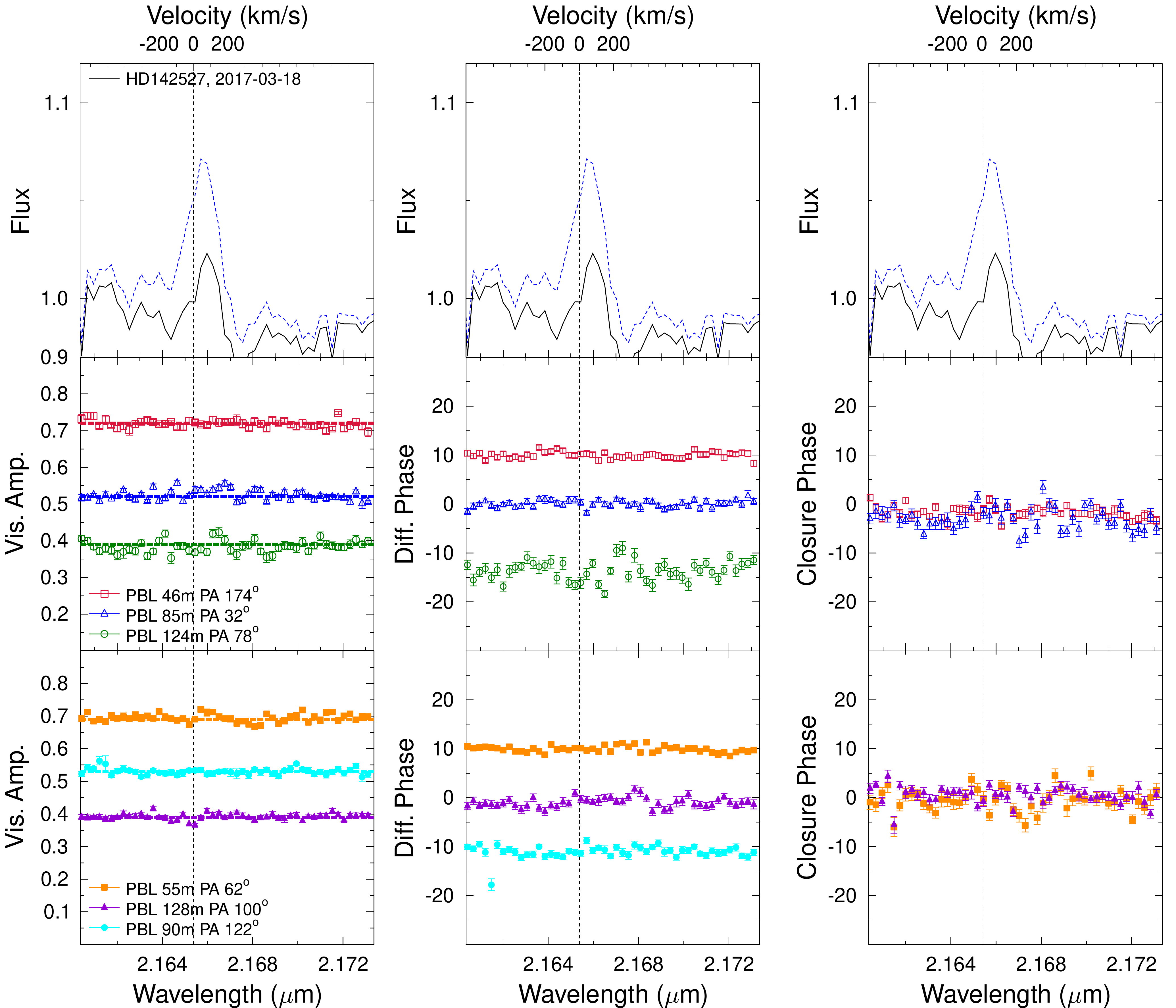}\\
    \includegraphics[width=0.75\textwidth]{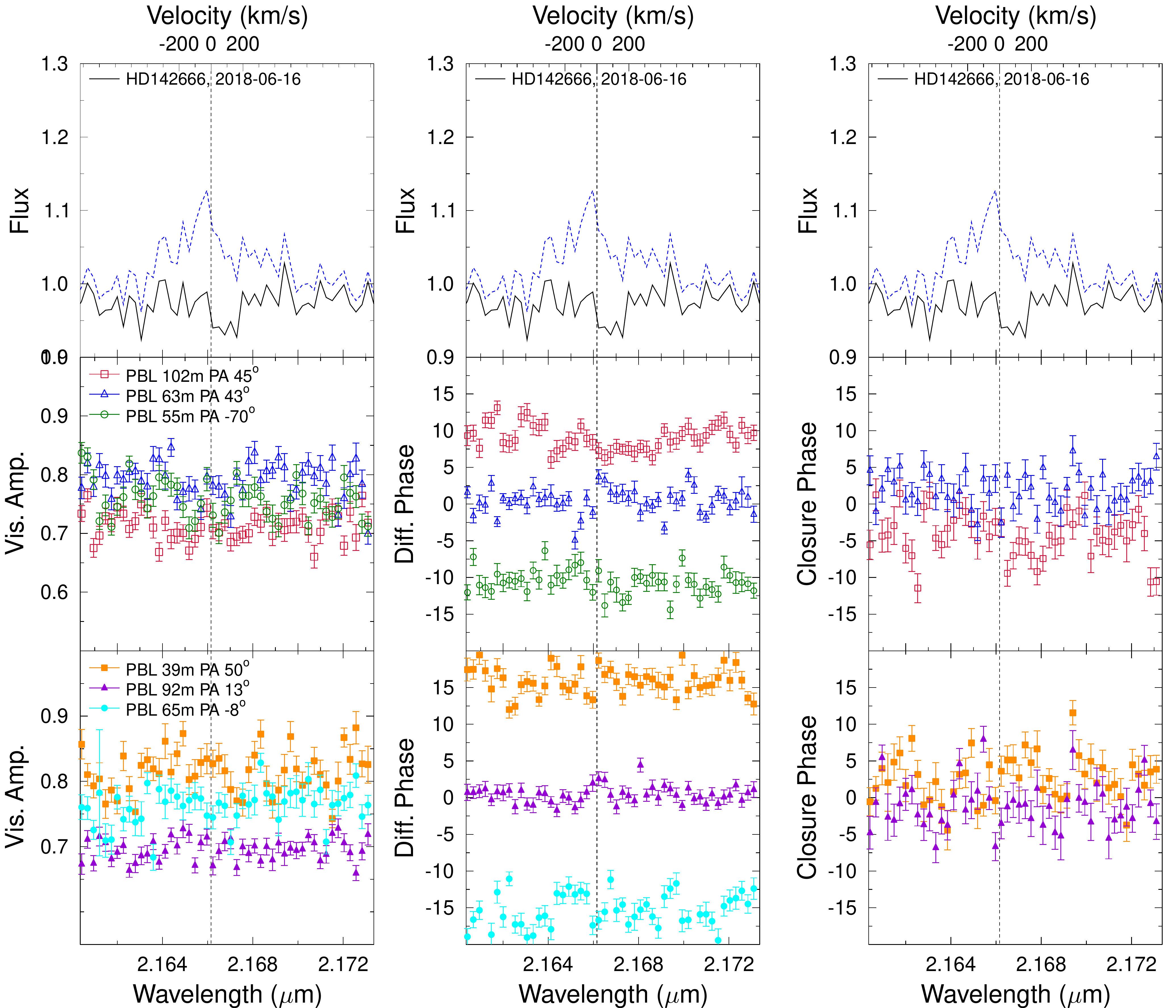}
    \caption{Same as Fig.\,\ref{fig:data1} but for HD\,142527 and HD\,142666.}
    \label{fig:data12}
\end{figure*}


\begin{figure*}
	\centering
    \includegraphics[width=0.75\textwidth]{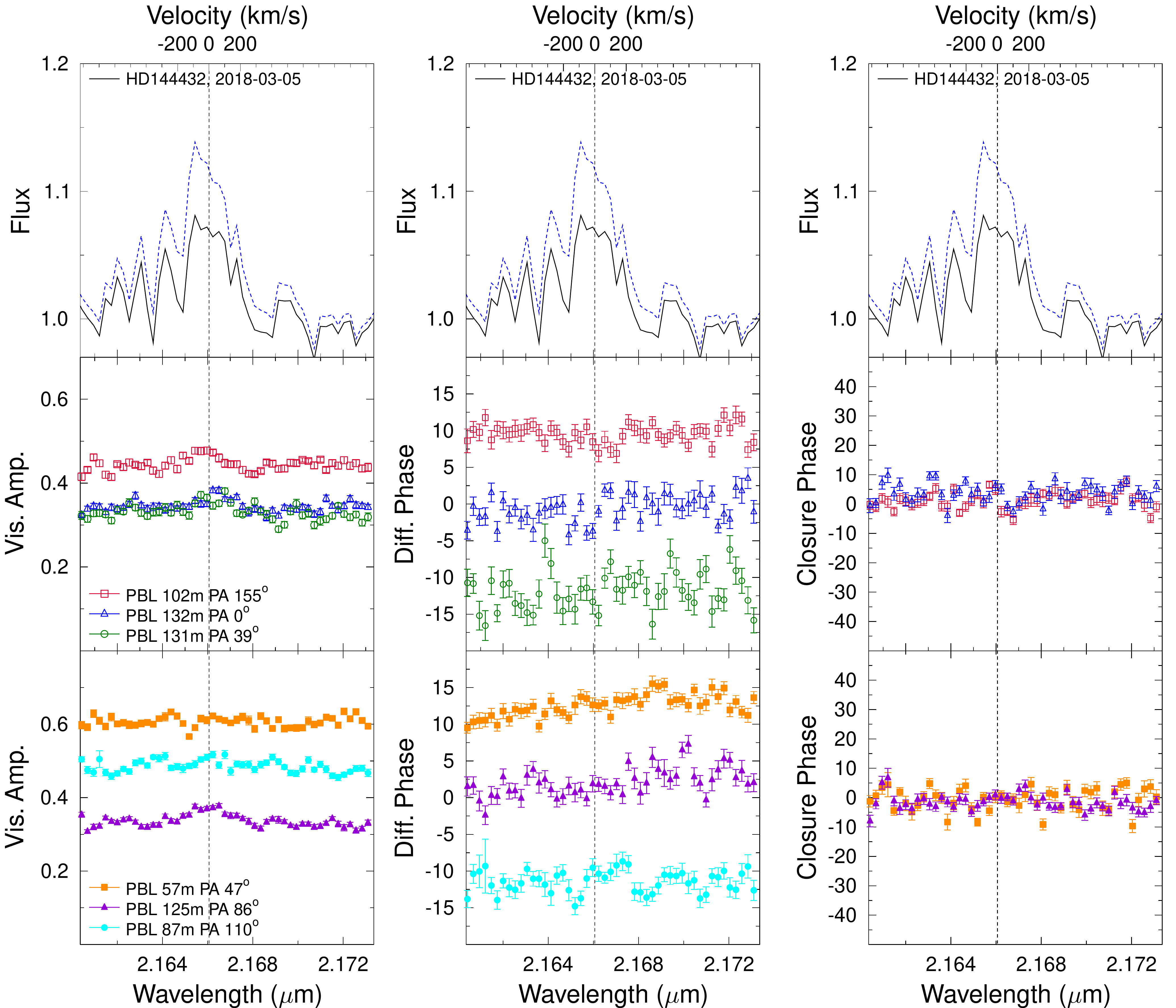}\\
    \includegraphics[width=0.75\textwidth]{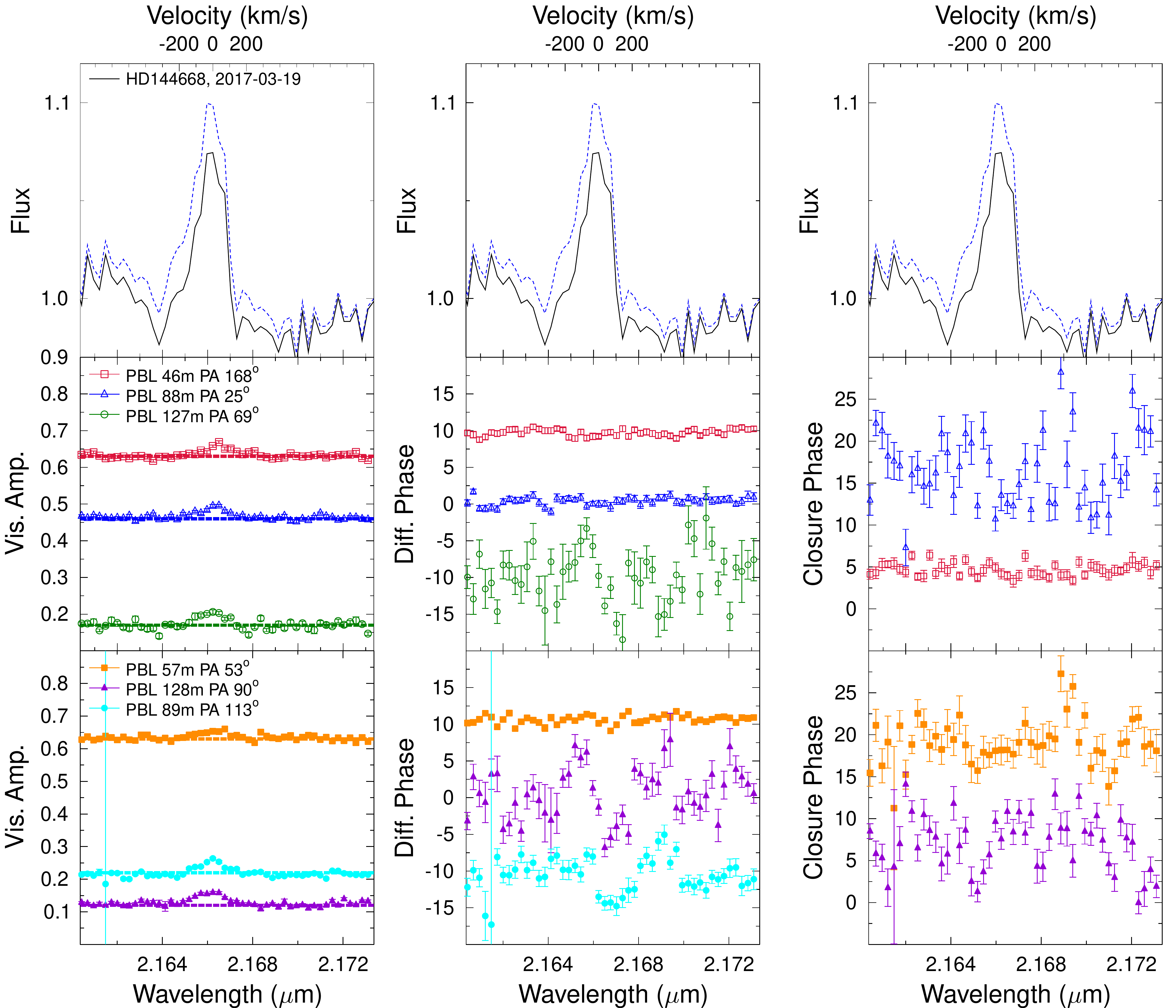}	
    \caption{Same as Fig.\,\ref{fig:data1} but for HD\,144432 and HD\,144668.}
    \label{fig:data13}
\end{figure*}

\begin{figure*}
    \centering
    \includegraphics[width=0.75\textwidth]{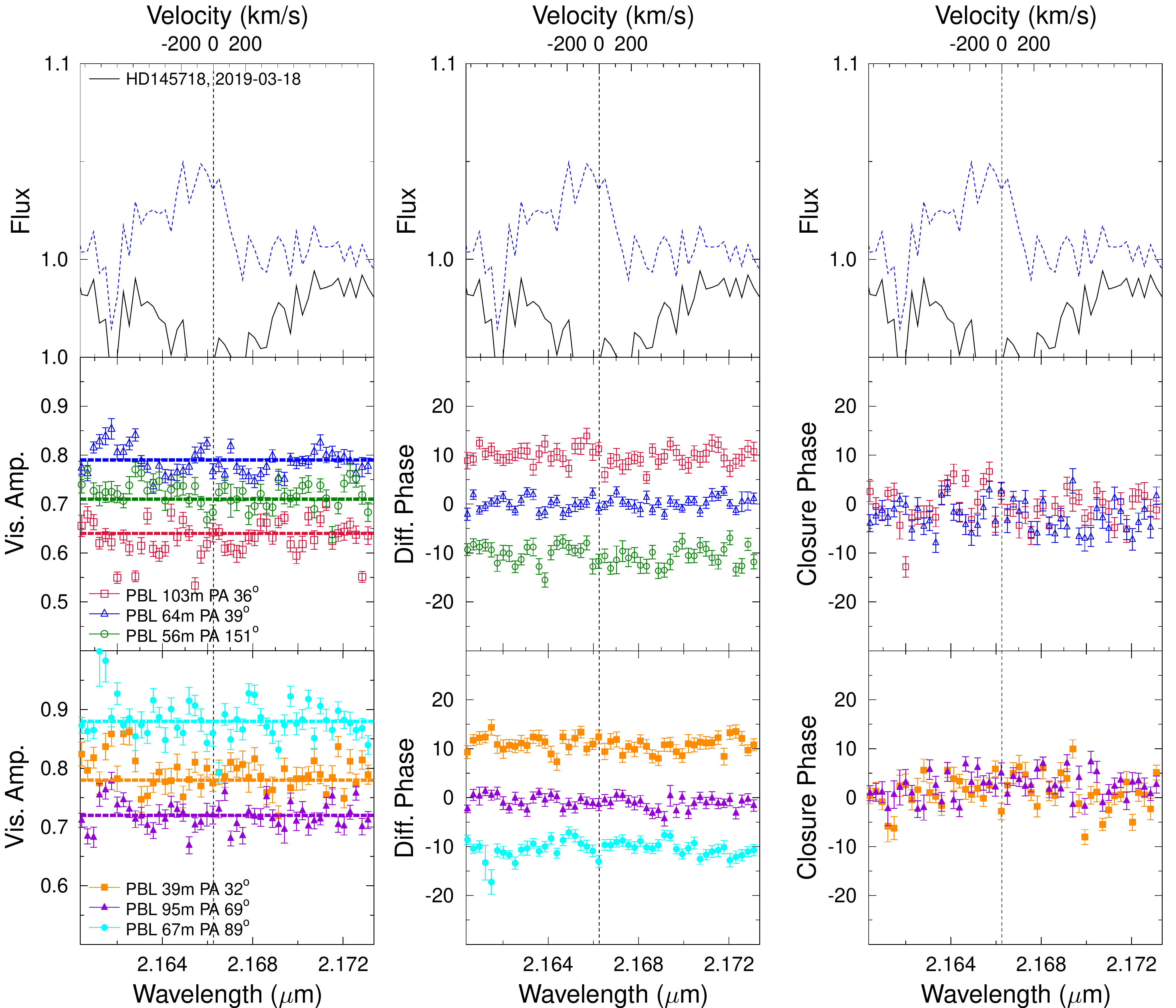}\\ 
    \includegraphics[width=0.75\textwidth]{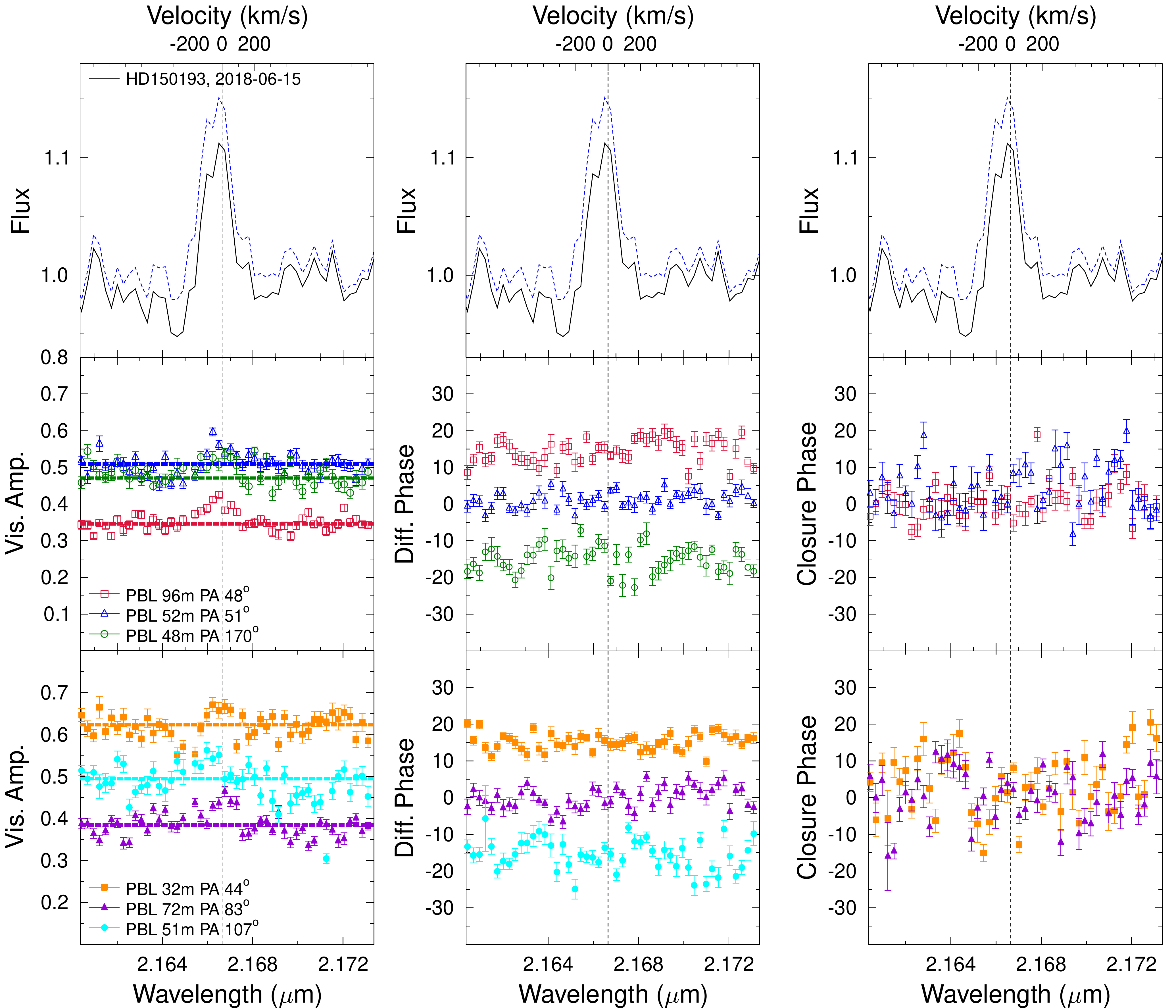}
    \caption{Same as Fig.\,\ref{fig:data1} but for HD\,145718 and HD\,150193.}
    \label{fig:data14}
\end{figure*}

\begin{figure*}
    \centering
    \includegraphics[width=0.75\textwidth]{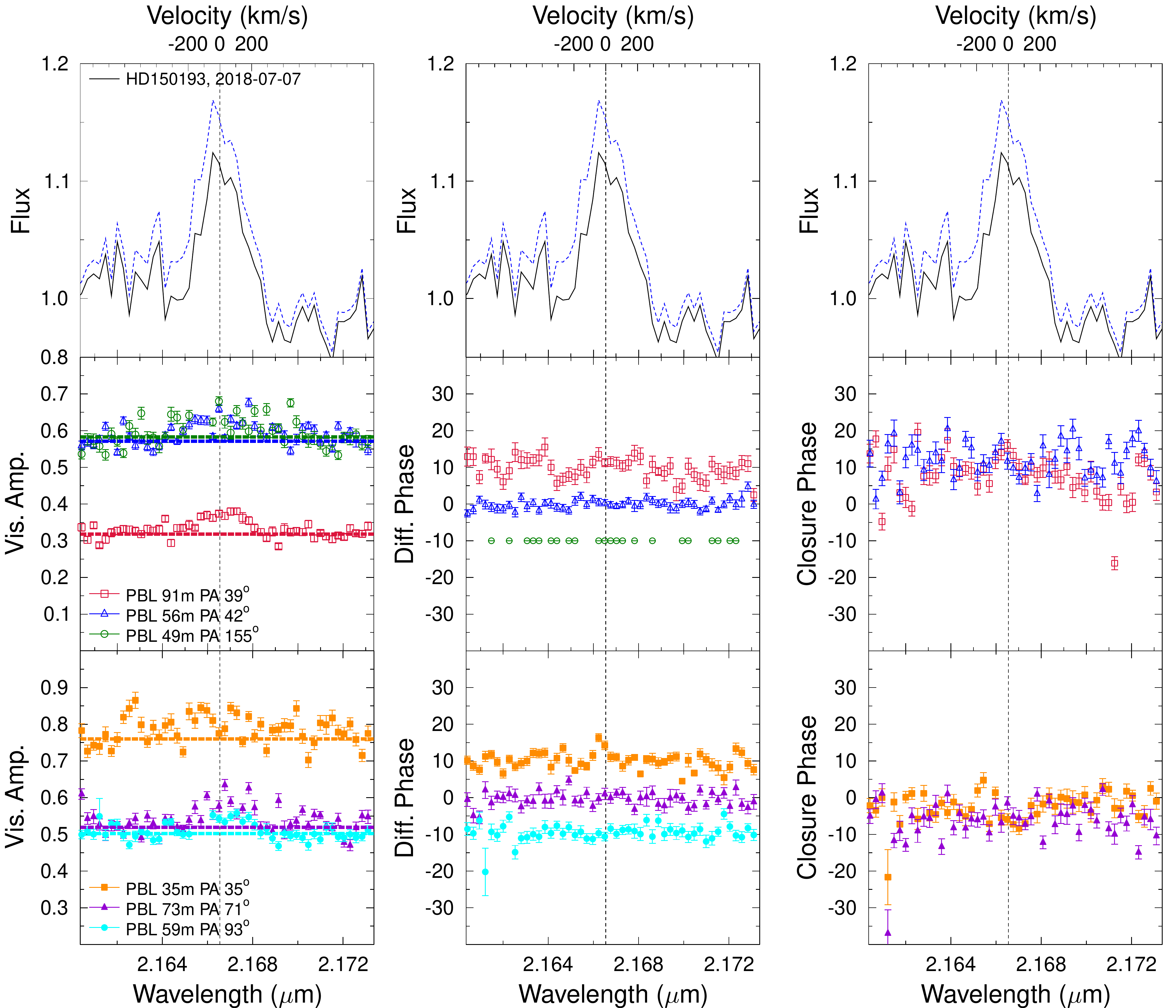}\\
    \includegraphics[width=0.75\textwidth]{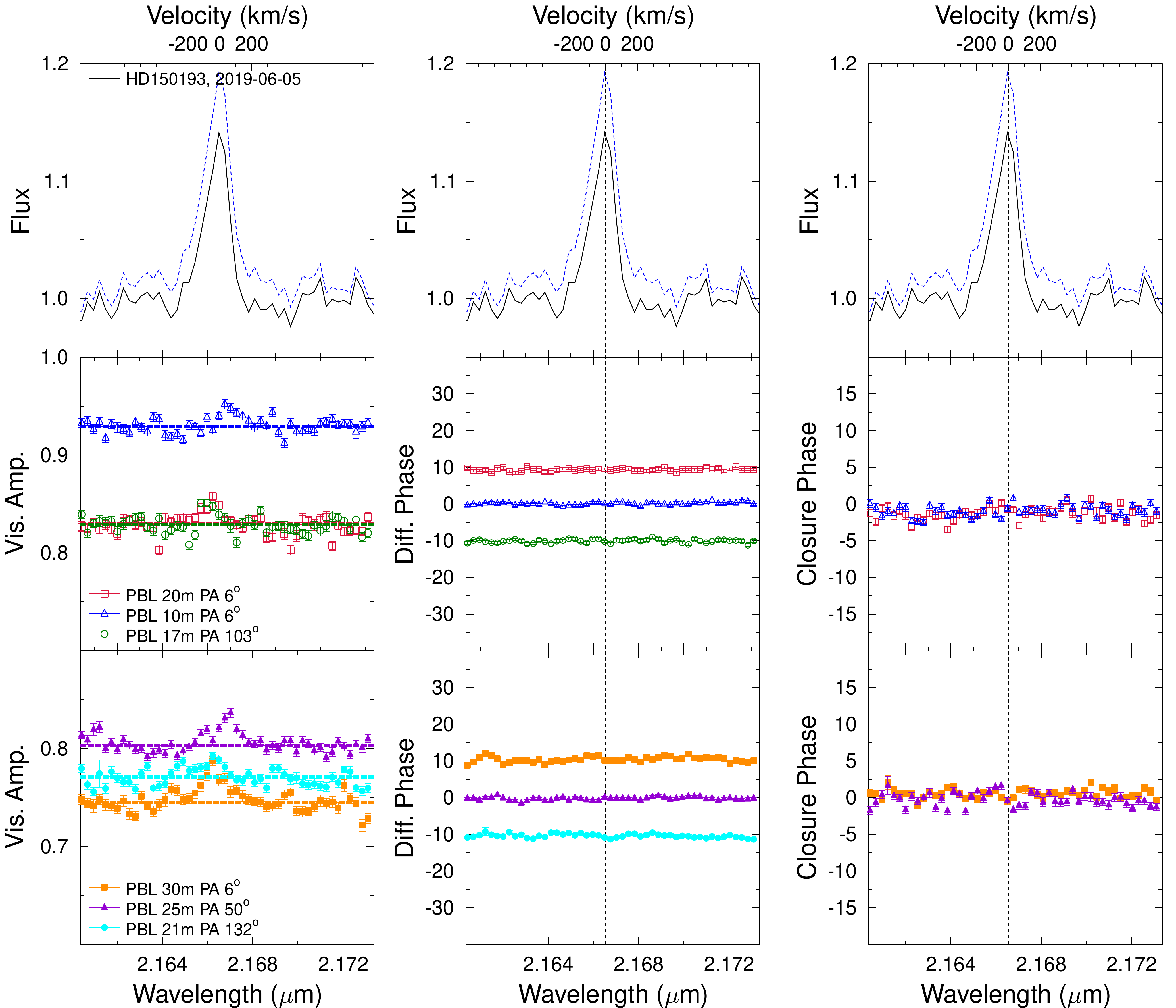}
    \caption{Same as Fig.\,\ref{fig:data1} but for HD\,150193.}
    \label{fig:data15}
\end{figure*}

\begin{figure*}
    \centering
    \includegraphics[width=0.75\textwidth]{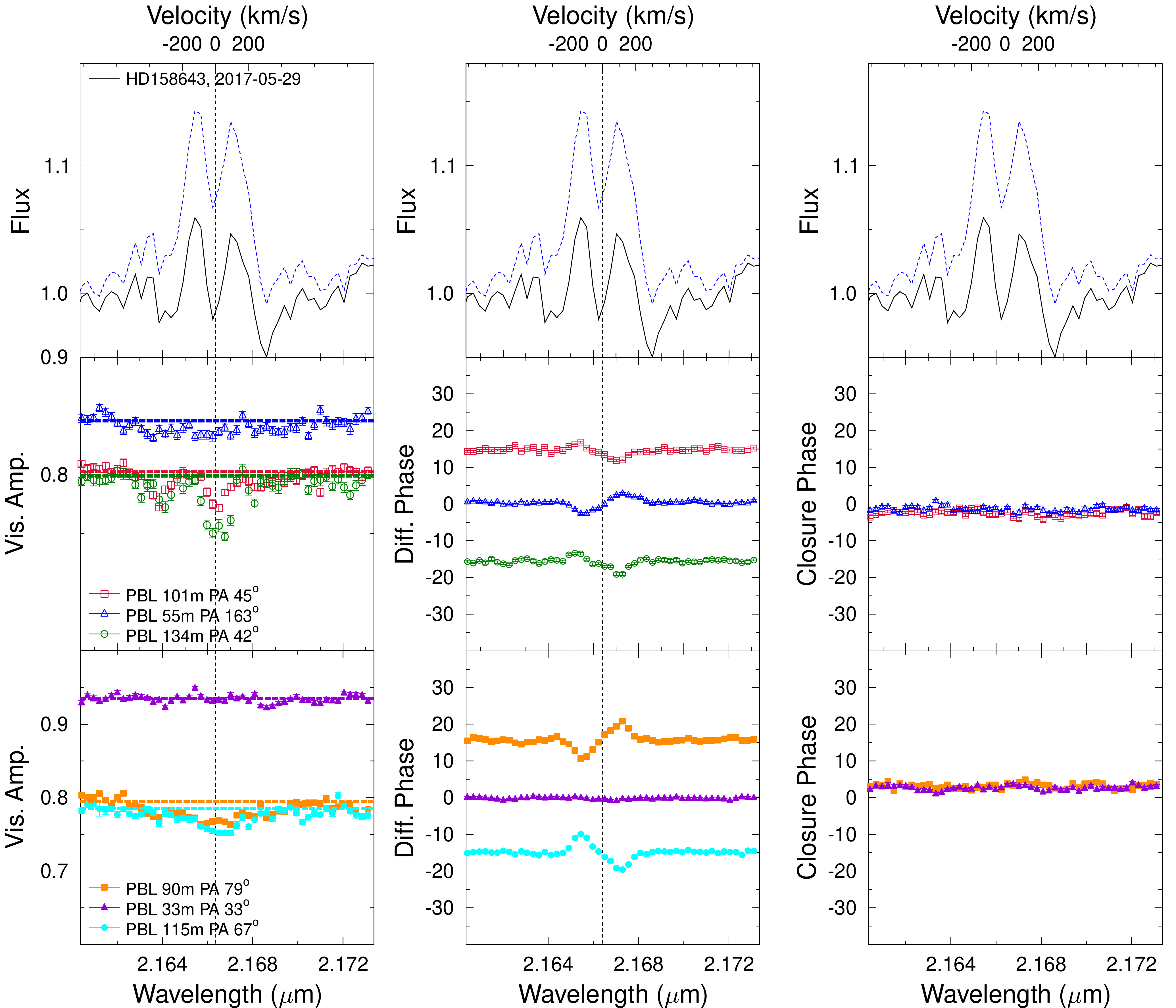}\\
    \includegraphics[width=0.75\textwidth]{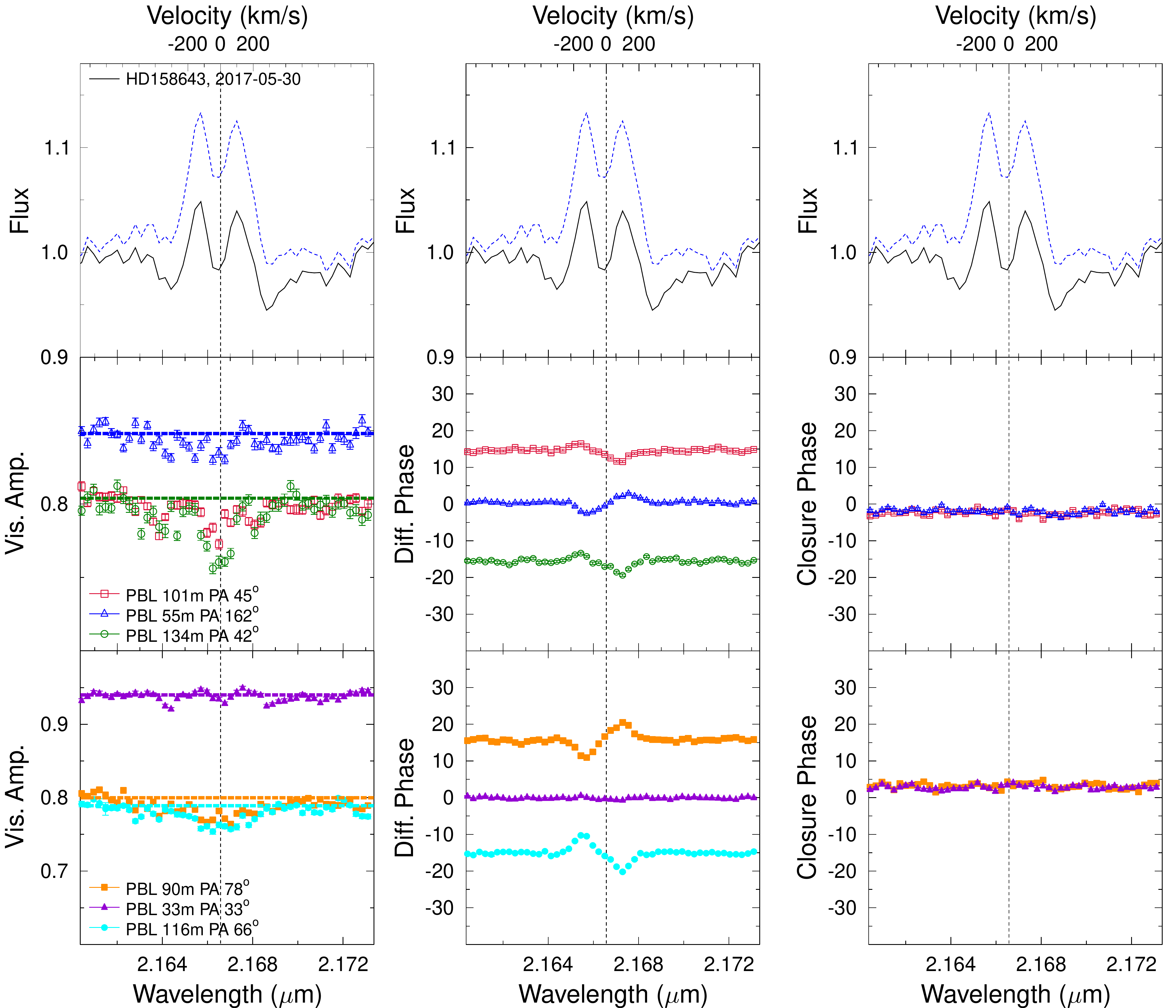}
    \caption{Same as Fig.\,\ref{fig:data1} but for HD\,158643.}
    \label{fig:data16}
\end{figure*}

\begin{figure*}
    \centering
    \includegraphics[width=0.75\textwidth]{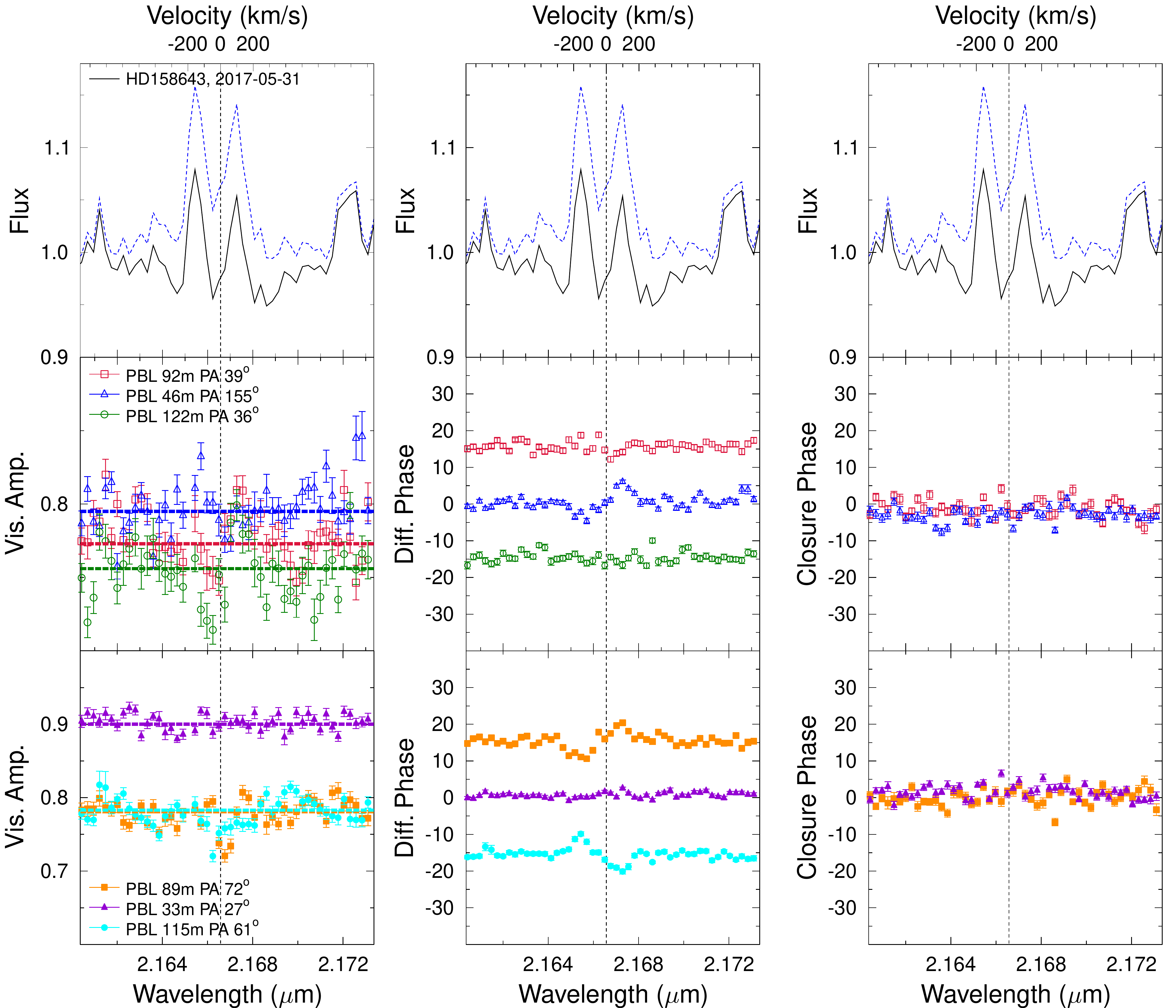} \\
     \includegraphics[width=0.75\textwidth]{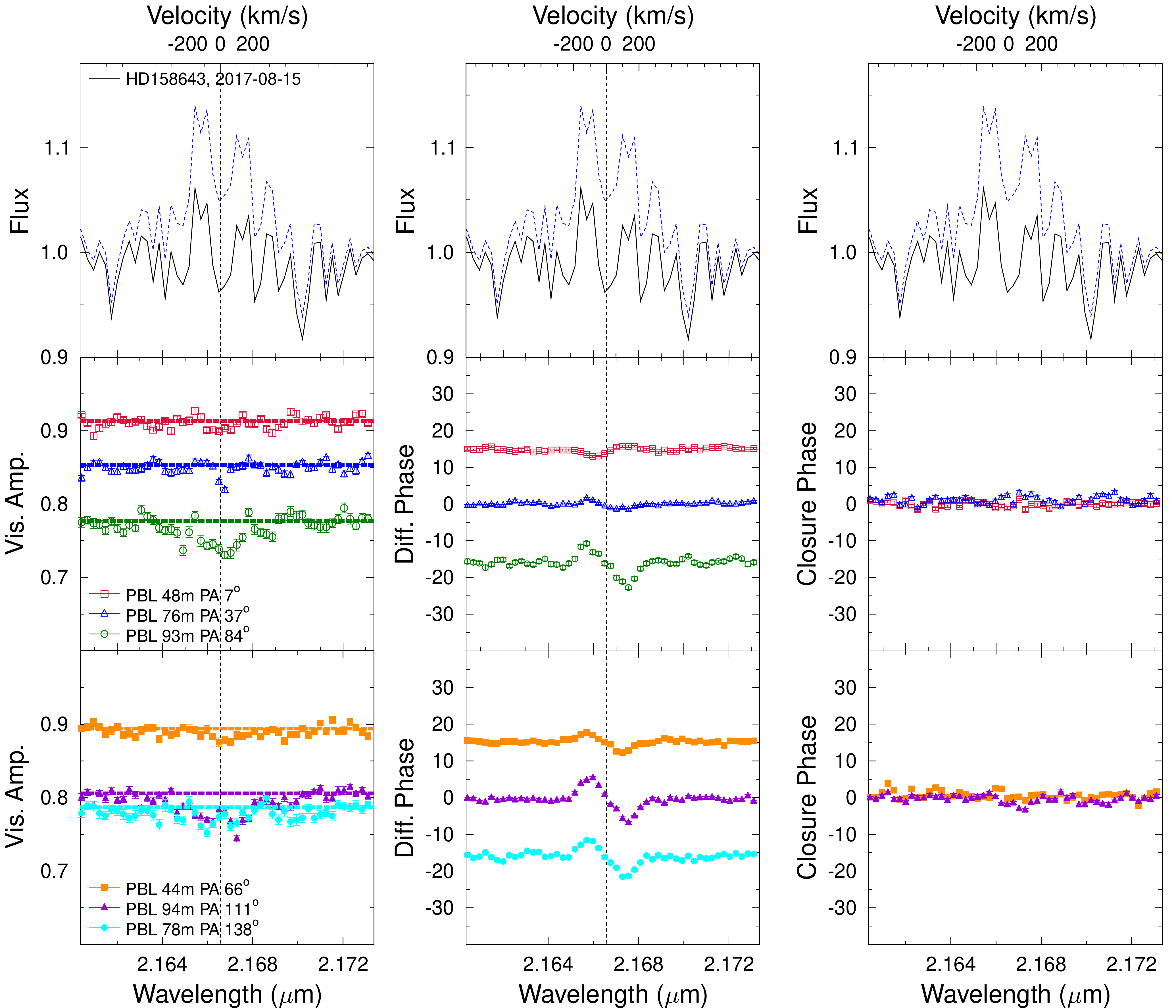}
      \caption{Same as Fig.\,\ref{fig:data1} but for HD\,158643.}
      \label{fig:data17}
\end{figure*}

\begin{figure*}
    \centering
    \includegraphics[width=0.75\textwidth]{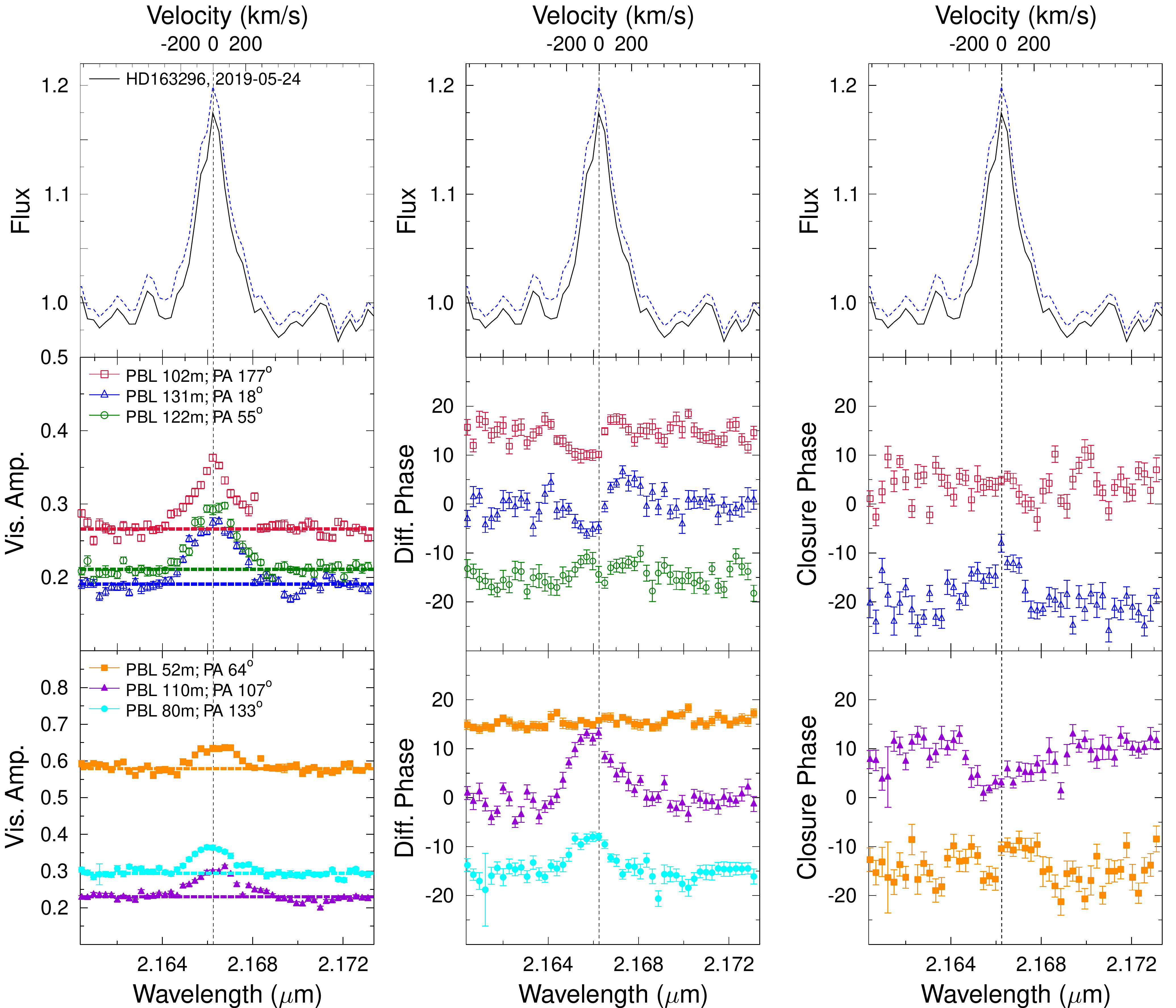}\\
    \includegraphics[width=0.75\textwidth]{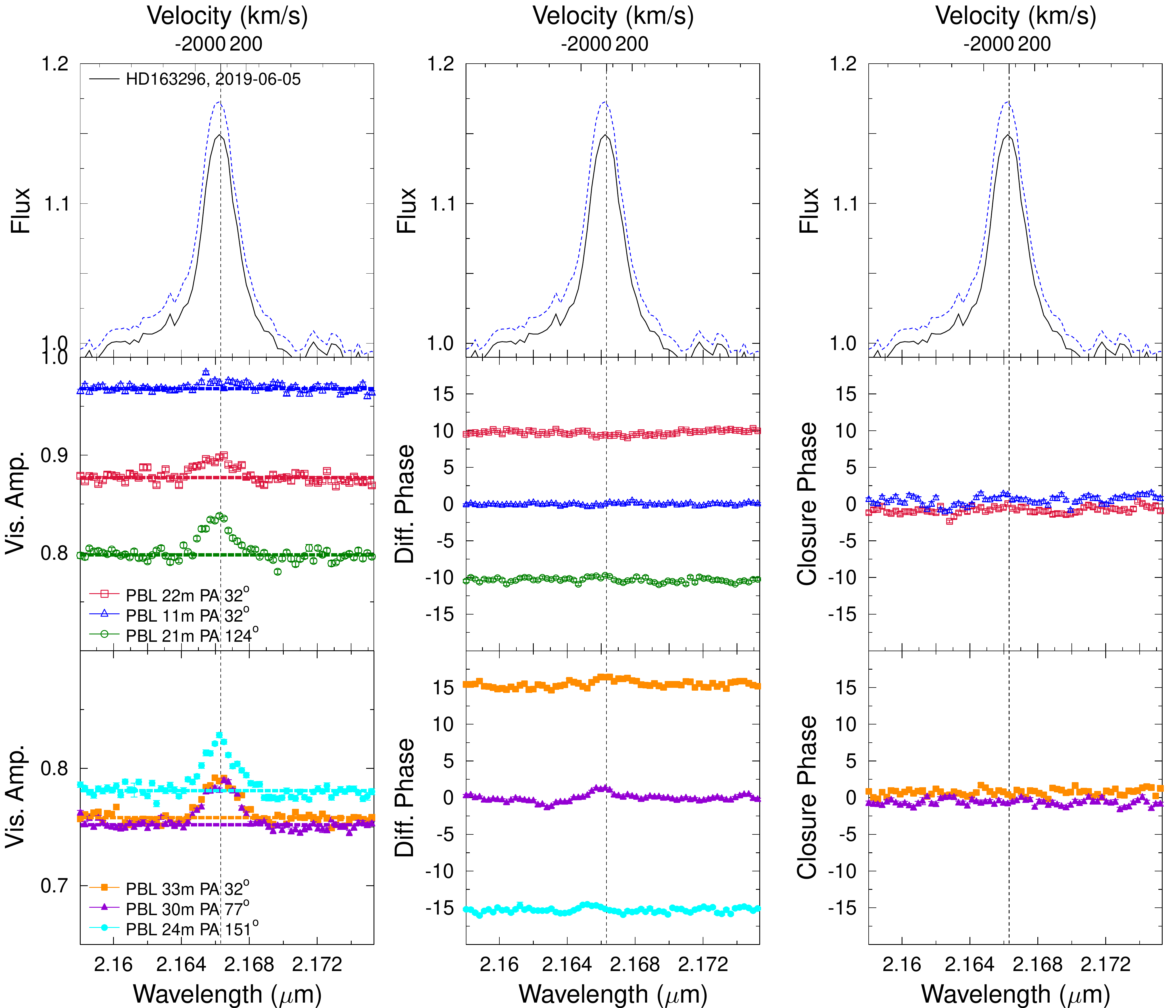}
     \caption{Same as Fig.\,\ref{fig:data1} but for HD\,163296.}
     \label{fig:data18}
\end{figure*}

\begin{figure*}
    \centering
\includegraphics[width=0.75\textwidth]{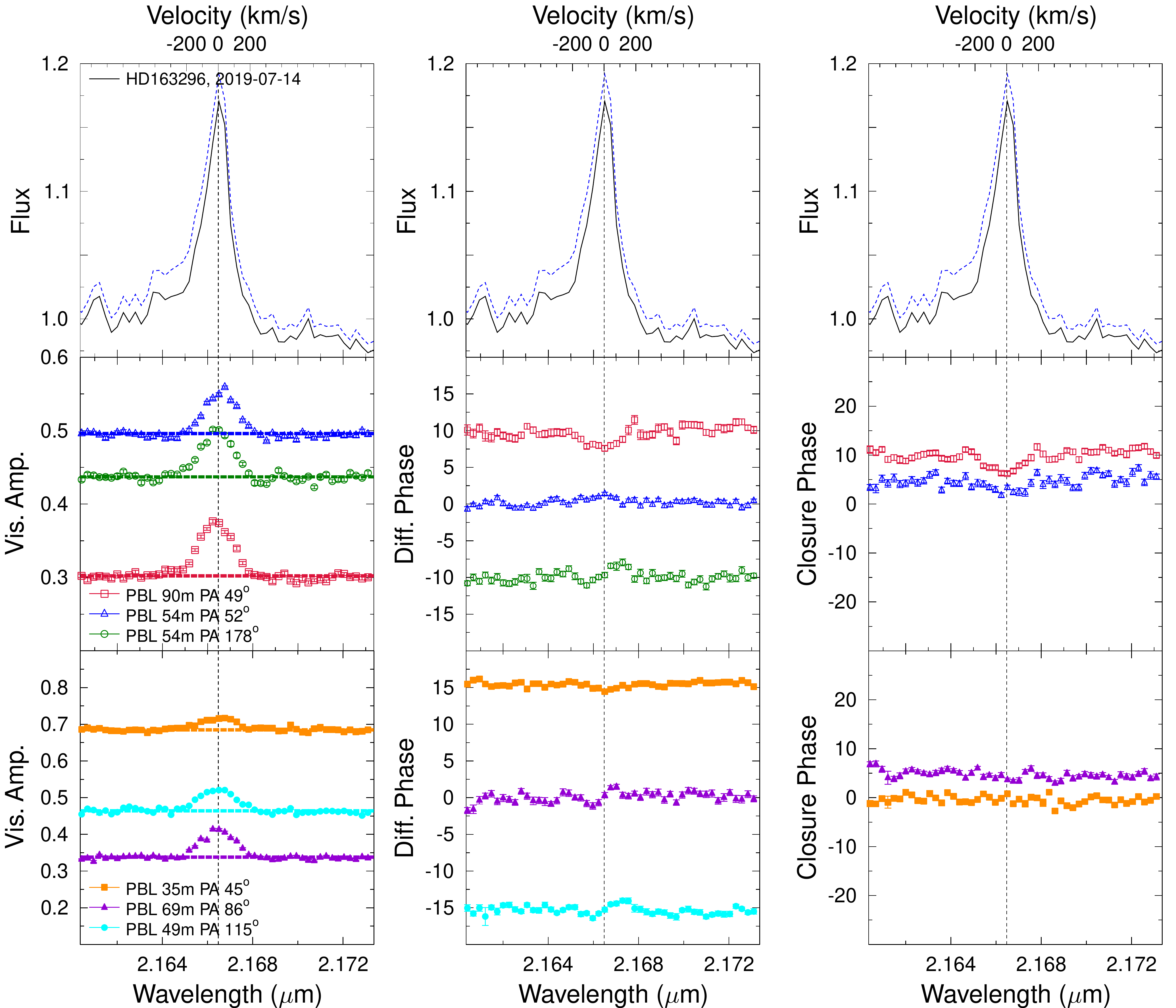}\\
\includegraphics[width=0.75\textwidth]{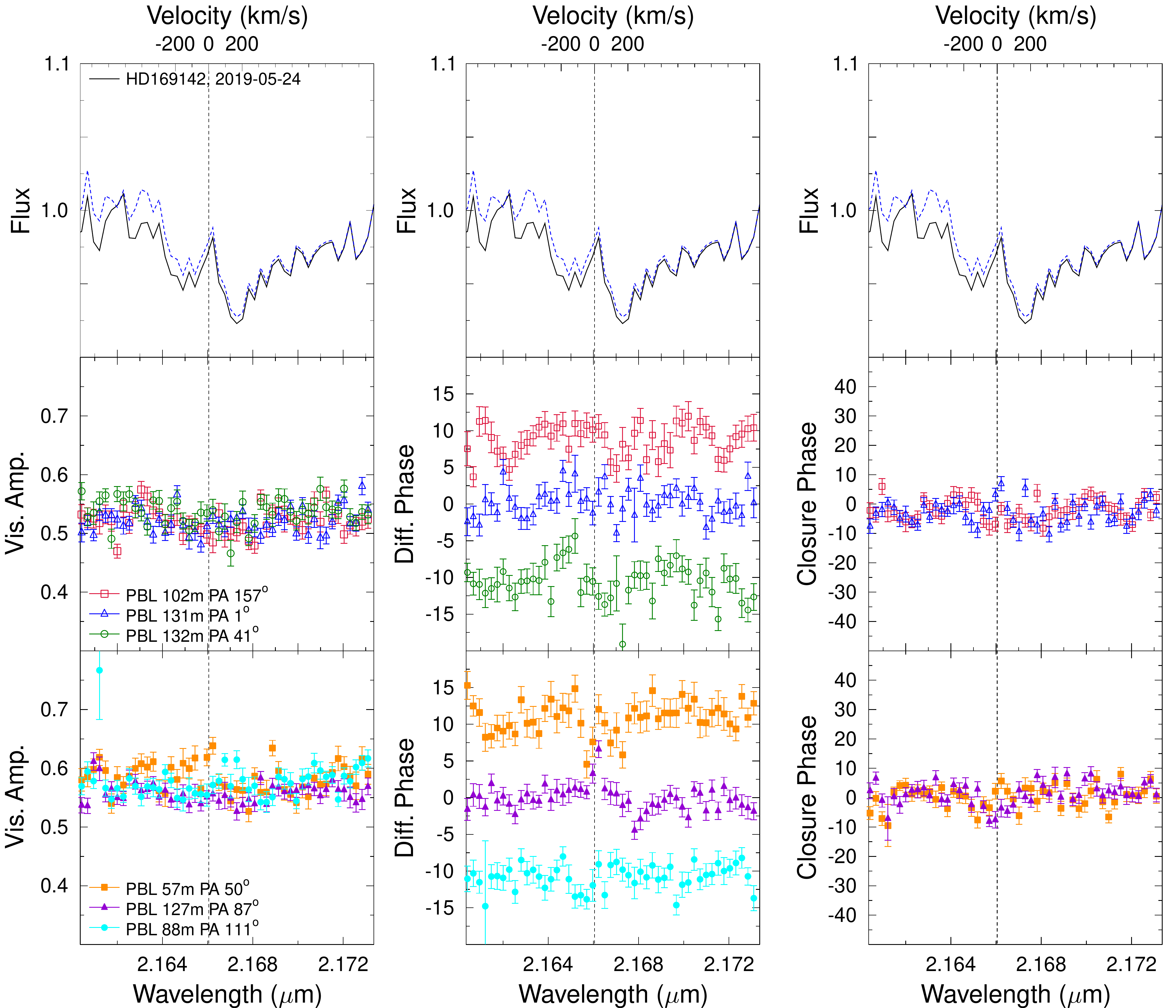}
    \caption{Same as Fig.\,\ref{fig:data1} but for HD\,163296 and HD\,169142.}
    \label{fig:data19}
\end{figure*}

\begin{figure*}
    \centering
    \includegraphics[width=0.75\textwidth]{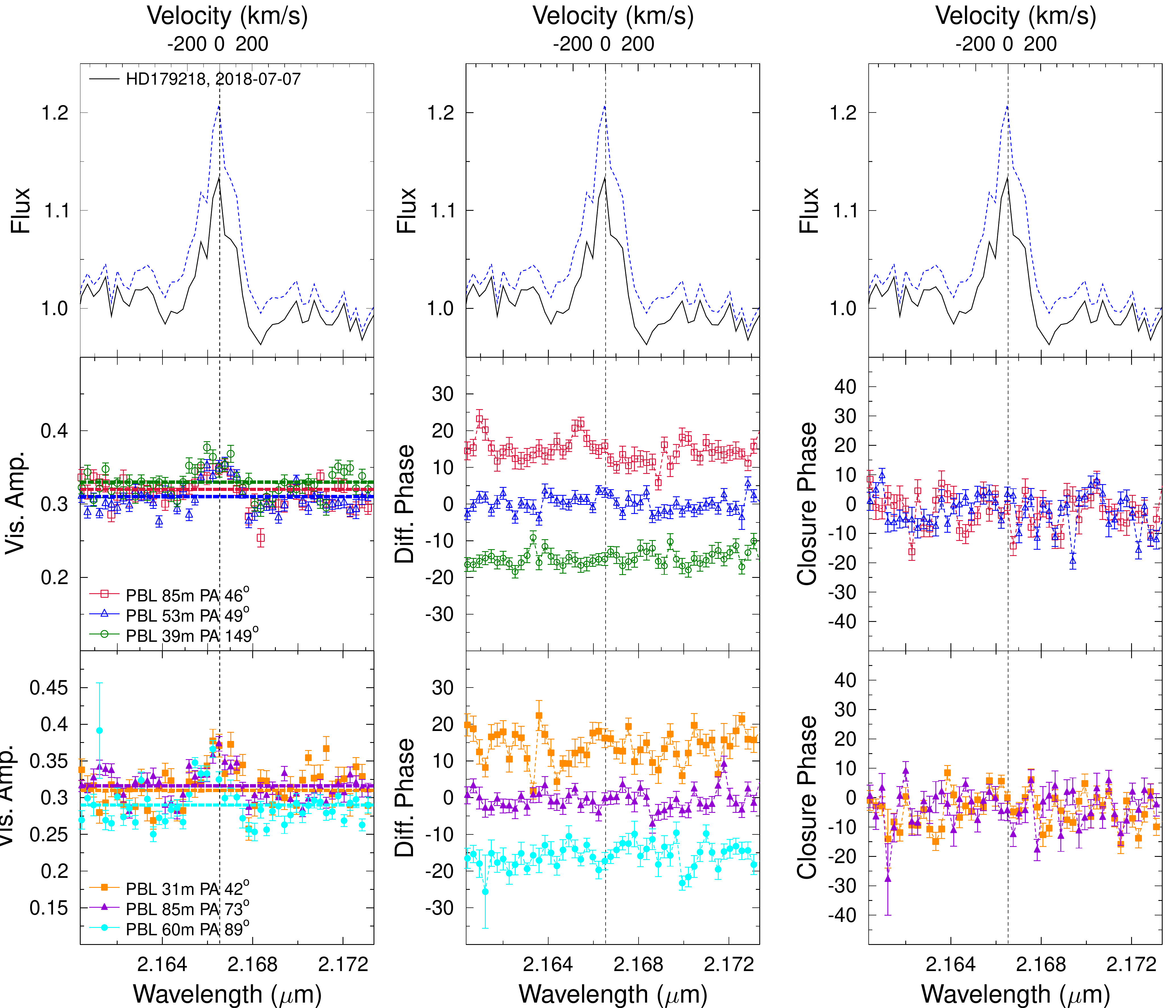}\\
     \includegraphics[width=0.75\textwidth]{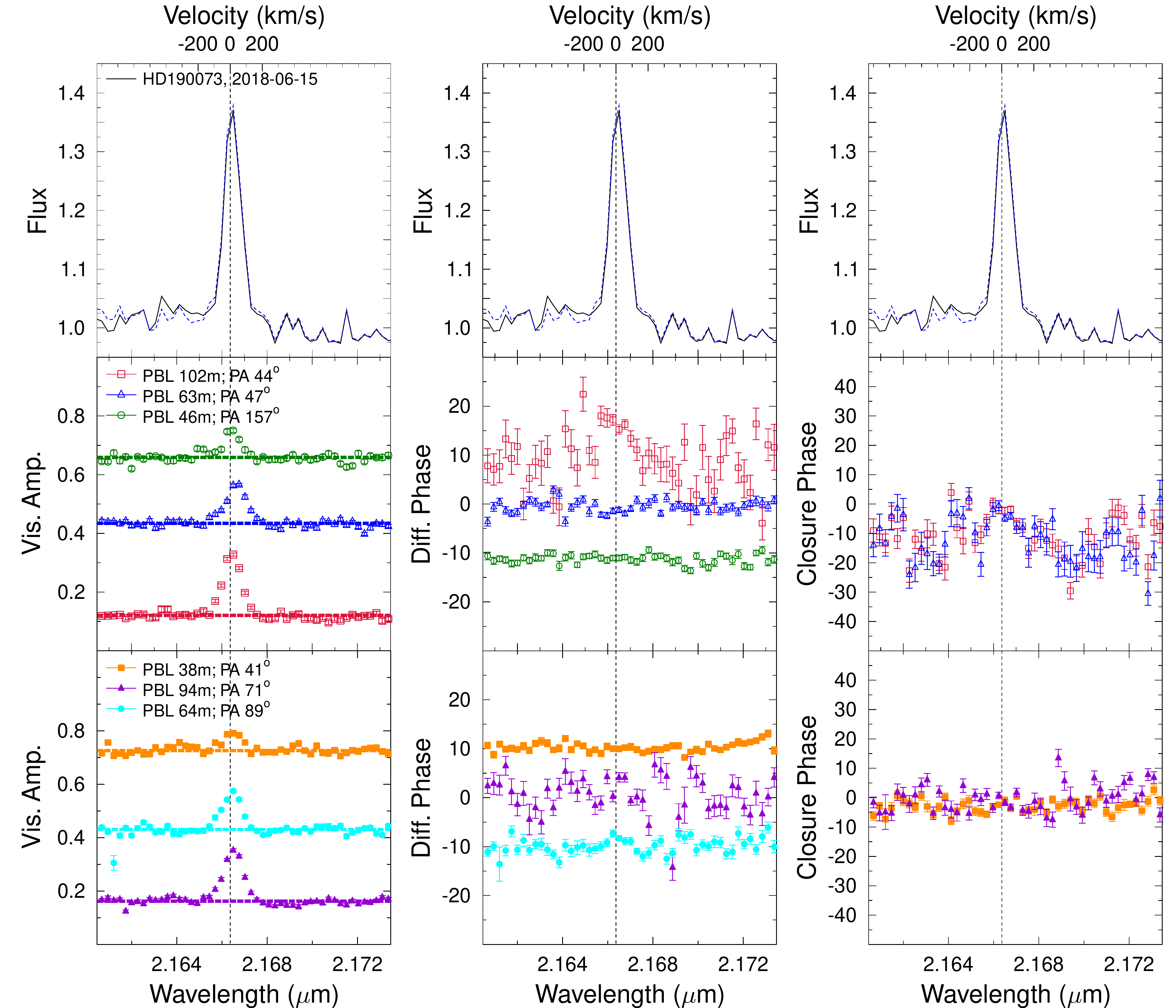}
    \caption{Same as Fig.\,\ref{fig:data1} but for HD\,179218 and HD\,190073.}
    \label{fig:data20}
\end{figure*}

\begin{figure*}
    \centering   
    \includegraphics[width=0.75\textwidth]{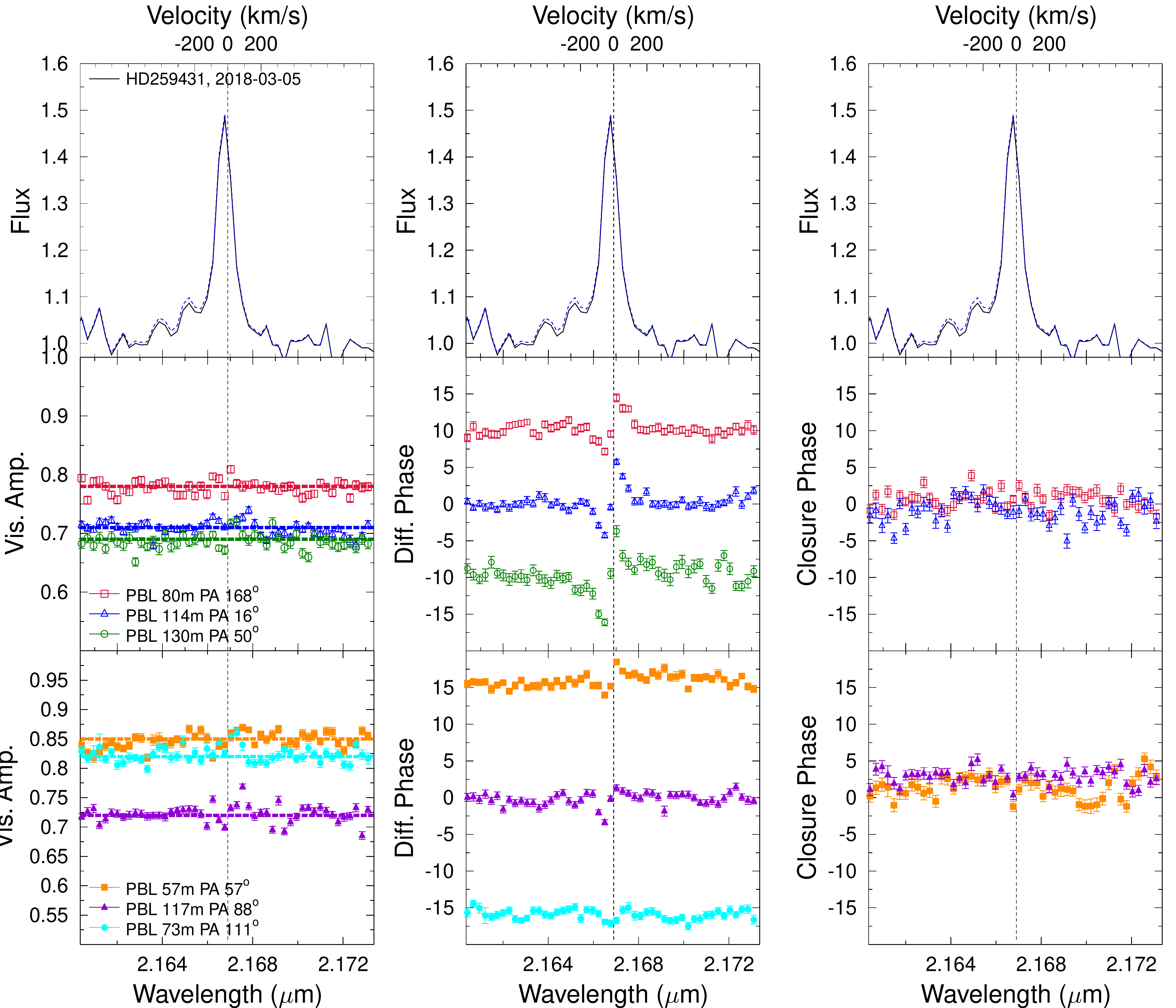}\\
     \includegraphics[width=0.75\textwidth]{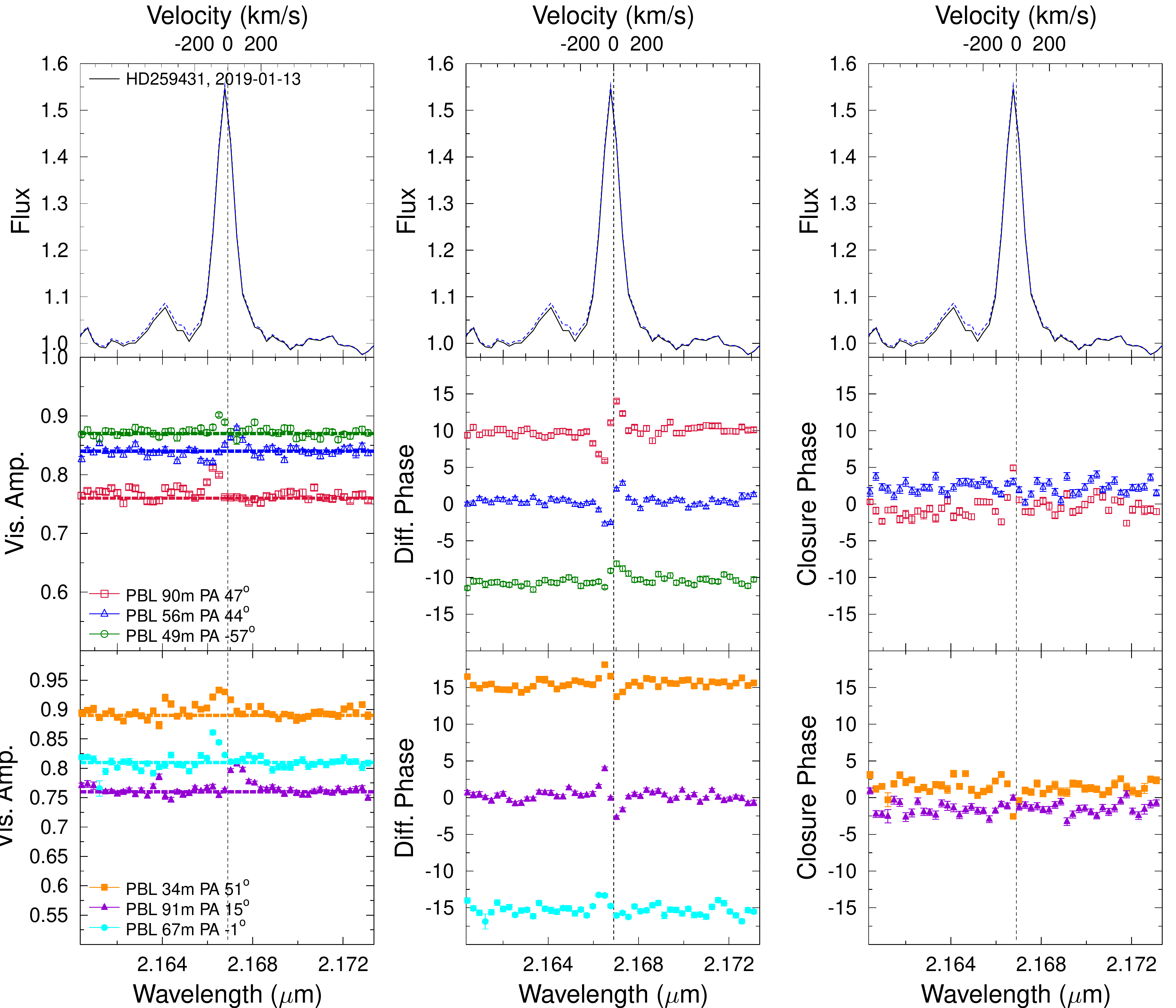}
     \caption{Same as Fig.\,\ref{fig:data1} but for HD\,259431.}
     \label{fig:data21}
\end{figure*}

\newpage
\section{Fits to continuum-subtracted visibilities}

\begin{figure*}
    \centering
    \includegraphics[width=\columnwidth]{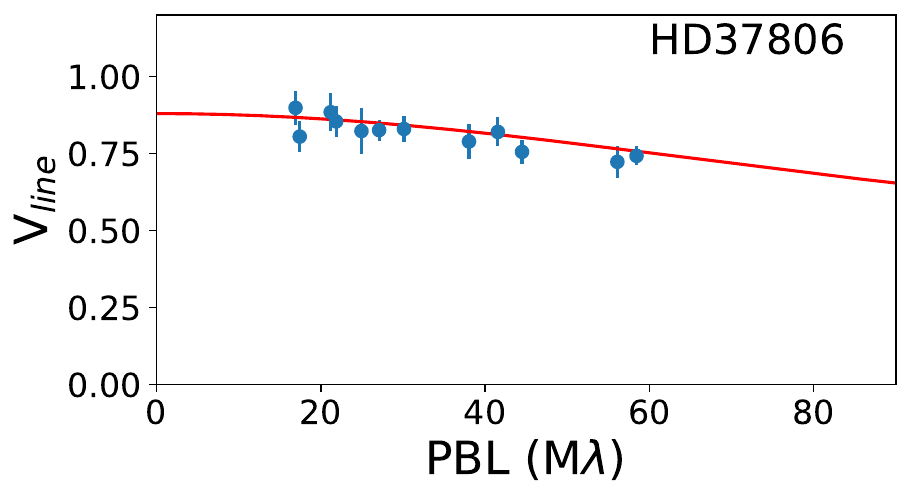}
    \includegraphics[width=\columnwidth]{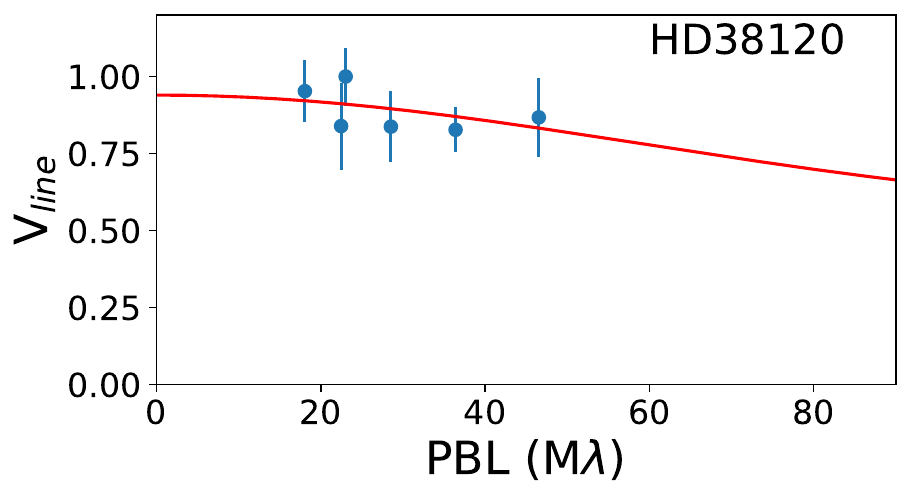}\\
    \includegraphics[width=\columnwidth]{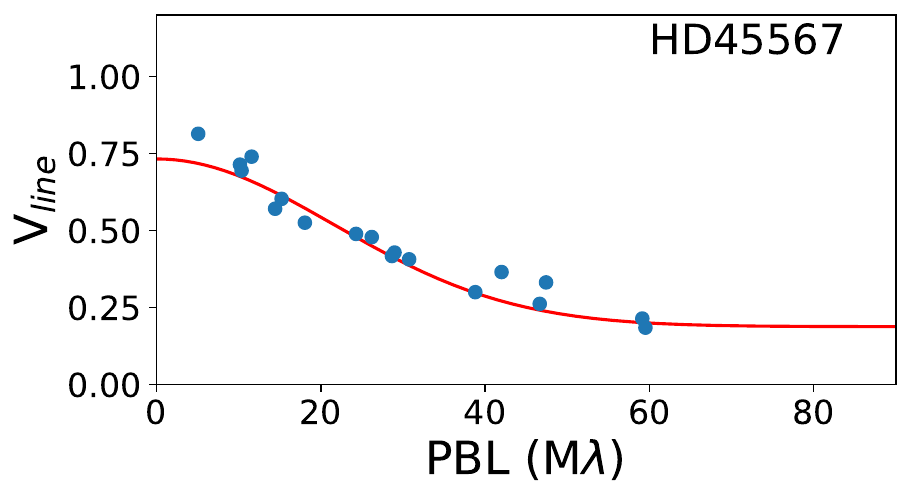}
    \includegraphics[width=\columnwidth]{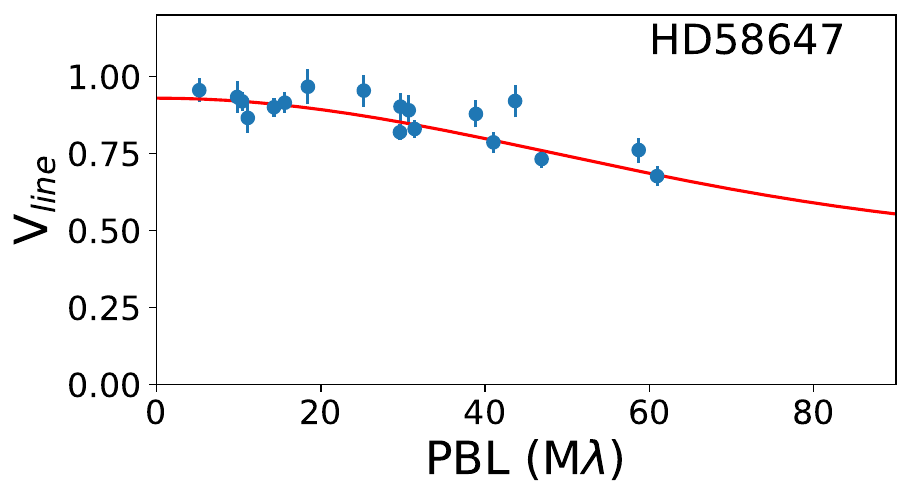}\\
    \includegraphics[width=\columnwidth]{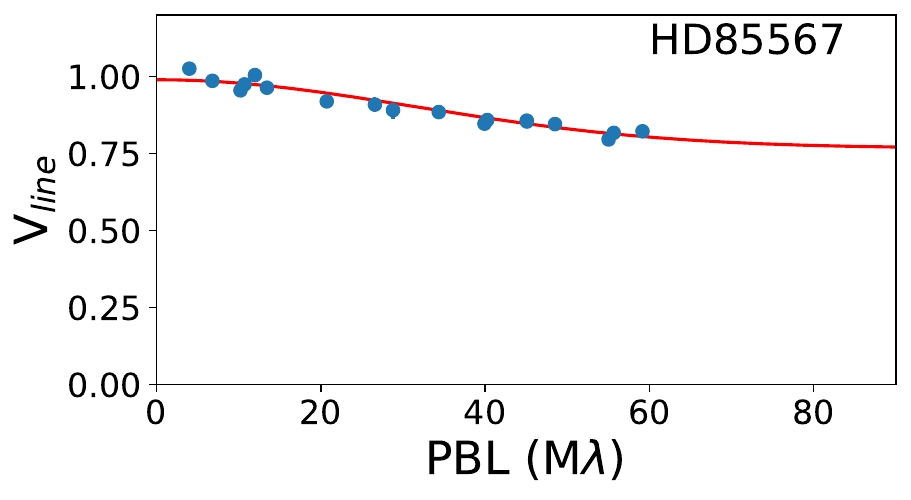}
     \includegraphics[width=\columnwidth]{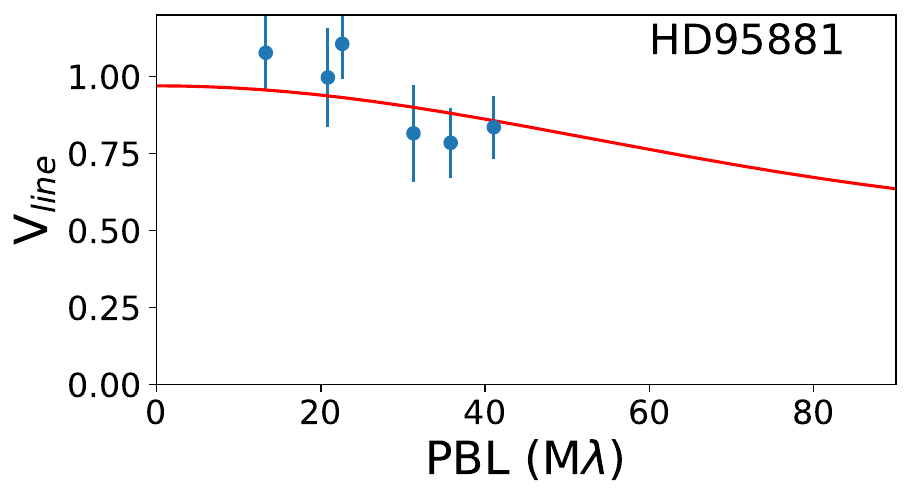}\\
     \includegraphics[width=\columnwidth]{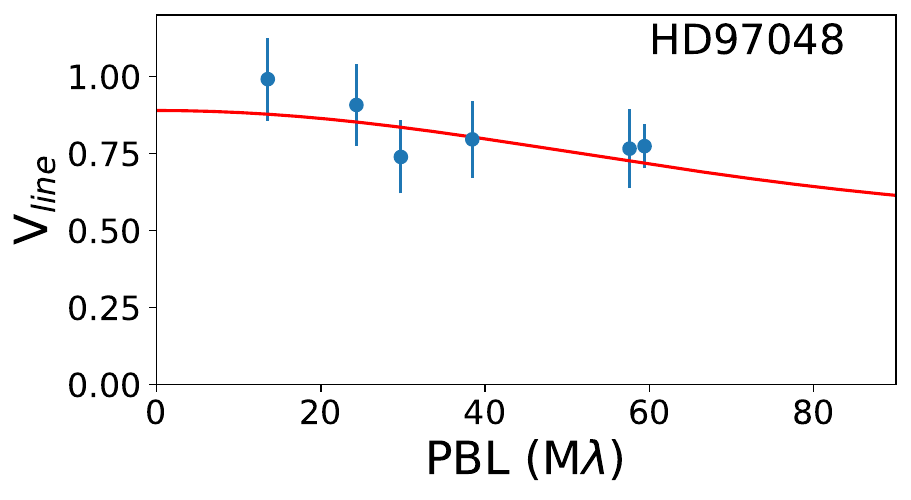}
     \includegraphics[width=\columnwidth]{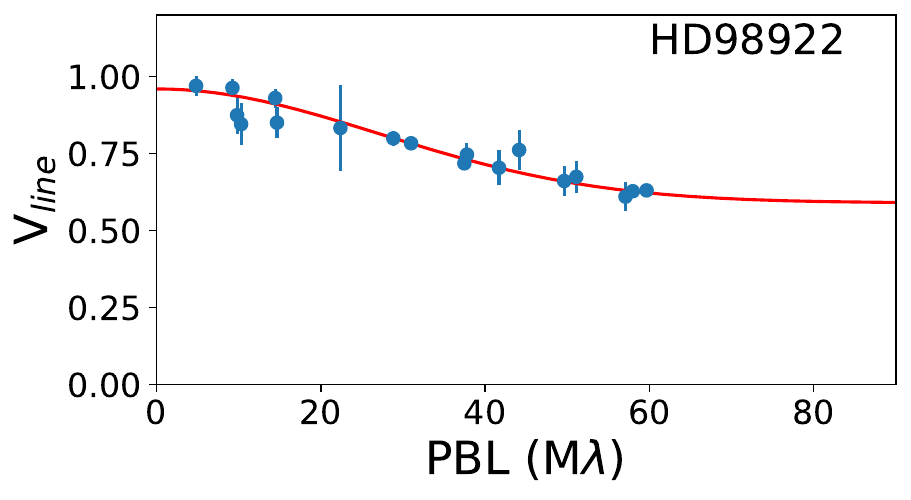}\\
     \sidecaption
     \includegraphics[width=\columnwidth]{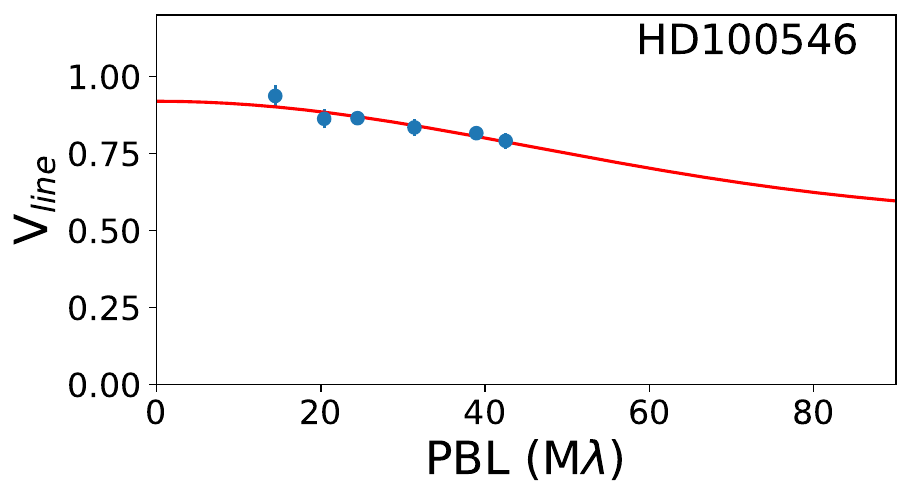}
    \caption{Continuum-subtracted \brg\ line visibilities for HD\,37806, HD\,38120, HD\,45567, HD\,58647, HD\,85567, HD\,95881, HD\,97048, and HD\,98922 as a function of the projected baseline (blue filled points). The best fit to the visibilities is overplotted as a red solid line (see Sect.\,\ref{sect:Brg_vis} and Table\,\ref{tab:results} for details). }
    \label{fig:visibility_fit1}
\end{figure*}

\begin{figure*}
    \centering
     \includegraphics[width=\columnwidth]{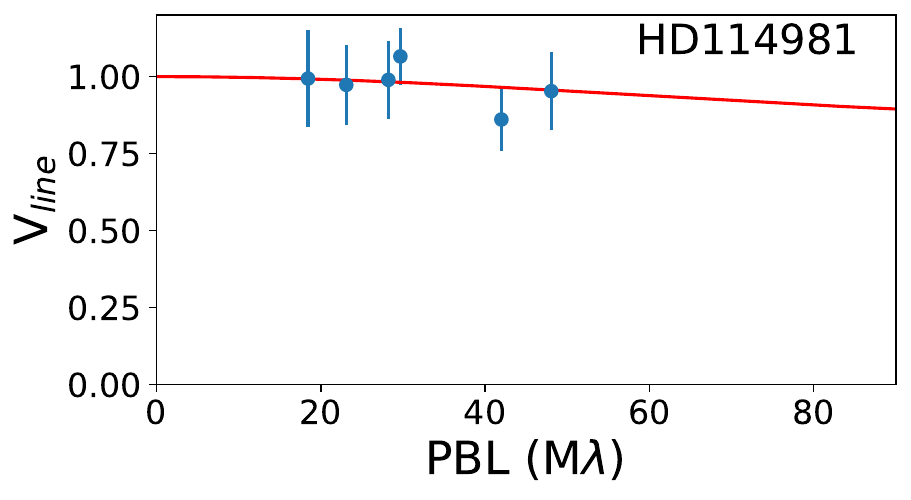}
     \includegraphics[width=\columnwidth]{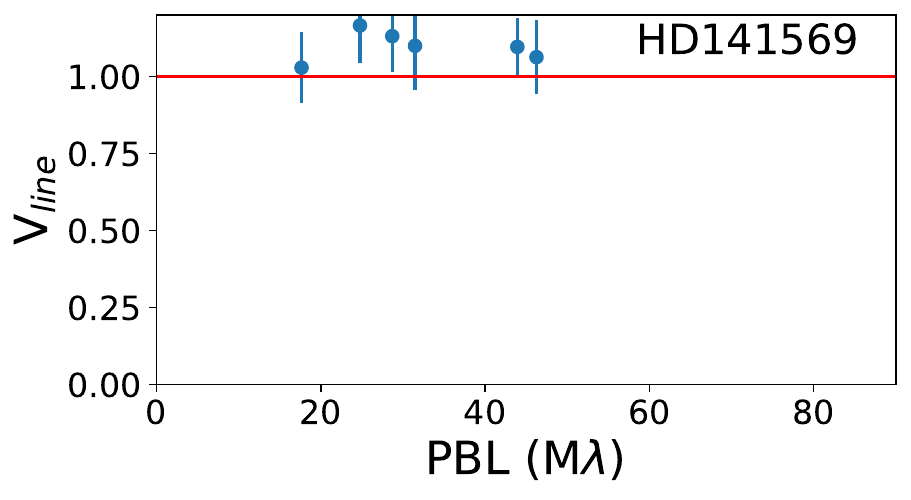}\\
     \includegraphics[width=\columnwidth]{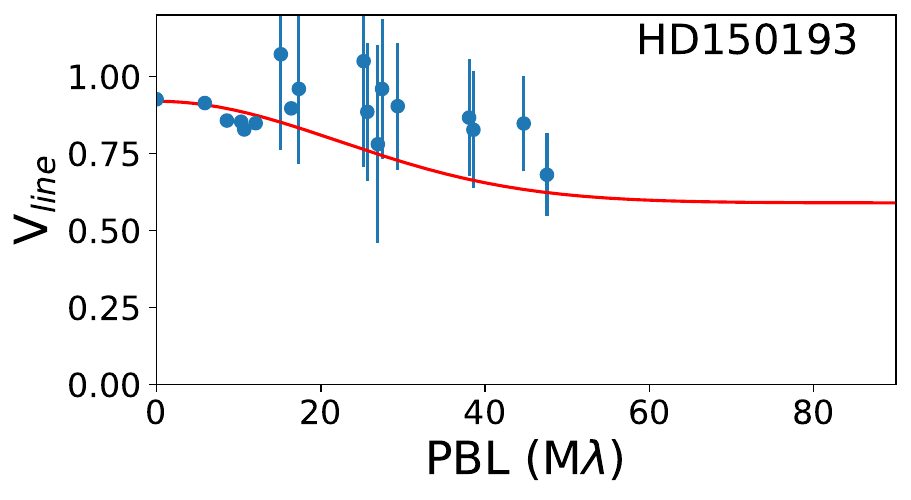}
     \includegraphics[width=\columnwidth]{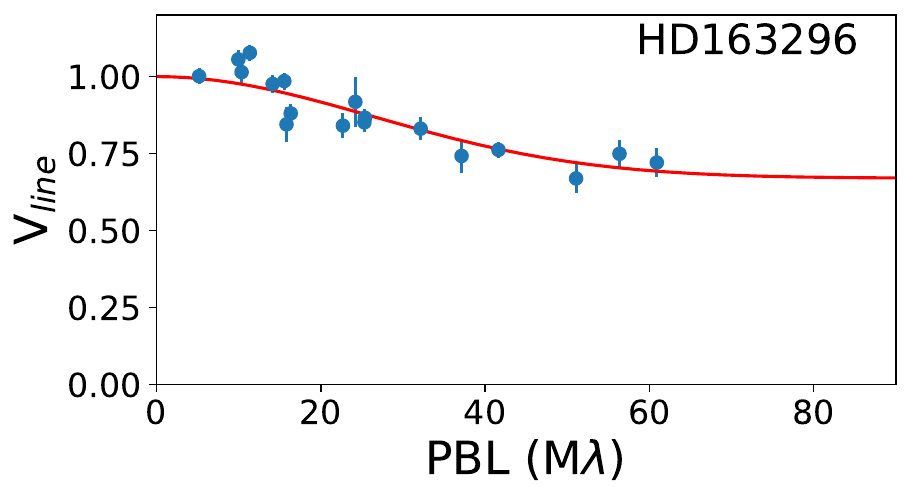}\\
     \includegraphics[width=\columnwidth]{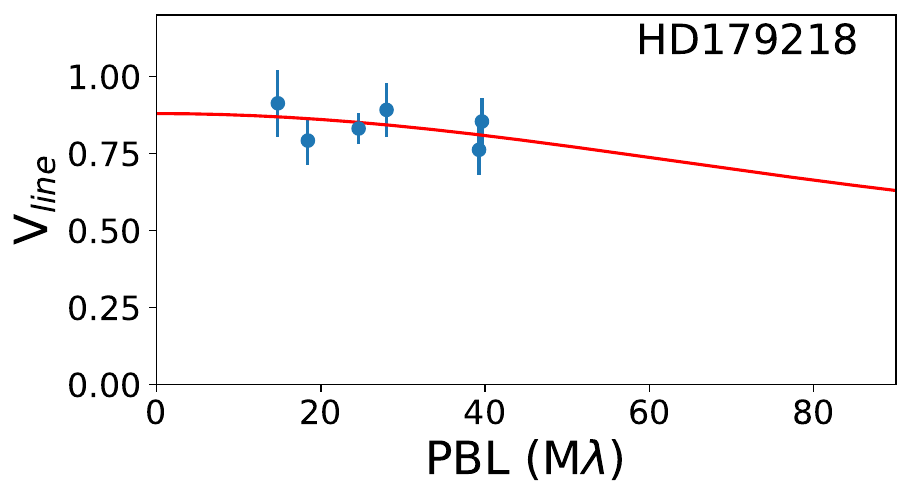}
     \includegraphics[width=\columnwidth]{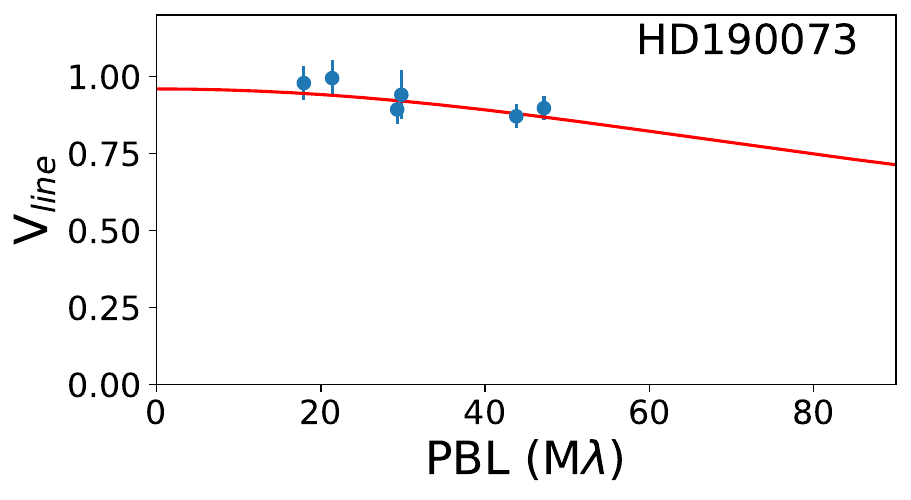}\\
     \sidecaption
     \includegraphics[width=\columnwidth]{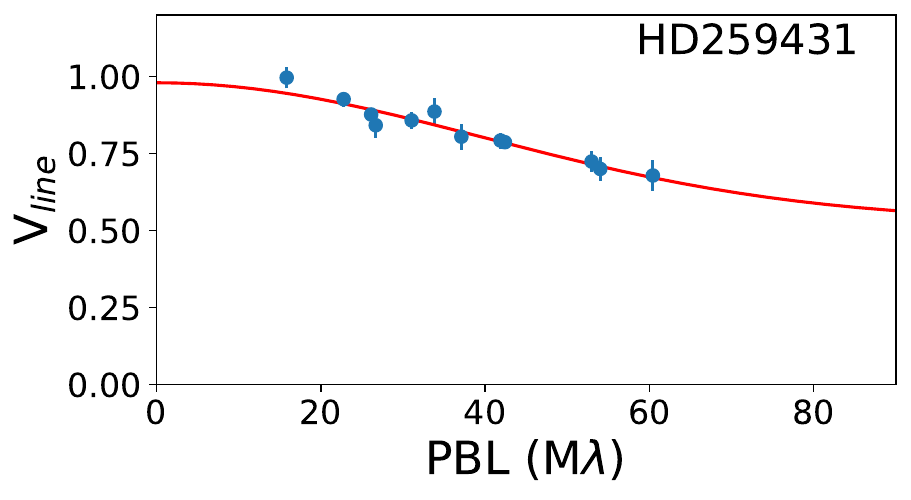}
    \caption{Same as Fig.\,\ref{fig:visibility_fit1} but for HD\,114981, HD\,141569, HD\,150193, HD\,163296, HD\,179218, HD\,190073, and HD\,259431.}
    \label{fig:visibility_fit2}
\end{figure*}

\end{appendix}

\end{document}